\documentclass[11pt,letterpaper]{article}

\usepackage{graphicx, amssymb, amstext, amsmath}
\usepackage{mathrsfs}
\usepackage{bbding}
\usepackage{subfigure}
\usepackage{float}
\usepackage{undertilde}
\usepackage{algorithm} 
\usepackage{algorithmic}
\usepackage{enumitem}
\usepackage{setspace}
\usepackage{color}
\usepackage[in]{fullpage}
\graphicspath{{Figures/}}

\newcommand{\cone}{$\mathscr{C}^1$}

\renewcommand{\newline}{\medskip}

\newcommand{\der}{\ensuremath{\rm d}}
\newcommand{\vect}[1]{\pmb{#1}}

\newcommand{\grad}{\ensuremath{\nabla}}

\newcommand{\austenite}{A}
\newcommand{\mOne}{M$_1$}
\newcommand{\mTwo}{M$_2$}
\newcommand{\mThree}{M$_3$}
\newcommand{\pooz}{[110] }

\newcommand{\lx}{$L_{x_1}$}
\newcommand{\ly}{$L_{x_2}$}
\newcommand{\lz}{$L_{x_3}$}
\begin{document}




\begin{center}
\Large{3D Coupled Thermo-Mechanical Phase-Field Modeling of \\
Shape Memory Alloy Dynamics via Isogeometric Analysis}
\end{center}

\begin{center}
R. Dhote$^{1,3}$, H. Gomez$^2$, R. Melnik$^3$, J. Zu$^1$

$^1$Mechanical and Industrial Engineering, University 
of Toronto, \\5 King's College Road, Toronto, ON, M5S3G8, Canada\\
$^2$Department of Applied Mathematics, University of A Coru\~{n}a \\ Campus de Elvina, s/n. 15192 A Coru\~{n}a, Spain\\
$^3$M$^2$NeT Laboratory, Wilfrid Laurier University, Waterloo, ON,  
N2L3C5, Canada
\end{center}

%
%
%
%

\begin{abstract}
The paper focuses on numerical simulation of the phase-field (PF) equations for modeling martensitic transformations in shape memory alloys (SMAs), their complex microstructures and thermo-mechanical behavior. The PF model is based on the  Landau-Ginzburg potential for the 3D cubic-to-tetragonal phase transformations in SMAs. The treatment of domain walls as diffuse interfaces, leads to a fourth-order differential equation in a strain-based order parameter PF model. The fourth-order equations introduce a number of unexplored numerical challenges because traditional numerical schemes have been primarily applied to second-order problems. We propose isogeometric analysis (IGA) as a numerical formulation for a straightforward solution to the fourth-order differential PF equations using continuously differentiable non-uniform rational B-splines (NURBS). We present microstructure evolution in different geometries of SMA nanostructures under temperature-induced phase transformations to illustrate the geometrical flexibility, accuracy and robustness of our approach. The simulations successfully capture the dynamic thermo-mechanical behavior of SMAs observed experimentally. 
\end{abstract}

Isogeometric analysis, phase-field model, Ginzburg-Landau theory, nonlinear thermo-elasticity



%


\section{Introduction} \label{sec:Introduction}

As a result of their interesting solid-to-solid phase transformations and coupled-physics (thermo-mechanical, magnetostrictive) properties, shape memory alloys (SMAs) have been used as micro- and nano- actuators and sensors for a broad spectrum of applications. Recently, there has been a major research focus on using SMA nanostructures \cite{vzuvzek2012electrochemical,bhattacharya2005material,koig2010micro,bayer2011carbon,clements2003wireless,san2008superelasticity,san2009nanoscale} for nanoelectromechanical (NEMS) and microelectromechanical systems (MEMS) and biomedical applications. These applications involve designing different geometries and using domain patterns for controlling distortions \cite{bhattacharya2005material}. All of these motivate the need for understanding domain patterns and their thermo-mechanical properties in realistic and complex geometries for better application development.

Several modeling approaches have been used to study the SMA behaviors \cite{Khandelwal2009,mamivand2013review,auricchio2009macroscopic,auricchio2010shape,auricchio2011three,sengupta2012multiscale}. In particular, phase-field (PF) models have been widely used to study the phase transformations in SMA meso- and nano- structures \cite{Ahluwalia2006,Bouville2008,Idesman2008,Bouville2009}. Broadly, PF models for SMAs can be divided into two approaches: the kinetic model using independent order parameter(s) (OPs) (see, for example, \cite{wang1997three,Levitas2002a}) and the strain-based OP PF models (e.g., \cite{Ahluwalia2006,Barsch1984}). The first approach often leads to a second-order differential equation for microstructure evolution, while the second approach typically leads to a fourth-order differential equation in space. 



Here, we focus on the second approach and use the PF methodology. Several 3D PF models for SMAs have been proposed in the literature. The majority of these models do not account for the dynamics of SMAs, but only relax the quasi-static microstructures using a dissipation potential or directly assume a quasi-static response. Moreover, most models assume isothermal conditions, which neglects the thermo-mechanical coupling of SMAs, a significant modeling limitation. 
The nucleation and growth of martensitic transformations have been widely studied by using the kinetic time-dependent Ginzburg-Landau models \cite{wang1997three,Levitas2002a,Levitas2002b,Levitas2003,ni2003three,seol2003computer,Idesman2008,man2011study,yeddu2012three}. Using the strain-based OP PF models, the temperature- and stress- induced phase transformations have been studied for SMAs \cite{jacobs2003simulations,Ahluwalia2006}. The full 3D dynamic model in its generality was first formulated by Melnik et al. \cite{melnik2000computing} and the first model-based explanation of thermally-induced hysteresis was discussed in \cite{melnik1999modeling,melnik2001coupled}. From a computational perspective, most of the above studies used traditional numerical methods, such as spectral collocation or the finite difference method. These algorithms typically lack geometrical flexibility, as the majority of the above studies were performed on a cubic domain with periodic boundary conditions. However, complex geometries exist in real life, and there is a need for more flexible methods which can allow to model geometrically complex and large domains with different boundary conditions. 
When geometrical flexibility is needed, the finite element method is the natural choice. However, if we do not want to include additional variables, solving fourth-order equations with the finite element method requires globally smooth basis functions, and this has proved very difficult to achieve with traditional finite elements. Due to its geometrical flexibility and the possibility of generating globally smooth basis functions, we propose isogeometric analysis (IGA) as an effective numerical method to solve the fourth-order PF model on non-trivial geometries. 

IGA is a new computational method originally developed to avoid mesh generation bottlenecks during engineering analysis \cite{hughes2005isogeometric,Hughes}. It was originally developed using non-uniform rational B-splines (NURBS), a backbone of CAD and animation technology, as basis functions, but it was later extended to accommodate other widely-used functions in the CAD community, such as, for example, T-Splines \cite{TS1,TS2,TS3,TS4}. IGA has been successfully applied to problems of fluid mechanics \cite{Evans2013141,Liu2013321,Liu201347,Bazilevs2007173}, solid mechanics \cite{Cottrell20065257,Cottrell20074160,Hughes20084104,Auricchio2007160,NME:NME3159,laura}, fluid-structure interaction \cite{fsi1,fsi2}, and condensed-matter physics \cite{cmp1,cmp2,cmp3,cmp4}. The use of rich basis functions provides IGA with a unique capability to model geometry exactly, in many instances, while field variables can be approximated with enhanced accuracy \cite{Evans20091726,Schillinger2013170}. IGA provides unique attributes of higher-order accuracy and robustness with the \cone- or higher-order continuity necessary for solving higher-order differential equations in a variational formulation. IGA has been successfully used to solve the PF theories and higher-order differential equations using Galerkin variational formulations \cite{borden2012phase,dede2012isogeometric,vilanovaisogeometric,ho1,ho2,ho3,ho4,anders2012computational,anders2012isogeometric}. Additionally, it has been recently shown by Gomez et al. \cite{Gomez2014} that the possibility of generating highly-smooth basis functions also permits deriving collocation methods that approximate directly the strong form of the equations, an approach that is not pursued in this work. 

We recently illustrated the flexibility of the IGA approach by applying it to a 2D PF model for SMAs \cite{dhote2013PCS,Dhote2013CM}. Here, we solve a three-dimensional theory for cubic-to-tetragonal phase transformations in nanostructured SMAs using IGA. The coupled equations of nonlinear thermoelasticity are developed using the PF model and the Ginzburg-Landau theory. The governing laws are introduced in the IGA framework using a variational formulation. Several numerical studies have been performed to illustrate the flexibility, accuracy and stability of the approach. Based on the above tasks, the paper is organized as follows. In Section \ref{sec:PF3DGL}, the governing coupled equations of nonlinear thermoelasticity and solid-solid phase transformations are presented. The details of the numerical implementation of the SMA governing equations in the IGA framework are given in Section \ref{sec:NumericalFormulation}. The developed methodology is exemplified with 3D numerical simulations on nanostructured SMA domains subjected to thermally-induced phase transformations in Section \ref{sec:NumericalSimulations}. Finally, the conclusions are given in Section \ref{sec:Conclusions}. 

\section{Mathematical Model of SMA Dynamics} \label{sec:PF3DGL}

\begin{figure}[H]
\centering
\subfigure[]
{
\includegraphics[width=0.3\linewidth]{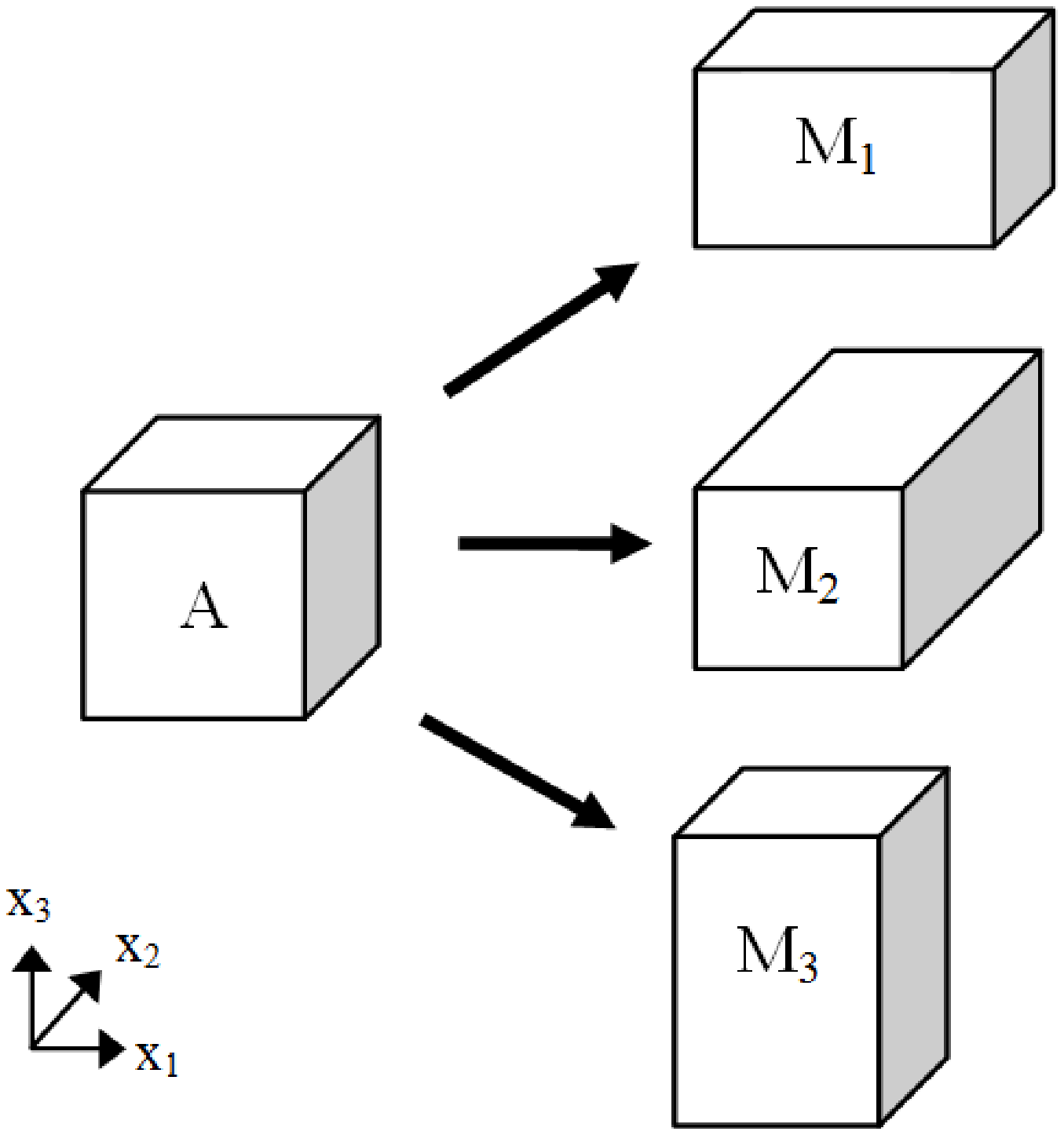}
\label{fig:Cubic2Tet1}
}
\subfigure[]
{
\includegraphics[width=0.4\textwidth]{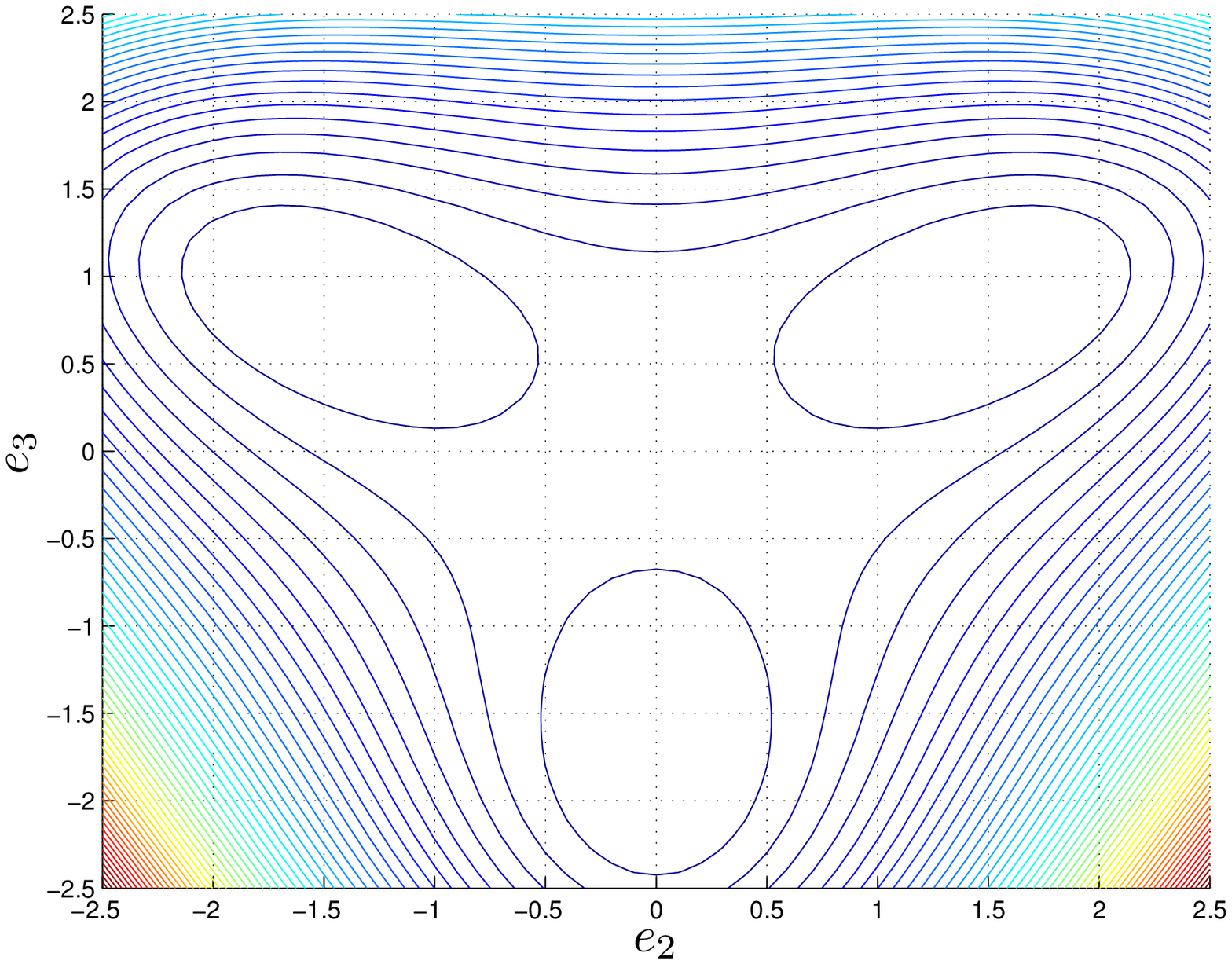}
\label{fig:ThreeDFunctional}
}
\caption{Cubic-to-tetragonal phase transformations (a) schematic of microstructures: austenite (\austenite), and martensite variants (\mOne, \mTwo, \mThree) (b) free energy function plot at $ \tau  = -1.2 $ (see Eq. (\ref{homog_free})).}
\label{fig:CubicToTetragonal}
\end{figure}

The cubic-to-tetragonal phase transformations occur in SMA alloys like NiAl, FePd or InTl. The cubic austenite phase is converted into tetragonal martensitic variants upon mechanical or thermal loadings as schematically shown in Fig. \ref{fig:Cubic2Tet1}. 

We have recently put forward a mathematical model for the 3D coupled thermo-mechanics of SMAs \cite{RPD_physics_paper}. Our model can be derived from a free-energy functional using Hamiltonian mechanics. The unknowns are the displacement field $\vect u=\{u_1,u_2, u_3\}^T$ and the temperature $\theta$. We assume that the problem takes place on the physical domain $\Omega\subset\mathbb{R}^3$, which is an open set parameterized by Cartesian coordinates $\vect x=\{x_1,x_2,x_3\}^T$. We will make use of the Cauchy-Lagrange infinitesimal strain tensor $\vect\epsilon=\{\epsilon_{ij}\}$, whose components are defined as $\epsilon_{ij} = \left( u_{i,j} +  u_{j,i} \right)/2,\, i,j\in\{1,2,3\}$, where an inferior comma denotes partial differentiation (e.g., $u_{i,j}=\partial u_i/\partial x_j$). Using the strain tensor, we define the strain measures $e_i$, for $i=1,\dots,6$ as follows:
\begin{equation}
\left\{\begin{matrix}
e_1 \\
e_2 \\
e_3 \\
e_4 \\
e_5 \\
e_6 \\
\end{matrix}\right\} = \left[\begin{array}{c|c}
\mathbb{D}_3 \hspace{.05cm}& \hspace{.05cm}\mathbb{O}_3 \\ [.1cm]
\hline \\[-.3cm]
\mathbb{O}_3 & \mathbb{I}_3
\end{array}\right] \left\{\begin{matrix}
\epsilon_{11} \\
\epsilon_{22} \\
\epsilon_{33} \\
\epsilon_{23} \\
\epsilon_{13} \\
\epsilon_{12} \\
\end{matrix}\right\}, 
\end{equation}
where $\mathbb{D}_3$, $\mathbb{O}_3$, $\mathbb{I}_3$ are $3\times3$ constant matrices. In particular,
\begin{equation}
\mathbb{D}_3=\left[\begin{matrix}
1/\sqrt{3} &  1/\sqrt{3} &  1/\sqrt{3} \\
1/\sqrt{2} & -1/\sqrt{2} &  0          \\
1/\sqrt{6} & -1/\sqrt{6} & -2/\sqrt{6}  
\end{matrix}\right],
\end{equation}
while $\mathbb{I}_3$, and $\mathbb{O}_3$ are, respectively, the $3\times3$ identity and zeros matrices. For future reference, we call $e_1$ hydrostatic strain, $e_2$ and $e_3$ deviatoric strains, and $e_4,e_5,e_6$ shear strains. The deviatoric strains are selected as the OPs to describe different phases in the domain. The free-energy functional $ \mathscr{F} $ for the cubic-to-tetragonal phase transformation is selected as \cite{Barsch1984,Ahluwalia2006}: 

\begin{equation}
\mathscr{F}[\vect u] = 
\int_{\Omega} \left[F_0(e_1,\dots,e_6,\theta) +\frac{k_g}{2} \left( |\grad e_2|^2 + |\grad e_3|^2 \right) \right] \der\Omega,
\label{eq:FEcub2tet}
\end{equation}
where $F_0$ is defined as,
\begin{equation}
\label{homog_free}
F_0(e_1,\dots,e_6,\theta)= \frac{a_1}{2} e_1^2 + \frac{a_2}{2} \left(e_4^2 + e_5^2 +e_6^2\right) + a_3 \tau  \left(e_2^2 + e_3^2\right) + a_4 e_3 \left(e_3^2 - 3 e_2^2\right)  + a_5 (e_2^2 + e_3^2)^2.
\end{equation}
Here $ a_i $, $i\in\{1,\dots,5\}$ are constants that define the mechanical properties of the material, $ k_g $ is the gradient energy coefficient, $ \tau $ is the dimensionless temperature defined as \mbox{$ \tau = \displaystyle (\theta - \theta_m)/(\theta_0 - \theta_m) $}, where $ \theta_0 $ and $ \theta_m $ are the material properties specifying the transformation start temperature and the temperature at which austenite becomes unstable, and $|\cdot|$ denotes the Euclidean norm of a vector. Using this notation, and the repeated-indices summation convention, our model can be written as
\begin{eqnarray}
& & \rho \ddot{u}_{i}=\sigma_{ij,j}+\eta\sigma'_{ij,j}+\mu_{ij,kkj}+f_i, \\
& & C_v\dot{\theta}=\kappa\theta_{,ii}+ \Xi\theta\left(u_{i,i}\dot{u}_{j,j}-3u_{i,i}\dot{u}_{i,i}\right) + g.
\end{eqnarray}
where a dot over a function denotes partial differentiation with respect to time, and $\rho$, $\eta$, $C_v$, $\kappa$, and $\Xi$ are positive constants that represent, respectively, the density, viscous dissipation, specific heat, thermal conductance coefficient, and strength of the thermo-mechanical coupling. The symmetric stress tensor $\vect \sigma=\{\sigma_{ij}\}$ is a nonlinear function of the strain measures $e_i$, and the temperature. In particular, 
\begin{subequations}
\renewcommand{\theequation}{\theparentequation.\arabic{equation}}
\label{eq:stresscomponents}
\begin{eqnarray}
\sigma_{11} &=& \!\!\!\frac{a_1 e_1}{\sqrt{3}}
+ \frac{e_2}{\sqrt{2}} \left[ 2 \tau a_3 - 6 a_4 e_3 + 4 a_5 (e_2^2+e_3^2)  \right] + \frac{1}{\sqrt{6}} \left[ e_3(2 \tau a_3 +  4 a_5 (e_2^2+e_3^2)) + 3 a_4 (e_3^2-e_2^2)  \right], \nonumber \\
&& \\
\sigma_{12} &=& \!\!\!\sigma_{21} = \frac{1}{2} a_2 e_6, \\
\sigma_{13} &=& \!\!\!\sigma_{31} = \frac{1}{2} a_2 e_5, \\
\sigma_{22} &=& \!\!\!\frac{a_1 e_1}{\sqrt{3}} - \frac{e_2}{\sqrt{2}} \left[ 2 \tau a_3 - 6 a_4 e_3 + 4 a_5 (e_2^2+e_3^2)  \right] + \frac{1}{\sqrt{6}} \left[ e_3(2 \tau a_3 +  4 a_5 (e_2^2+e_3^2)) + 3 a_4 (e_3^2-e_2^2)  \right], \nonumber \\
&& \\
\sigma_{23} &=& \!\!\!\sigma_{32} = \frac{1}{2} a_2 e_4, \\
\sigma_{33} &=& \!\!\!\frac{1}{\sqrt{3}} a_1 e_1
- \frac{2}{\sqrt{6}} \left[ 2 \tau a_3 e_3 +  3 a_4 (e_3^2-e_2^2) + 4 a_5 e_3 (e_2^2+e_3^2)   \right].
\end{eqnarray}
\end{subequations}
We call $\vect\sigma'=\{\sigma'_{ij}\}$ dissipational stress tensor, which is a linear function of the strain rates $\dot{e}_i$, $i=1,\dots,6$. Since $\vect\sigma'$ is a second-rank symmetric tensor, it may be defined by giving only six of its entries. Thus, we define
\begin{equation}
\left\{\begin{matrix}
\sigma'_{11} \\
\sigma'_{22} \\
\sigma'_{33} \\
\sigma'_{23} \\
\sigma'_{13} \\
\sigma'_{12} \\
\end{matrix}\right\} = \left[\begin{array}{c|c}
\mathbb{D}_3^T \hspace{.05cm}& \hspace{.05cm}\mathbb{O}_3 \\ [.1cm]
\hline \\[-.3cm]
\mathbb{O}_3 & \frac{1}{2}\mathbb{I}_3 
\end{array}\right]  \left\{\begin{matrix}
\dot{e}_{1} \\
\dot{e}_{2} \\
\dot{e}_{3} \\
\dot{e}_{4} \\
\dot{e}_{5} \\
\dot{e}_{6} \\
\end{matrix}\right\}.
\end{equation}
The second-rank tensor $\vect \mu=\{\mu_{ij}\}$, which we will call microstress tensor, is a non-symmetric tensor defined as $\vect\mu=\frac{k_g}{3}\left(\grad^T\vect u-3\nabla_d\vect u\right)$, where $\grad^T\vect u$ denotes the transpose of the displacement gradient (i.e., $\grad^T\vect u=\{u_{j,i}\}$), and $\grad_d\vect u=\hbox{\rm diag}(u_{1,1},u_{2,2},u_{3,3})$ where $\hbox{\rm diag}(a,b,c)$ is a $3\times3$ diagonal matrix whose diagonal entries starting in the upper left corner are $a,b,c$. Finally, $\vect f=\{f_1,f_2,f_3\}^T$ and $g$ represent, respectively, mechanical and thermal loads.

\subsection{Continuous Problem in Strong Form} \label{sec:SF3D}

Let us denote by $\Gamma$ the boundary of $\Omega$, and its outward normal by $\pmb{n}$. We will assume that $\Gamma$ is sufficiently smooth (e.g., Lipschitz). For the temperature field, we will consider insulated boundary conditions on $\Gamma$, that is, 
\begin{equation}
\theta_{,i}n_i=0, \,\, \qquad \qquad \text{on}\,\,\Gamma\times(0,T),
\end{equation}
where $(0,T)$ is the time interval of interest. We note that for the displacement field we need two boundary conditions at each point of the boundary, because this field is governed by a fourth-order partial-differential equation. The first boundary condition, that we will impose on the entire boundary, states that the normal component of the gradient of the microstress tensor vanishes, that is,
\begin{equation}
\mu_{ij,k}n_k=0,\,\, \qquad \qquad \text{on}\,\,\Gamma\times(0,T).
\end{equation}
For the remaining boundary condition we consider either imposed displacements or stress-free conditions. Thus, we assume that $\Gamma$ admits decompositions
\begin{equation}
\left.\begin{array}{ccc}
\Gamma    & = & \overline{\Gamma_{Di}\cup\Gamma_{Si}} \\
\emptyset & = & \Gamma_{Di}\cap\Gamma_{Si}
\end{array}\right\};\,\, i=1,2,3.
\end{equation}
Then, for each spatial direction $i$, the boundary condition takes on the form:
\begin{eqnarray}
\left(\sigma_{ij}+\eta\sigma'_{ij}+\Delta\mu_{ij}\right)n_j& =0, & \quad\text{on}\quad\Gamma_{Si}\times(0,T),\\
u_i & = u^D_i, & \quad\text{on}\quad\Gamma_{Di}\times(0,T),
\end{eqnarray}
where the $u_i^D$'s are known functions that prescribe the displacements on the boundary. At this point, we are ready to state our problem in strong form as follows: Find the displacement field $\pmb{u}:\overline{\Omega}\times(0,T)\mapsto\mathbb{R}^3$, and temperature $ \theta $: $ \overline{\Omega} \times (0, T) \mapsto \mathbb{R}$ such that  

\begin{subequations}
\renewcommand{\theequation}{\theparentequation.\arabic{equation}}
\label{eq:StrongForm3D}
\begin{alignat}{2}
& \rho \ddot{u}_{i} = \sigma_{ij,j}+\eta\sigma'_{ij,j}+\mu_{ij,kkj}+f_i, &\qquad\text{in  }& \Omega \times (0,T), 
\label{eq:structuralconservation3D}\\
& C_v\dot{\theta} = \kappa\theta_{,ii}+ \Xi\theta\left(u_{i,i}\dot{u}_{j,j}-3u_{i,i}\dot{u}_{i,i}\right) + g, &\qquad\text{in }& \Omega \times (0,T), \label{eq:thermalconservation3D} \\
& \mu_{ij,k}n_k = 0, &\qquad\text{on  }& \Gamma\times(0,T), \\
& \left(\sigma_{ij}+\eta\sigma'_{ij}+\Delta\mu_{ij}\right)n_j = 0, &\qquad\text{on }& \Gamma_{Si}\times(0,T),\\
& u_i = u_i^D,   &\qquad\text{on  }& \Gamma_{Di} \times (0,T), \label{eq:bc12D}\\
& \theta_{,i}n_i  = 0, &\qquad\text{on  }& \Gamma \times (0,T), \label{eq:thermalfluxbc3D}\\
& u_i(\pmb{x},0) = u_i^0(\pmb{x}), &\qquad\text{in }& \overline{\Omega}, \label{eq:structuralIC3D}\\
& \theta(\pmb{x},0) = \theta^0(\pmb{x}), &\qquad\text{in }& \overline{\Omega},
\end{alignat}
\end{subequations}
where $u_i^0 : \overline{\Omega} \mapsto \mathbb{R}$,  $ \theta_0 : \overline{\Omega} \mapsto \mathbb{R}$ are given functions which represent the initial displacements, and temperature in the closed domain $ \overline{\Omega} $. 


\section{Numerical Formulation} \label{sec:NumericalFormulation}
To use IGA, we first derive a weak form of the governing equations. We discretize the computational domain using  \cone-continuous functions required for the discretization of fourth-order PDEs in a primal form. We integrate in time using the generalized-$\alpha$ method, which was originally developed for the structural dynamics equations \cite{chung1993time}, and subsequently applied to fluid mechanics problems \cite{Jansen2000305}.

\subsection{Continuous Problem in the Weak Form} \label{sec:Weak3DExp}
To perform the space discretization of the problem we begin by deriving a weak form of Eqs. (\ref{eq:StrongForm3D}). Let us define the following trial solution spaces
\begin{eqnarray}
\mathcal{S}_{i}    & = & \left\{ u_i\in\mathcal{H}^2\,\,|\,\,u_i=u_i^D\;\; \text{on} \;\; \Gamma_{Di} \right\},\; i=1,2,3, \\
\mathcal{S}_{\theta} & = & \left\{ \theta\in\mathcal{H}^1 \right\},
\end{eqnarray}
where $ \mathcal{H}^k $ is the Sobolev space of square-integrable functions with square-integrable derivatives up to order $k$. The variation spaces are defined as
\begin{eqnarray}
\mathcal{W}_{i}    & = & \left\{ w_i\in\mathcal{H}^2\,\,|\,\,w_i=0\;\; \text{on} \;\; \Gamma_{Di} \right\},\; i=1,2,3, \\
\mathcal{W}_{q} & = & \left\{ q\in\mathcal{H}^1 \right\}.
\end{eqnarray}
The weak form of the structural equations is obtained by multiplying them with $w_i$ and integrating by parts multiple times. The thermal equation is then multiplied with $q$ and integrated by parts once. Taking into account all of this, the variational form of the problem can be stated as: Find $\vect{\mathsf{S}}=\{\vect u,\theta\}\in\mathcal{S}_{1}\times \mathcal{S}_{2}\times\mathcal{S}_{3}\times\mathcal{S}_{\theta}$ such that $\mathsf{B}(\vect{\mathsf{S}},\vect{\mathsf{W}})=0$ for all $\vect{\mathsf{W}}=\{\vect w,\theta\}\in\mathcal{W}_{1}\times \mathcal{W}_{2}\times \mathcal{W}_{3}\times \mathcal{W}_{\theta}$, where
%
\begin{eqnarray}
\mathsf{B}(\vect{\mathsf{S}},\vect{\mathsf{W}}) = & \displaystyle{\left(w_i,\left(\rho \ddot{u}_{i}-f_i\right)\right)+\left(w_{i,j},\sigma_{ij}+\eta\sigma'_{ij}\right)-\left(w_{i,jk},\mu_{ij,k} \right)} \nonumber\\
                   + & \displaystyle{\left(q,\left(C_v\dot{\theta} -\Xi\theta(u_{i,i}\dot{u}_{j,j} -3u_{i,i}\dot{u}_{i,i})-g\right) \right) + \left(\kappa q_{,i},\theta_{,i}\right)}.
\label{eq:WeakFormThreeD} 
\end{eqnarray}
Here, the operator $(\cdot,\cdot)$ denotes the $\mathcal{L}^2$ inner product on the domain $\Omega$.

\subsection{The Semi-Discrete Formulation}  \label{sec:SemiDiscrete3D}

To derive the semi-discrete formulation of Eq. (\ref{eq:WeakFormThreeD}), we define the conforming trial solution spaces $\mathcal{S}_{i}^h\subset\mathcal{S}_{i}$  and $\mathcal{S}_{\theta}^h\subset \mathcal{S}_{\theta}$. Let us also define the conforming weighting function spaces $\mathcal{W}_{i}^h\subset\mathcal{W}_{i}$ and $\mathcal{W}_{\theta}^h\subset \mathcal{W}_{\theta}$. We will use the Galerkin method, so a member of $\mathcal{S}_{i}^h$ is constructed by taking a member of $\mathcal{W}_{i}^h$ and adding a sufficiently smooth function that verifies the Dirichlet boundary conditions. The variational problem over the finite-dimensional spaces may be stated as follows:
\begin{equation}
\mathsf{B}(\vect{\mathsf{S}^h},\vect{\mathsf{W}}^h) = 0,
\end{equation}
where $\vect{\mathsf{W}}^h$ is defined as 
\begin{equation}
\label{eq:grp1}
\vect{\mathsf{W}}^h = \{ \pmb{w}^h,q^h\},\quad w_{i}^h(\vect x,t) = \sum\limits_{A=1}^{n_b} w_{iA}(t) N_A(\vect x),\quad  q^h(\vect x,t) = \sum\limits_{A=1}^{n_b} q_A(t) N_A(\vect x),
\end{equation}
where the $ N_A $'s are the basis functions, and $ n_b $ is the dimension of the discrete space. In Eq. \eqref{eq:grp1} the $w_{iA}$'s are the coordinates of $w_i^h$ in the space $\mathcal{W}_{i}$. In the context of isogeometric analysis, these coordinates are called control variables. Note that the condition $\mathcal{W}_{i}^h \subset \mathcal{W}_{i} $ mandates our displacements discrete space to be at least $ \mathcal{H}^2 $-conforming. In practice, to simplify our implementation, we will use an $ \mathcal{H}^2 $-conforming space also for the temperature field, even if an $ \mathcal{H}^1 $-conforming space would suffice. Note that $ \mathcal{H}^2 $ spaces can be generated using the NURBS basis functions with $\mathscr{C}^k$ global continuity for $k\geq 1$. For a detailed description of how NURBS functions are defined, we refer the reader to \cite{Hughes}.

\subsection{Time Discretization and Implementation}
We use the generalized-$\alpha$ method for time integration. This method finds a  wide range of applications in the computations where control over high frequency dissipation is useful, such as, for example, nonlinear structural dynamics and turbulence \cite{Jansen2000305,akkerman2008role,bazilevs2007variational}. Recently, this method has been applied in the IGA framework to the  Cahn-Hillard equation \cite{Gomez2008} and the isothermal Navier-Stokes-Korteweg equations \cite{gomez2010isogeometric}. Here, we take advantage of the fact that generalized-$\alpha$ permits a straightforward one-step discretization of a coupled system of first- and second-order ordinary differential equations, which is precisely the structure of our semi-discrete form.

\subsection{Time Stepping Scheme} 

Let us assume that the time interval $(0,T)$ is divided into $N$ subintervals $\mathcal{I}_n=(t_n,t_{n+1})$, $n=0,\dots,N-1$. We call $\vect{\mathsf{U}}_n=\{\vect u_A\}_{A=1,\dots,n_b}$ and $\vect{\mathsf{\Theta}}_n=\{\theta_A\}_{A=1,\dots,n_b}$ the vectors associated to displacements and temperature global degrees of freedom (control variables) at time $t_n$. We define $\vect{\mathsf{Y}}_n=\{\vect{\mathsf{U}}_n,\vect{\mathsf{\Theta}}_n\}^T$. The first and second time derivatives of $\vect{\mathsf{Y}}_n$ are denoted by $\dot{\vect{\mathsf{Y}}}_n$ and $\ddot{\vect{\mathsf{Y}}}_n$, respectively. We now define the following residual vectors:
\begin{subequations}
\renewcommand{\theequation}{\theparentequation.\arabic{equation}}
\label{eq:grp23d}
\begin{align} 
\vect{R} &= \{ \vect{R}^{\vect u}, \vect{R}^\theta \}^T, \\
\vect{R}^{\vect u} &= \{ R_{iA}^{\vect u} \},  \\
R_{iA}^{\vect u} &= \mathsf{B} \left( \{ N_A \vect{\rm e}_i, 0\}, \{ \vect{u}^h , \theta^h \} \right), \label{eq:eq4} \\
\vect{R}^\theta &= \{ R_{A}^\theta \}, \label{eq:eq53d} \\
R_{A}^\theta &= \mathsf{B} \left( \{  0, N_A\}, \{ \vect{u}^h , \theta^h \} 
\right), \label{eq:eq6}
\end{align}
\end{subequations}
where $\{\vect{\rm e}_1, \vect{\rm e}_2, \vect{\rm e}_3\}$ is the canonical basis of the space $\mathbb{R}^3$. Our time-integration algorithm may be formulated as follows: Given $\vect{\mathsf{Y}}_n$, 
$\dot{\vect{\mathsf{Y}}}_n$ and $\ddot{\vect{\mathsf{Y}}}_n$  and 
$ \Delta t_n = t_{n+1} - t_n $, find 
$\vect{\mathsf{Y}}_{n+1}$, 
$\dot{\vect{\mathsf{Y}}}_{n+1}$, 
$\ddot{\vect{\mathsf{Y}}}_{n+1}$, such that
\begin{subequations}
\renewcommand{\theequation}{\theparentequation.\arabic{equation}}
\label{eq:grp2}
\begin{align}
&\vect{R}^{\vect u} \left( \vect{\mathsf{Y}}_{n+\alpha_f}, \dot{\vect{\mathsf{Y}}}_{n+\alpha_f}, \ddot{\vect{\mathsf{Y}}}_{n+\alpha_m} \right) = 
0, \label{eq:eq22} \\
&\vect{R}^\theta \left( \vect{\mathsf{Y}}_{n+\alpha_f}, \dot{\vect{\mathsf{Y}}}_{n+\alpha_f}, \ddot{\vect{\mathsf{Y}}}_{n+\alpha_m} \right) = 
0, \label{eq:eq33} \\
&\vect{\mathsf{Y}}_{n+\alpha_f} = \vect{\mathsf{Y}}_n + \alpha_f \left( \vect{\mathsf{Y}}_{n+1} - \vect{\mathsf{Y}}_n \right), 
\label{eq:eq44} \\
&\dot{\vect{\mathsf{Y}}}_{n+\alpha_f} = \dot{\vect{\mathsf{Y}}}_n + \alpha_f \left( \dot{\vect{\mathsf{Y}}}_{n+1} - \dot{\vect{\mathsf{Y}}}_n \right), 
\label{eq:eq55} \\
&\ddot{\vect{\mathsf{Y}}}_{n+\alpha_m} = \ddot{\vect{\mathsf{Y}}}_n + \alpha_m \left( \ddot{\vect{\mathsf{Y}}}_{n+1} - \ddot{\vect{\mathsf{Y}}}_n \right), 
\label{eq:eq66} \\
&\dot{\vect{\mathsf{Y}}}_{n+1} = \dot{\vect{\mathsf{Y}}}_n + \Delta t_n 
\left[ \left(1- \gamma \right)\ddot{\vect{\mathsf{Y}}}_{n} + \gamma \ddot{\vect{\mathsf{Y}}}_{n+1} \right], \label{eq:eq77} \\
&\vect{\mathsf{Y}}_{n+1} = \vect{\mathsf{Y}}_n + \Delta t_n \dot{\vect{\mathsf{Y}}}_{n} + 
\frac{\left( \Delta t \right)^2}{2} 
\left[ \left(1- 2 \beta \right)\ddot{\vect{\mathsf{Y}}}_{n} + 2 \beta \ddot{\vect{\mathsf{Y}}}_{n+1} \right].
 \label{eq:eq88}
\end{align}
\end{subequations}
Note that, although the vectors of degrees of freedom $\vect{\mathsf{Y}}_{n+1}$, 
$\dot{\vect{\mathsf{Y}}}_{n+1}$ and $\ddot{\vect{\mathsf{Y}}}_{n+1}$ are treated independently, they must honor Eqs. \eqref{eq:eq77}--\eqref{eq:eq88}, and, as a consequence, only one of them is independent. The authors of \cite{chung1993time} showed that the algortihm \eqref{eq:grp2} achieves second-order accuracy for a second-order ordinary differential equation if
\begin{equation}
\label{gamma}
\gamma = \frac{1}{2} + \alpha_m - \alpha_f, \qquad 
\end{equation}
and
\begin{equation}
\label{beta}
\beta = \frac{1}{4} \left( 1 - \alpha_f + \alpha_m  \right)^2,
\end{equation}
while unconditional stability (for a linear problem) requires 
\begin{eqnarray}\label{stab}
\alpha_m \ge \alpha_f \ge \frac{1}{2}. 
\end{eqnarray}
Jansen et al. \cite{Jansen2000305} showed that the generalized-$\alpha$ algorithm can also be applied to first-order ordinary differential equations just dropping the $\vect{\mathsf{Y}}$ variables ($\dot{\vect{\mathsf{Y}}}$ and $\ddot{\vect{\mathsf{Y}}}$ are kept) in Eqs. \eqref{eq:grp2}. Additionally, they showed that conditions \eqref{gamma} and \eqref{stab} for accuracy and stability will still hold true (condition \eqref{beta} does not apply to the first-order problem because when the $\vect{\mathsf{Y}}$ variables are dropped, the parameter $\beta$ plays no role in the algorithm).

The generalized-$\alpha$ algorithm permits optimal high-frequency dissipation by parameterizing $\alpha_m$ and $\alpha_f$ in terms of the spectral radius of the amplification matrix as $\Delta t\rightarrow\infty$, namely $\varrho_{\infty}$. Optimal high-frequency dissipation is achieved when all the eigenvalues of the amplification matrix take on the value $\varrho_{\infty}$. For the first-order problem this may be achieved by taking \cite{Jansen2000305}
\begin{equation}\label{alphas}
\alpha_m^{(1)} = \frac{1}{2} \left( \frac{3-\varrho_{\infty}^{(1)}}{1+\varrho_{\infty}^{(1)}} \right), \qquad
\alpha_f^{(1)} = \frac{1}{1+\varrho_{\infty}^{(1)}}, 
\end{equation}
while for the second-order problem we need \cite{chung1993time}
\begin{equation}\label{alphas_2}
\alpha_m^{(2)} = \frac{2-\varrho_{\infty}^{(2)}}{1+\varrho_{\infty}^{(2)}} , \qquad
\alpha_f^{(2)} = \frac{1}{1+\varrho_{\infty}^{(2)}}. 
\end{equation}
Since we want to equate the residuals to zero at the same $\alpha$-levels for both the first- and the second-order equations, we need $\alpha_m^{(1)}=\alpha_m^{(2)}$, and this can only be achieved with optimal high-frequency damping if $\varrho_\infty^{(1)}=\varrho_\infty^{(2)}=1$. The case $\varrho_\infty=1$ corresponds to the midpoint rule, which leads to vanishing high-frequency damping and, in our opinion, this is not robust enough for nonlinear computations. In the calculations presented in this paper, we selected $\alpha_m$ and $\alpha_f$ by taking $\varrho_\infty=1/2$ in Eqs. \eqref{alphas}. This makes the temperature equation optimally damped, while the structural equations have sub-optimally dissipation. Note that with this choice, second-order accuracy and unconditional stability of a linear system of coupled first- and second-order ordinary differential equations is still guaranteed. We note, however, that the problem of interest in this paper is nonlinear and unconditional stability cannot be expected \cite{provably}, but we believe that this choice will permit taking relatively large time steps retaining stability. 

In what follows, we approximate the solution to the equation
\begin{equation}
\vect R\left( \vect{\mathsf{Y}}_{n+\alpha_f}, \dot{\vect{\mathsf{Y}}}_{n+\alpha_f}, \ddot{\vect{\mathsf{Y}}}_{n+\alpha_m} \right) = 0,
\end{equation}
using Newton's method. Note that as indicated in Eqs. \eqref{eq:eq44}--\eqref{eq:eq66}, the $\alpha$-levels are nothing else but linear interpolations of the variables at times $t_n$ and $t_{n+1}$. In addition, Eqs. \eqref{eq:eq77}--\eqref{eq:eq88} must be satisfied, which indicates that $\vect{\mathsf{Y}}_{n+\alpha_f}, \dot{\vect{\mathsf{Y}}}_{n+\alpha_f}, \ddot{\vect{\mathsf{Y}}}_{n+\alpha_m}$ can all be written in terms of $\ddot{\vect{\mathsf{Y}}}_{n+1}$ (they can also be written in terms of $\vect{\mathsf{Y}}_{n+1}$ and $\dot{\vect{\mathsf{Y}}}_{n+1}$), which is the variable we select to perform the linearization. Before we start the nonlinear iterative process, we need a prediction for $\ddot{\vect{\mathsf{Y}}}_{n+1}$ that will be denoted by $\ddot{\vect{\mathsf{Y}}}_{n+1,(0)}$. The prediction is based on the equal-velocity approximation, namely $\dot{\vect{\mathsf{Y}}}_{n+1,(0)}=\dot{\vect{\mathsf{Y}}}_{n}$. Then, our predictions are
\begin{subequations}
\renewcommand{\theequation}{\theparentequation.\arabic{equation}}
\label{eq:grp3}
\begin{align}
&\dot{\vect{\mathsf{Y}}}_{n+1,(0)} = \dot{\vect{\mathsf{Y}}}_n, \label{eq:eq1sol} \\
&\ddot{\vect{\mathsf{Y}}}_{n+1,(0)} = \frac{\gamma-1}{\gamma} \ddot{\vect{\mathsf{Y}}}_n, \label{eq:eq23d2}\\
&\vect{\mathsf{Y}}_{n+1,(0)} = \vect{\mathsf{Y}}_n + 
\Delta t_n \dot{\vect{\mathsf{Y}}}_n +
\frac{\left( \Delta t_n \right)^2}{2} \left[ \left(1- 2 \beta \right)\ddot{\vect{\mathsf{Y}}}_{n} + 2 \beta \ddot{\vect{\mathsf{Y}}}_{n+1,(0)} \right]\label{eq:eq23d3},
\end{align}
\end{subequations}
where we have used \eqref{eq:eq77} and \eqref{eq:eq88} to derive \eqref{eq:eq23d2} and \eqref{eq:eq23d3}, respectively. After computing the predictions, we repeat the following steps for $ i = 1, 2,\cdots, i_{max} $ or until convergence is reached:
\begin{enumerate}
\item Evaluate iterates at the $ \alpha $-levels\\
\begin{subequations}
\renewcommand{\theequation}{\theparentequation.\arabic{equation}}
\label{eq:grp4}
\begin{align}
\vect{\mathsf{Y}}_{n+\alpha_f,(i-1)}  = \vect{\mathsf{Y}}_n + \alpha_f \left( \vect{\mathsf{Y}}_{n+1,(i-1)} - \vect{\mathsf{Y}}_n \right), \label{eq:eq23d} \\
\dot{\vect{\mathsf{Y}}}_{n+\alpha_f,(i-1)}  = \dot{\vect{\mathsf{Y}}}_n + \alpha_f \left( \dot{\vect{\mathsf{Y}}}_{n+1,(i-1)} - \dot{\vect{\mathsf{Y}}}_n \right), \label{eq:eq1mul} \\
\ddot{\vect{\mathsf{Y}}}_{n+\alpha_m,(i-1)}  = \ddot{\vect{\mathsf{Y}}}_n + \alpha_m \left( \ddot{\vect{\mathsf{Y}}}_{n+1,(i-1)} - \ddot{\vect{\mathsf{Y}}}_n \right). \label{eq:eq1mu2}
\end{align}
\end{subequations}

\item If the convergence conditions
\begin{equation}
\frac{\left|\vect R^{\vect u}\left( \vect{\mathsf{Y}}_{n+\alpha_f,(i-1)}, \dot{\vect{\mathsf{Y}}}_{n+\alpha_f,(i-1)}, \ddot{\vect{\mathsf{Y}}}_{n+\alpha_m,(i-1)} \right)\right|}{\left|\vect R^{\vect u}\left( \vect{\mathsf{Y}}_{n+\alpha_f,(0)}, \dot{\vect{\mathsf{Y}}}_{n+\alpha_f,(0)}, \ddot{\vect{\mathsf{Y}}}_{n+\alpha_m,(0)} \right)\right|}<\hbox{\rm tol}_{\vect u} \label{tolc}
\end{equation}
and
\begin{equation}
\frac{\left|\vect R^\theta\left( \vect{\mathsf{Y}}_{n+\alpha_f,(i-1)}, \dot{\vect{\mathsf{Y}}}_{n+\alpha_f,(i-1)}, \ddot{\vect{\mathsf{Y}}}_{n+\alpha_m,(i-1)} \right)\right|}{\left|\vect R^\theta\left( \vect{\mathsf{Y}}_{n+\alpha_f,(0)}, \dot{\vect{\mathsf{Y}}}_{n+\alpha_f,(0)}, \ddot{\vect{\mathsf{Y}}}_{n+\alpha_m,(0)} \right)\right|}<\hbox{\rm tol}_\theta \label{tolt}
\end{equation}
are simultaneously satisfied, then we set the variables at time $t_{n+1}$ as,
\begin{subequations}
\renewcommand{\theequation}{\theparentequation.\arabic{equation}}
\begin{align}
&\ddot{\vect{\mathsf{Y}}}_{n+1} = \ddot{\vect{\mathsf{Y}}}_{n+1,(i-1)}, \\
& \dot{\vect{\mathsf{Y}}}_{n+1} =  \dot{\vect{\mathsf{Y}}}_{n+1,(i-1)}, \\
&\vect{\mathsf{Y}}_{n+1}        = \vect{\mathsf{Y}}_{n+1,(i-1)},
\end{align}
\end{subequations}
and exit the nonlinear iteration algorithm. If Eqs. \eqref{tolc} and \eqref{tolt} are not satisfied, then we proceed to the next step.

\item Use the solutions at the $ \alpha$-levels to assemble the residual and the tangent matrix of the linear system
\begin{equation}
\vect{K}_{(i-1)} \Delta \ddot{\vect{\mathsf{Y}}}_{n+1,(i)} = 
-\vect{R}_{(i-1)}.
\label{eq:residualmatrix}
\end{equation}
Solve this linear system using a preconditioned GMRES algorithm to a specified tolerance. 
\item Use $\Delta \ddot{\vect{\mathsf{Y}}}_{n+1,(i)}$ to update the iterates as
\begin{subequations}
\renewcommand{\theequation}{\theparentequation.\arabic{equation}}
\label{eq:grp5}
\begin{align}
&\ddot{\vect{\mathsf{Y}}}_{n+1,(i)} = \ddot{\vect{\mathsf{Y}}}_{n+1,(i-1)} + \Delta \ddot{\vect{\mathsf{Y}}}_{n+1,(i)},  \label{eq:eq111} \\
&\dot{\vect{\mathsf{Y}}}_{n+1,(i)} = \dot{\vect{\mathsf{Y}}}_{n+1,(i-1)} + \gamma \Delta t_n\Delta \ddot{\vect{\mathsf{Y}}}_{n+1,(i)},  \label{eq:eq112} \\
&\vect{\mathsf{Y}}_{n+1,(i)} = \vect{\mathsf{Y}}_{n+1,(i-1)} + \beta \left( \Delta t_n \right)^2\Delta \ddot{\vect{\mathsf{Y}}}_{n+1,(i)}.  \label{eq:eq113} 
\end{align}
\end{subequations}
The tangent matrix $\vect {K}_{(i-1)}$ in Eq. \eqref{eq:residualmatrix} is a block matrix with the structure
\begin{equation}
\vect K=\left(\begin{array}{cc}
\vect K_{11} & \vect K_{12} \\
\vect K_{21} & \vect K_{22} 
\end{array}\right),
\end{equation}
where the sub-index $(i-1)$ has been omitted for notational simplicity. We also note that
\begin{equation}
\vect K_{11}=\frac{\partial\vect R^{\vect u}}{\partial\ddot{\vect{\mathsf{U}}}_{n+1}},\; \vect K_{12}=\frac{\partial\vect R^{\vect u}}{\partial\ddot{\vect{\mathsf{\Theta}}}_{n+1}},\; \vect K_{21}=\frac{\partial\vect R^\theta}{\partial\ddot{\vect{\mathsf{U}}}_{n+1}},\;  \vect K_{22}=\frac{\partial\vect R^\theta}{\partial\ddot{\vect{\mathsf{\Theta}}}_{n+1}}.
\end{equation}
To compute the blocks of the tangent matrix $\vect K$ we use the chain rule because the residuals are not functions of $\ddot{\vect{\mathsf{Y}}}_{n+1}$, but of $\vect{\mathsf{Y}}_{n+\alpha_f}$, $\dot{\vect{\mathsf{Y}}}_{n+\alpha_f}$, and $\ddot{\vect{\mathsf{Y}}}_{n+\alpha_f}$. Let us illustrate this by computing $\vect K_{11}$ as follows:
%
%
%
\begin{eqnarray}
\vect{K}_{11} &&= \frac{\partial\vect R^{\vect u}}{\partial\ddot{\vect{\mathsf{U}}}_{n+1}} = 
\frac{\partial \vect{R}^{\vect u}}{\partial \vect{\mathsf{U}}_{n+\alpha_f}}
\frac{\partial \vect{\mathsf{U}}_{n+\alpha_f}}{\partial \vect{\mathsf{U}}_{n+1}}
\frac{\partial \vect{\mathsf{U}}_{n+1}}{\partial\ddot{ \vect{\mathsf{U}}}_{n+1}} 
+ \frac{\partial \vect{R}^{\vect u}}{\partial \dot{\vect{\mathsf{U}}}_{n+\alpha_f}} \frac{\partial \dot{\vect{\mathsf{U}}}_{n+\alpha_f}}{\partial \dot{\vect{\mathsf{U}}}_{n+1}} 
\frac{\partial \dot{\vect{\mathsf{U}}}_{n+1}}{\partial \ddot{\vect{\mathsf{U}}}_{n+1}}
+ \frac{\partial \vect{R}^{\vect u}}{\partial \ddot{\vect{\mathsf{U}}}_{n+\alpha_m}} \frac{\partial \ddot{\vect{\mathsf{U}}}_{n+\alpha_m}}{\partial \ddot{\vect{\mathsf{U}}}_{n+1}}  \nonumber \\
&&= \frac{\partial \vect{R}^{\vect u}}{\partial \vect{\mathsf{U}}_{n+\alpha_f}}\alpha_f \beta  \left( \Delta t_n \right)^2
+ \frac{\partial \vect{R}^{\vect u}}{\partial \dot{\vect{\mathsf{U}}}_{n+\alpha_f}}  \alpha_f \gamma \Delta t_n  +
\frac{\partial \vect{R}^{\vect u}}{\partial \ddot{\vect{\mathsf{U}}}_{n+\alpha_m}} \alpha_m.
\end{eqnarray}
The remaining blocks of $\vect K$ can be obtained analogously.
\end{enumerate}


\section{Numerical Simulations} \label{sec:NumericalSimulations}

In this section, we present numerical studies on nanostructured SMAs subjected to temperature-induced loadings. The  microstructure evolution and in turn its effect on thermo-mechanical properties of SMA specimens have been investigated. For convenience, the developed thermo-mechanical model described by Eqs. (\ref{eq:StrongForm3D}) is rescaled in the spatio-temporal domain as described in \ref{app:Rescaling}. The rescaled equations are then converted into the rescaled weak formulation of Eqs. (\ref{eq:WeakFormThreeD}). The Fe$_{70}$Pd$_{30}$ material parameters \cite{Ahluwalia2006} used for the simulations have been summarized in Table \ref{tab:MatProperties}. 

\begin{figure}[H]
\centering
\subfigure[]
{
\includegraphics[width=0.375\linewidth]{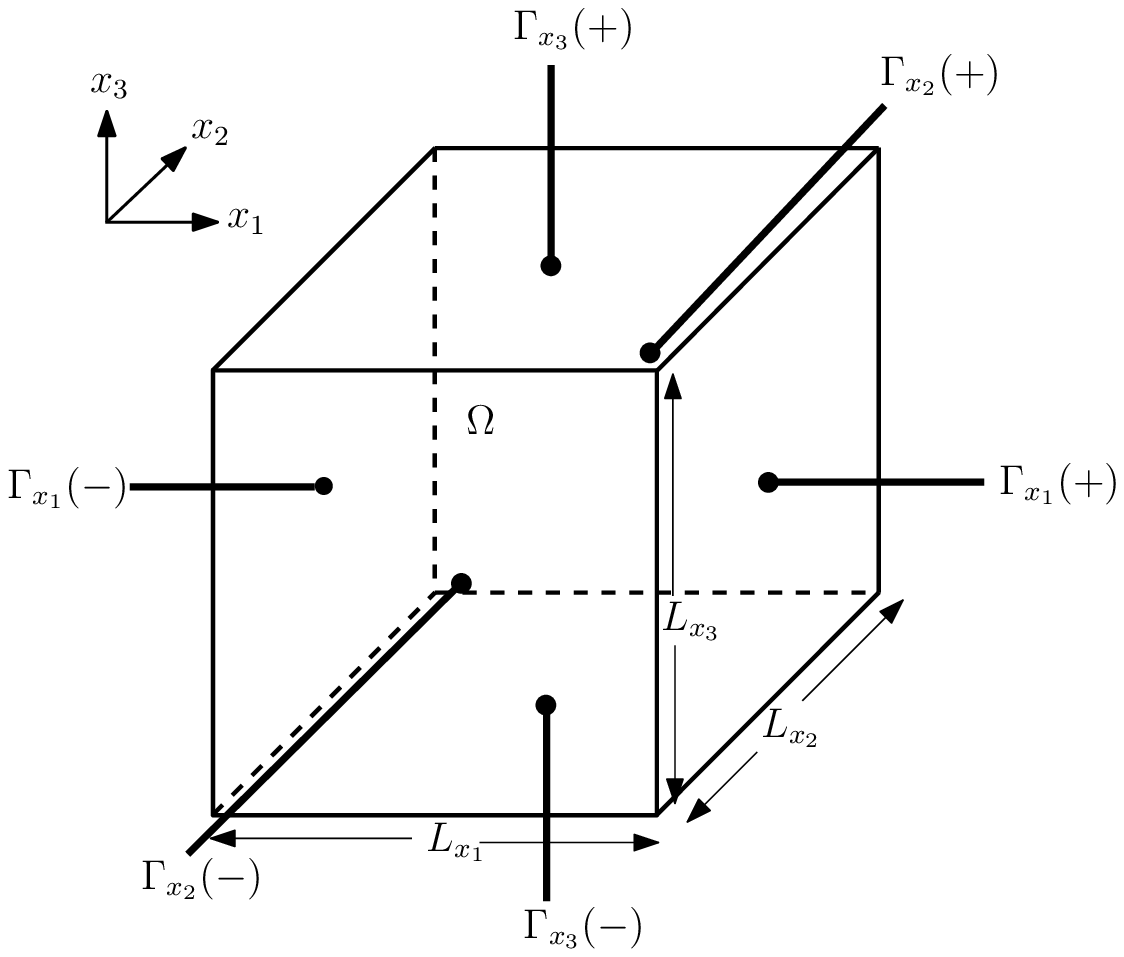}
\label{fig:SchematicRectangularPrism}
}
\subfigure[]
{
\includegraphics[width=0.25\linewidth]{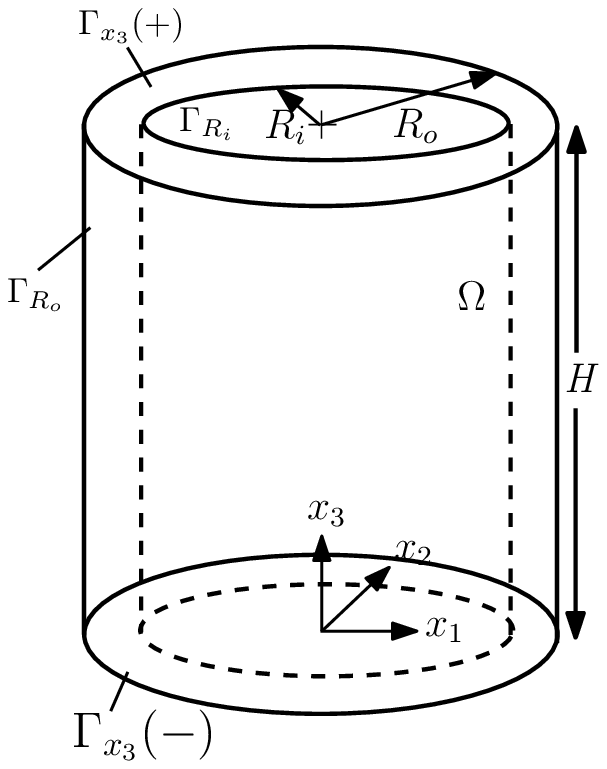}
\label{fig:SchematicCylindricalTube}
}
\subfigure[]
{
\includegraphics[width=0.3\linewidth]{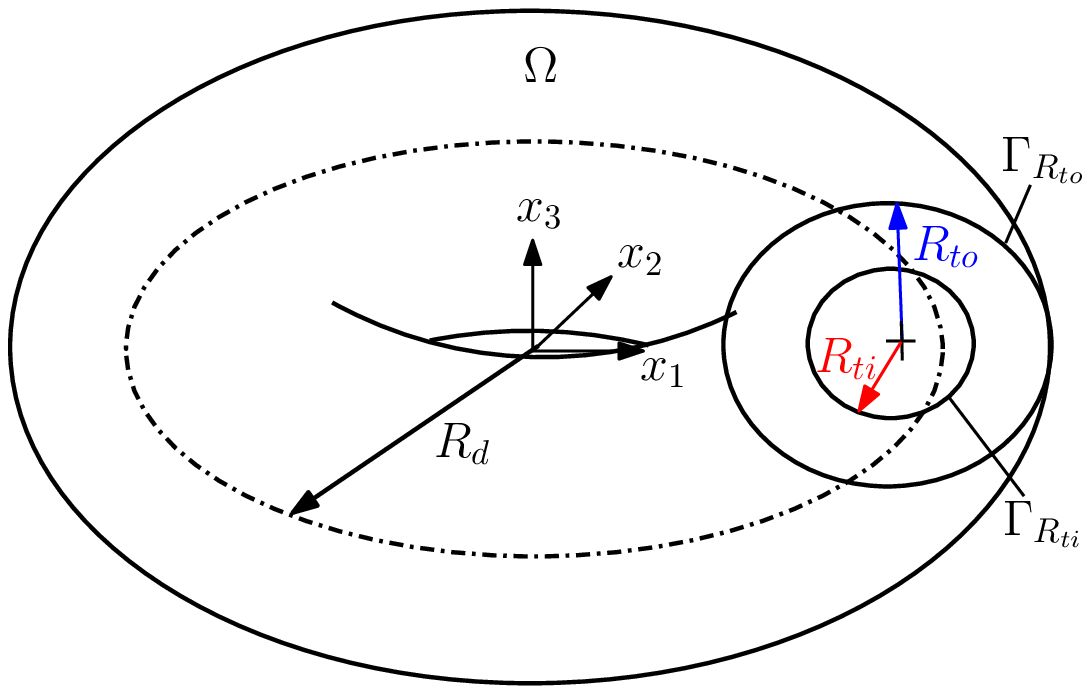}
\label{fig:SchematicTorus}
}
\caption{Schematic and nomenclature of the (a) rectangular prism, (b) cylindrical tube, and (c) tubular torus domain ($\Omega$) and boundaries ($ \Gamma $).}
\label{fig:SchematicDomains}
\end{figure}

As mentioned in Section \ref{sec:Introduction}, applications exist where different geometries are required, often beyond simple cube domains; here we conduct numerical experiments on different nanostructured SMA geometries. The schematic and nomenclature of different geometries and their boundaries are described in Fig. \ref{fig:SchematicDomains}. The mechanical boundary conditions for different simulations have been described in the respective subsections. For the thermal physics, insulated boundary conditions have been used for all the simulations. We discretize the domains using B-spline or NURBS basis functions with $\mathscr{C}^k$ global continuity for $k\geq 1$.

\begin{table}[htbp]
  \centering
  \caption{Fe$_{70}$Pd$_{30}$ material constants}
    \begin{tabular}{ccccccc}
$a_1$ & $a_2 $ & $a_3 $ & $a_4 $ & $a_5 $ & $ \eta $ \\ \hline
192.3 GPa & 280 GPa & 19.7 GPa & 2.59 $ \times $  10$^3$ GPa & 8.52 $ \times $  10$^4$ Gpa & 0.25 N-s m$^2$   \\ \hline
$k_g $ & $\theta_m$ & $\theta_0$ & $C_v $ & $\kappa$ & $\rho$ \\ \hline
3.15 $ \times $ 10$^{-8}$ N & 270 K & 295 K & 350 J kg$^{-1}$ K$^{-1}$ & 78 W m$^{-1}$ K$^{-1}$ & 10000 kg m$^{-3}$ \\
\hline \\
    \end{tabular}%
  \label{tab:MatProperties}%
\end{table}%
 
In the following subsections, first a mesh convergence and time step studies will be conducted to show the stability and effectiveness of the approach. Next, we analyze temperature-induced phase transformations and study microstructure morphology on SMA domains of different geometries. 

\subsection{Mesh and Time Step Refinement Studies} \label{sec:MeshConvergence}

\begin{figure}[H]
\centering
\subfigure[]
{
\includegraphics[width=0.4\linewidth]{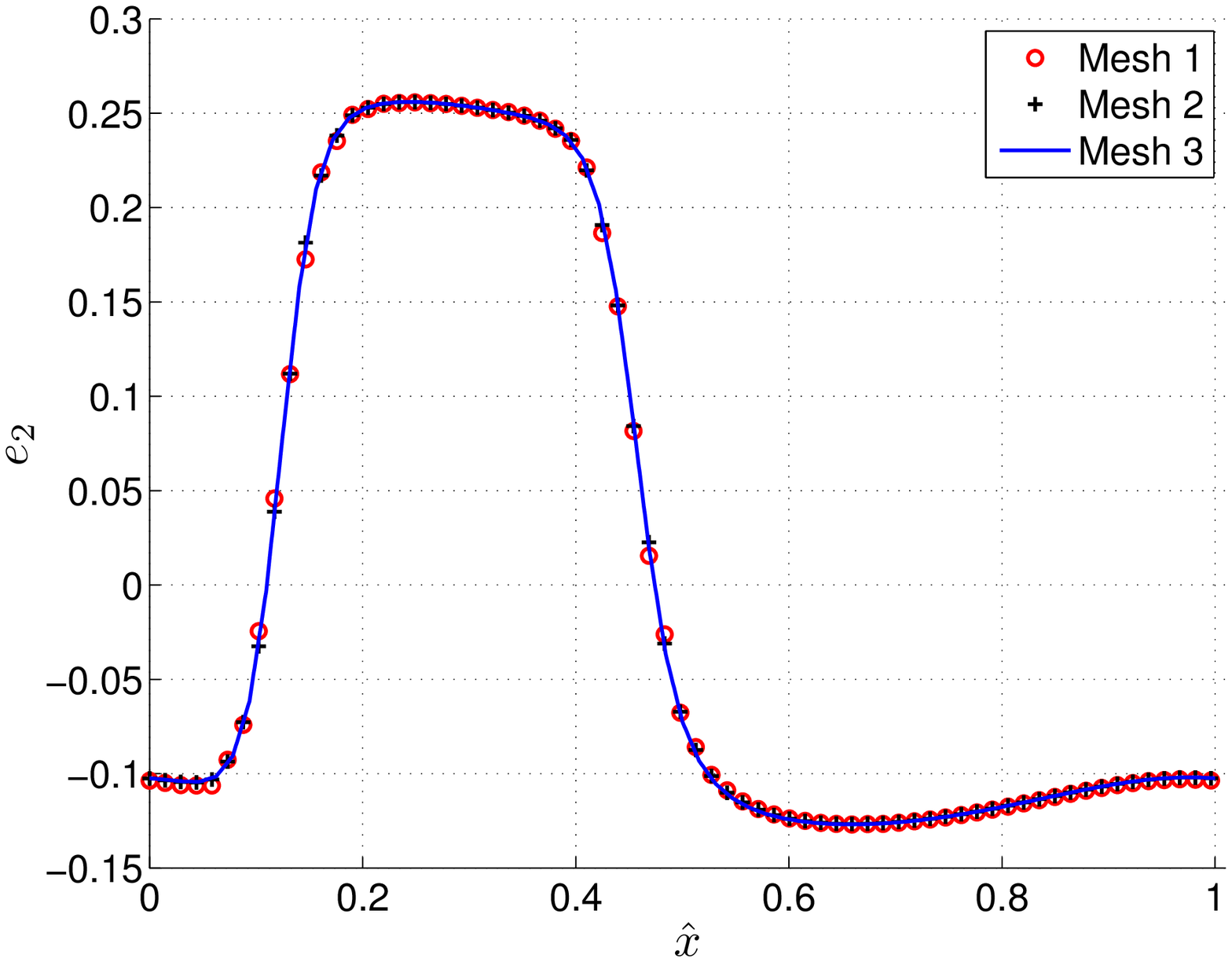}
\label{fig:MeshConvergencee2}
}
\subfigure[]
{
\includegraphics[width=0.4\textwidth]{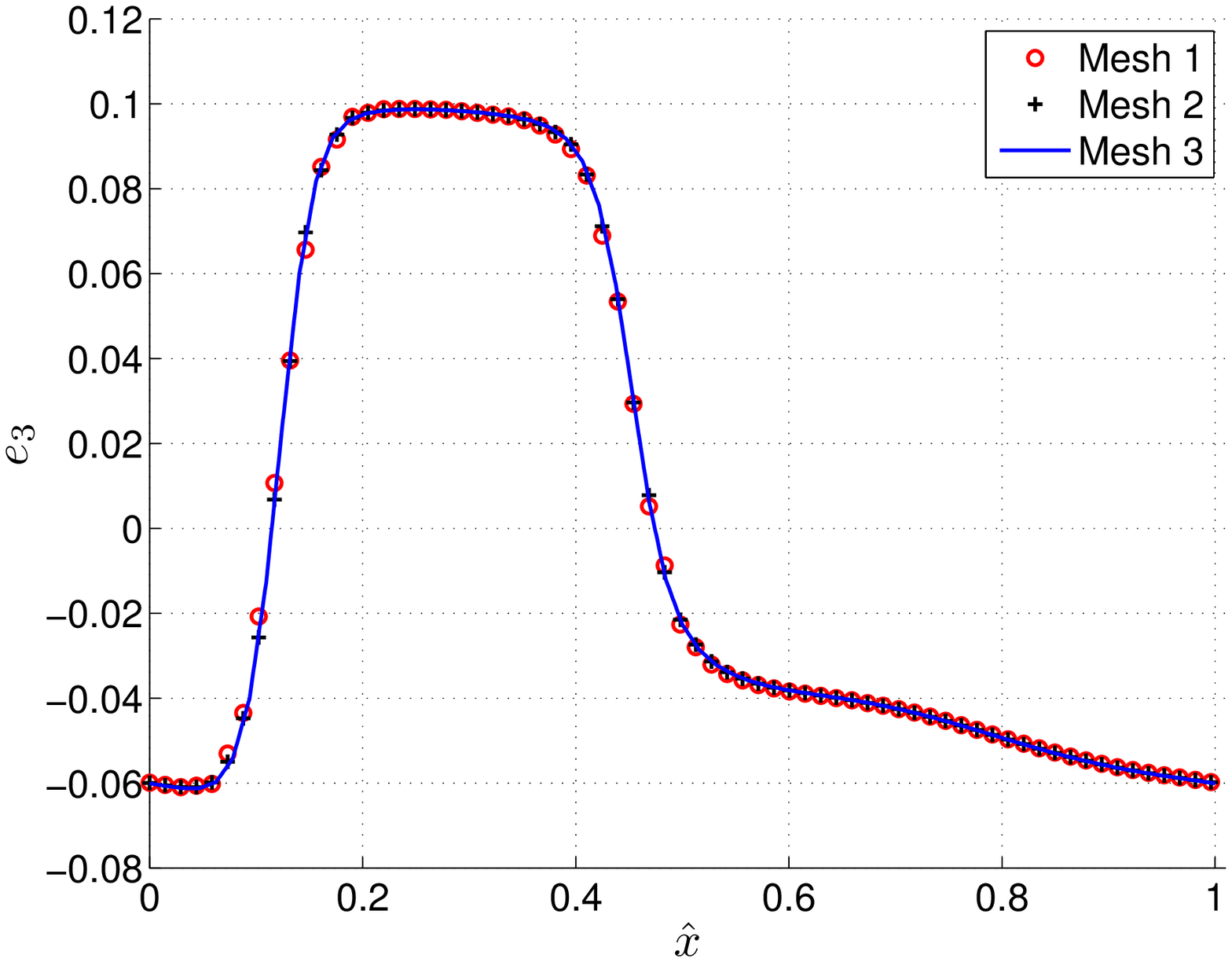}
\label{fig:MeshConvergencee3}
}
\caption{(Color online) Mesh refinement studies: cut-line plot of OPs (deviatoric strains) (a) $ e_2 $ and (b) $ e_3 $ along the normalized distance $ \hat{x} $ between points (0,15,15) nm and (32,15,15) nm on a 32 nm side cube.}
\label{fig:MeshConvergence}
\end{figure}

We begin by performing a refinement study of the spatial mesh holding the time step fixed at a sufficiently small value. The spatial mesh is refined using the classical $h$-refinement and the new paradigm for mesh refinement introduced by IGA, $k$-refinement, in which the order of the basis functions is elevated, but their global continuity is likewise increased. The mesh convergence studies have been carried out on a \lx = \ly = \lz = 32 nm cube with periodic boundary conditions and starting with initial random conditions. The cube is discretized using three meshes: two meshes with uniform \cone-continuous quadratic B-splines with 34 (Mesh 1), and 50 (Mesh 2) basis functions in each direction and a third mesh with uniform cubic $\mathscr{C}^2$-continuous B-splines with 69 (Mesh 3) basis functions in each direction. The cube is quenched to temperature corresponding to $ \tau  = -1.2$ and allowed to evolve for a sufficiently long time till it is stabilized. We plot the  cut lines of OP deviatoric strains $ e_2 $ and $ e_3 $ along the normalized distance $ \hat{x} $ between the two points (0,15,15) nm and (32,15,15) nm on the opposite surfaces of the cube, for all three meshes as shown in Fig. \ref{fig:MeshConvergence}. The maximum error of the coarsest mesh (Mesh 1) is 2 \% with respect to the fine mesh (Mesh 3), thus indicating that good results can be obtained by using IGA even with the coarsest mesh. 

%

\begin{figure}[H]
\centering
\subfigure[]
{
\includegraphics[width=0.3\linewidth]{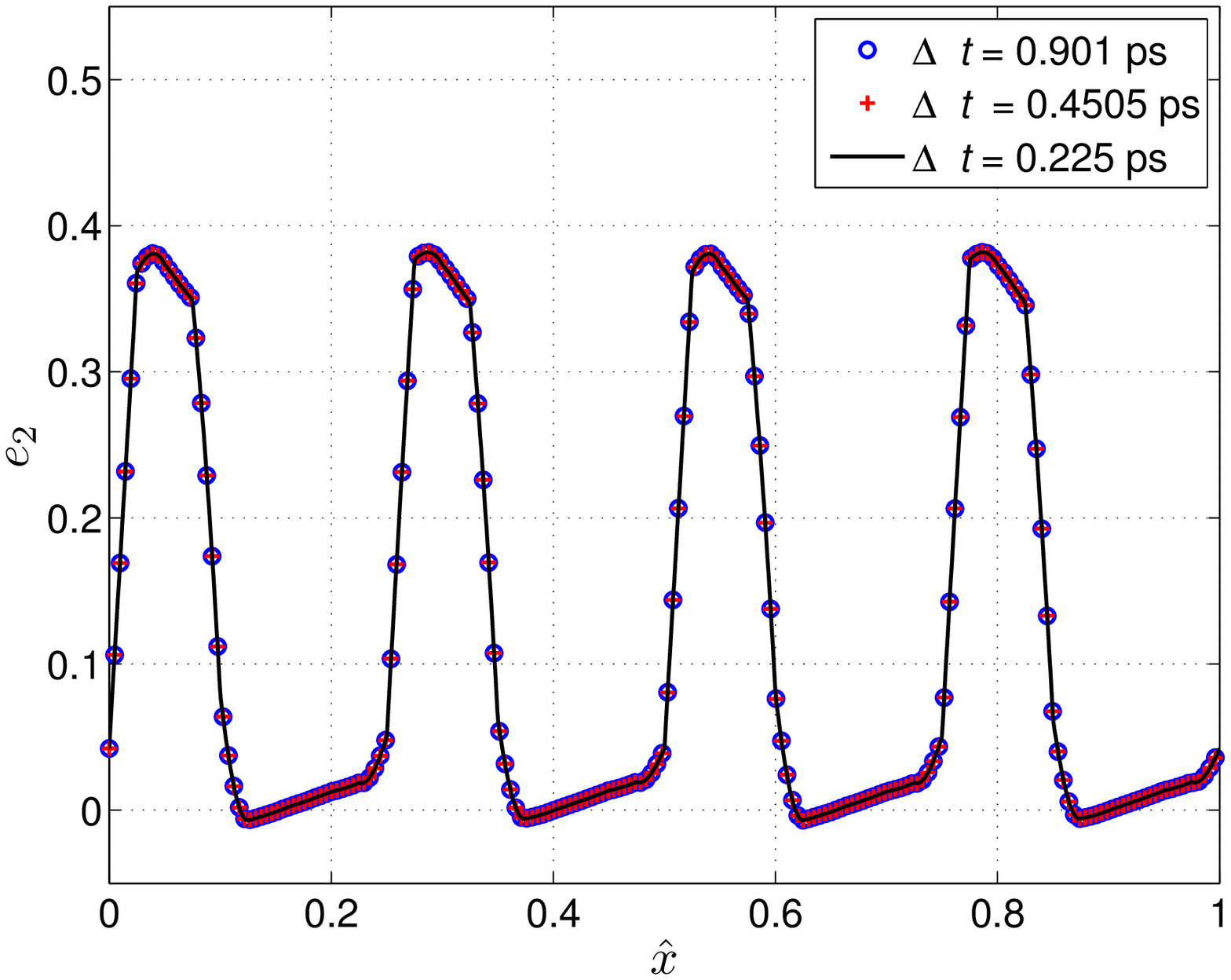}
\label{fig:TimeStepStudye2}
}
\subfigure[]
{
\includegraphics[width=0.3\textwidth]{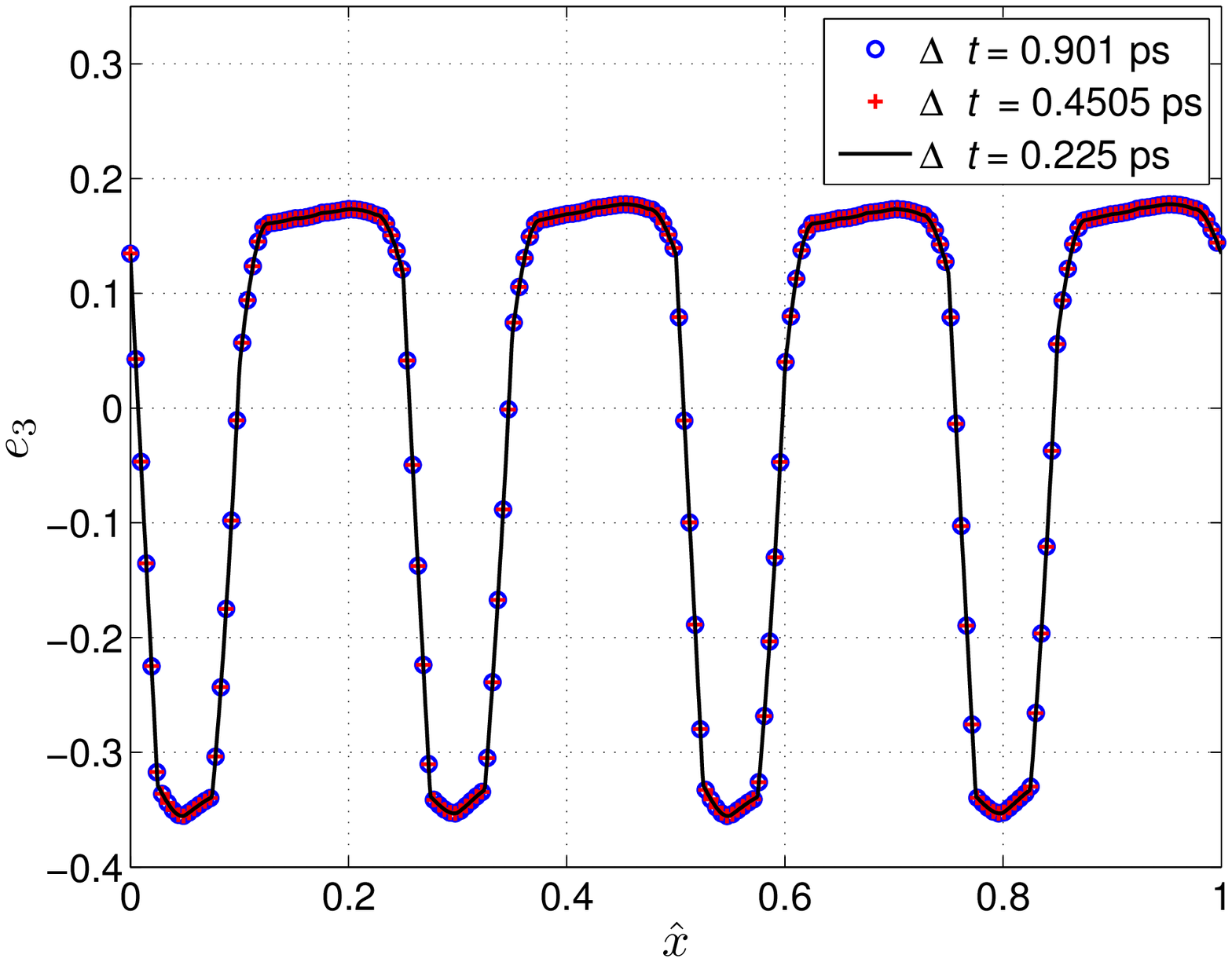}
\label{fig:TimeStepStudye3}
}
\subfigure[]
{
\includegraphics[width=0.3\textwidth]{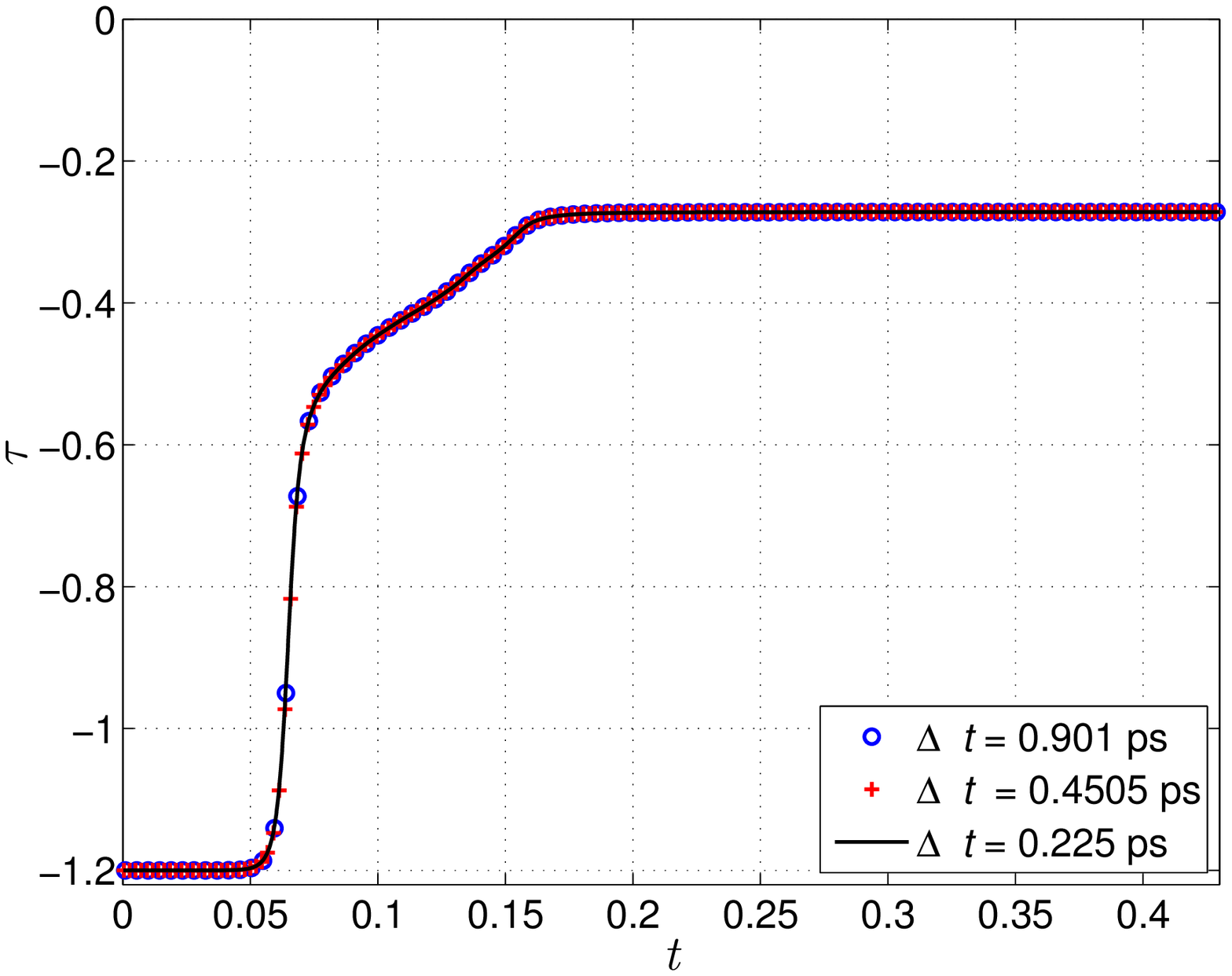}
\label{fig:TimeVsTempTSstudy}
}
\caption{(Color online) Time step refinement studies: cut-line plot of the OPs (deviatoric strains) (a) $ e_2 $ and (b) $ e_3 $ along the normalized diagonal $ \hat{x} $ between diagonal points (0,0,0)--(50,50,50) nm and (c) time (ns) evolution of average temperature coefficient $ \tau $.}
\label{fig:TimeStepStudy}
\end{figure}

We now study the impact of the time step size on the solution, holding the spatial discretization fixed. The simulations have been performed on a cube with sides \lx = \ly = \lz = 50 nm by using periodic boundary conditions and starting with a random initial condition. We have used three different fixed time steps 0.225 ps, 0.4505 ps, and 0.901 ps. Cut lines of the OPs $ e_2 $ and $ e_3 $ are shown in Figs. \ref{fig:TimeStepStudye2}--\ref{fig:TimeStepStudye3}, along the diagonal (0,0,0) -- (50,50,50) nm, at time 0.427 ns. The time evolution of the average temperature coefficient $ \tau $ is plotted in Fig. \ref{fig:TimeVsTempTSstudy}.  The maximum error in the OPs of the largest time step (0.901 ps) with respect to the smallest one (0.225 ps) is 0.02 \%. The plots show little sensitivity of the solution to the time step, and that for this numerical example, the largest time step can be safely used. 

\subsection{Microstructure Morphology on Different Geometries} \label{sec:DiffGeom}

Most PF simulations in the literature have been conducted on cubic specimens with periodic or stress-free boundary conditions. However, applications exist where complex geometries are required. In this section, we generate a thermally-induced microstructure in SMA specimens of different geometries, by quenching them to the temperature corresponding to $ \tau = -1.2$ and allowing the system to evolve. All the simulations have been started with a small-amplitude random initial condition, corresponding to the austenite phase, for the displacement vector $\vect u$. Periodic, constrained or stress-free boundary conditions have been used on different surfaces for the structural physics. In all the following simulations, the  $x_1$, $x_2$, and $x_3$ directions in the microstructure morphology plots are indicated by $ x $, $  y $ and $ z $ directions, respectively. The cube and slab SMA geometries are modeled using uniform \cone-quadratic B-spline basis functions, while the cylindrical tube and tubular torus geometries require the use of NURBS functions. 

\subsubsection{Cube Geometry} \label{sec:DiffGeomCube}
\begin{figure}[h!]
\centering
\subfigure[\textit{t} = 0.036 ns]
{
\includegraphics[trim=20mm 20mm 20mm 40mm,clip, width=0.16\linewidth]{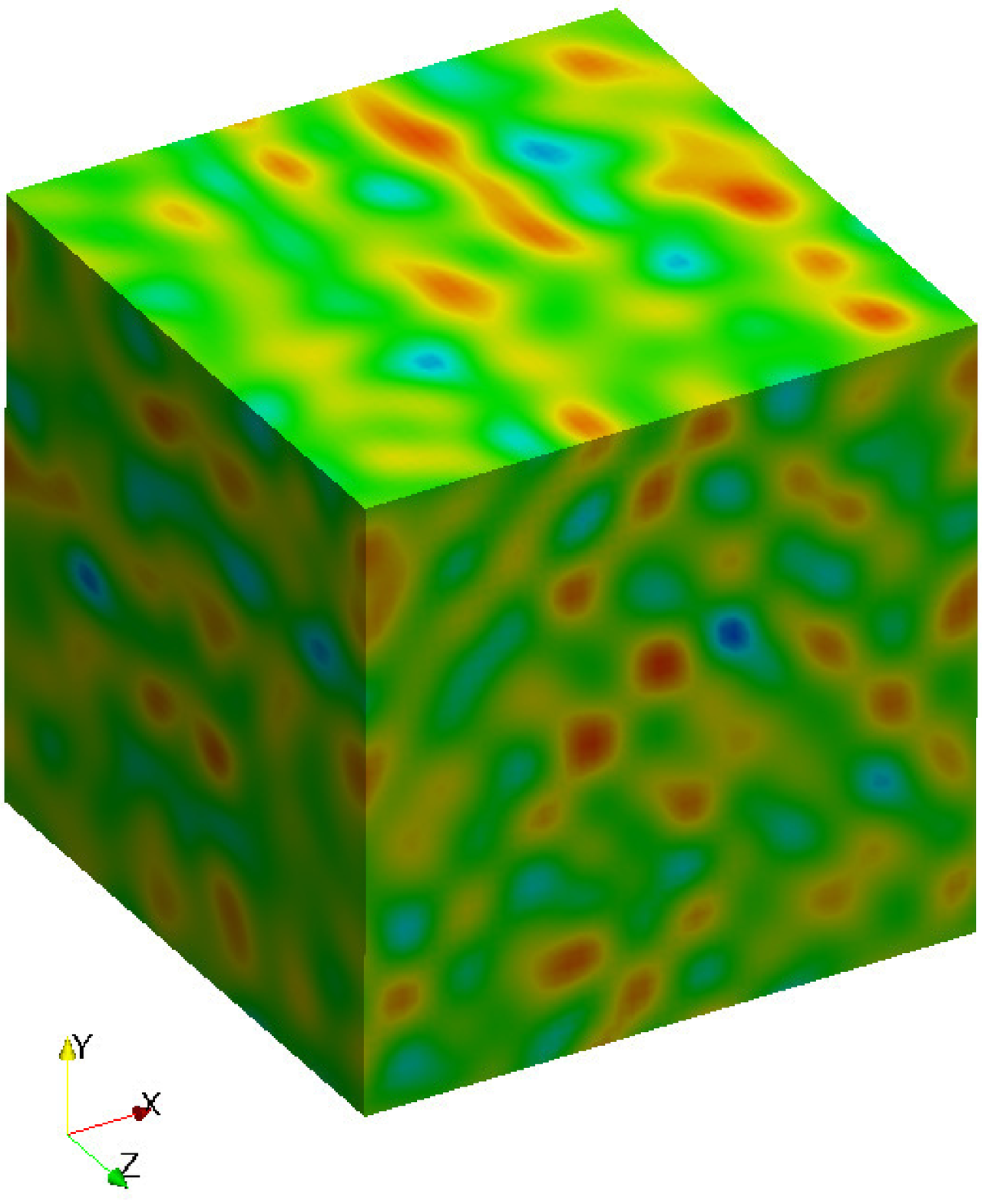}
}
\subfigure[\textit{t} = 0.09 ns]
{
\includegraphics[trim=20mm 20mm 20mm 40mm,clip, width=0.16\textwidth]{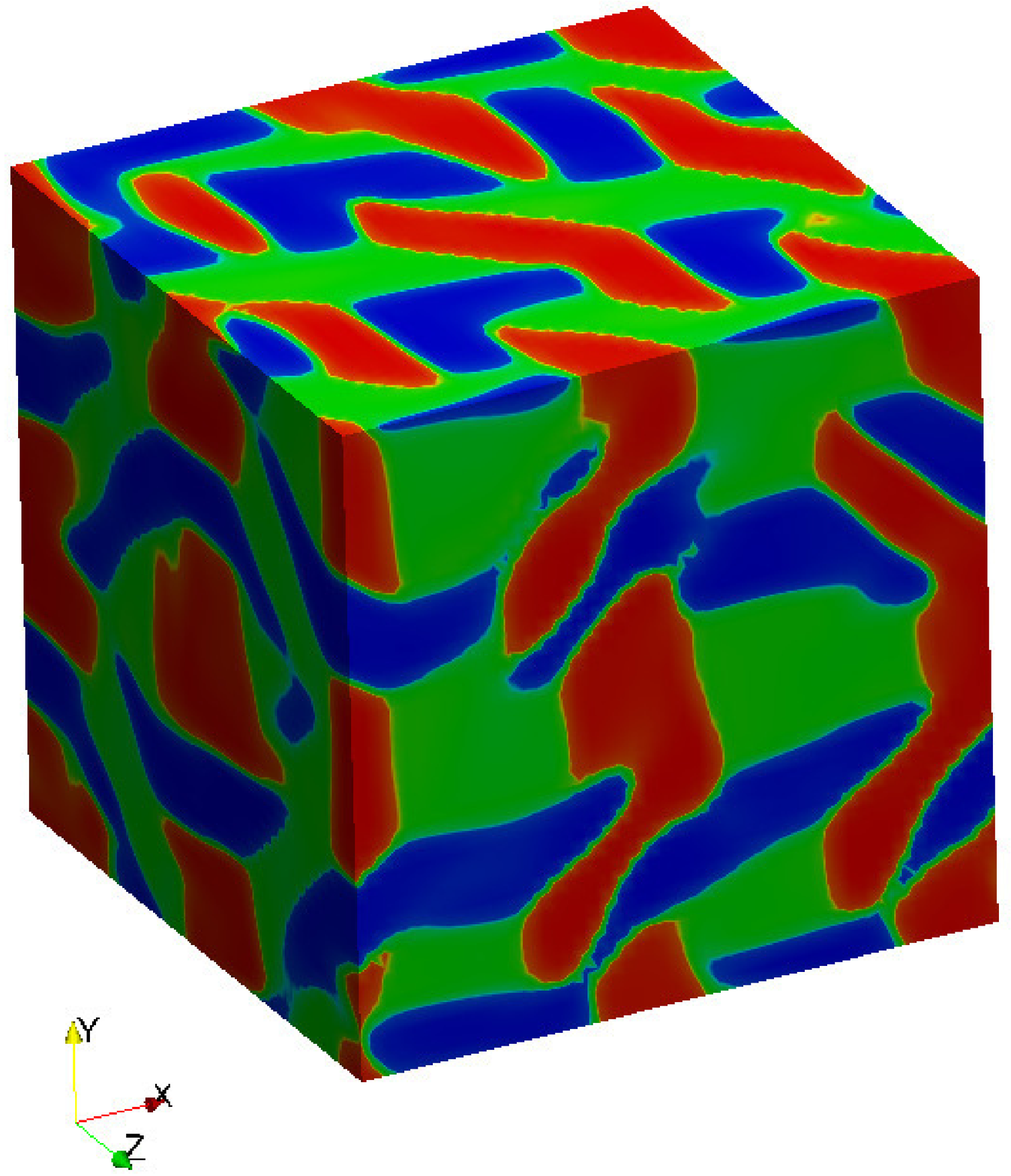}
}
\subfigure[\textit{t} = 0.18 ns]
{
\includegraphics[trim=20mm 20mm 20mm 40mm,clip, width=0.16\textwidth]{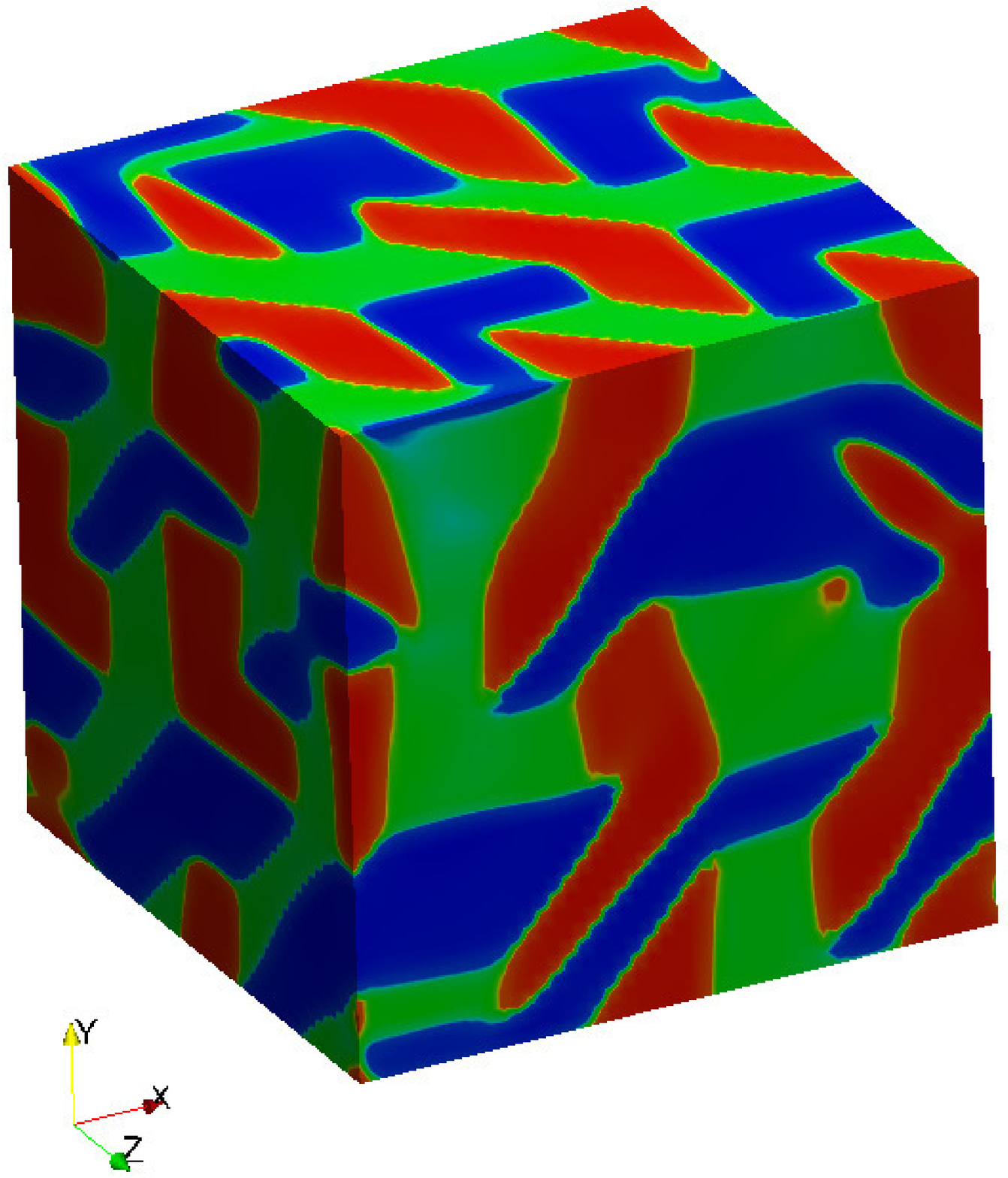}
}
\subfigure[\textit{t} = 0.27 ns]
{
\includegraphics[trim=20mm 20mm 20mm 40mm,clip, width=0.16\textwidth]{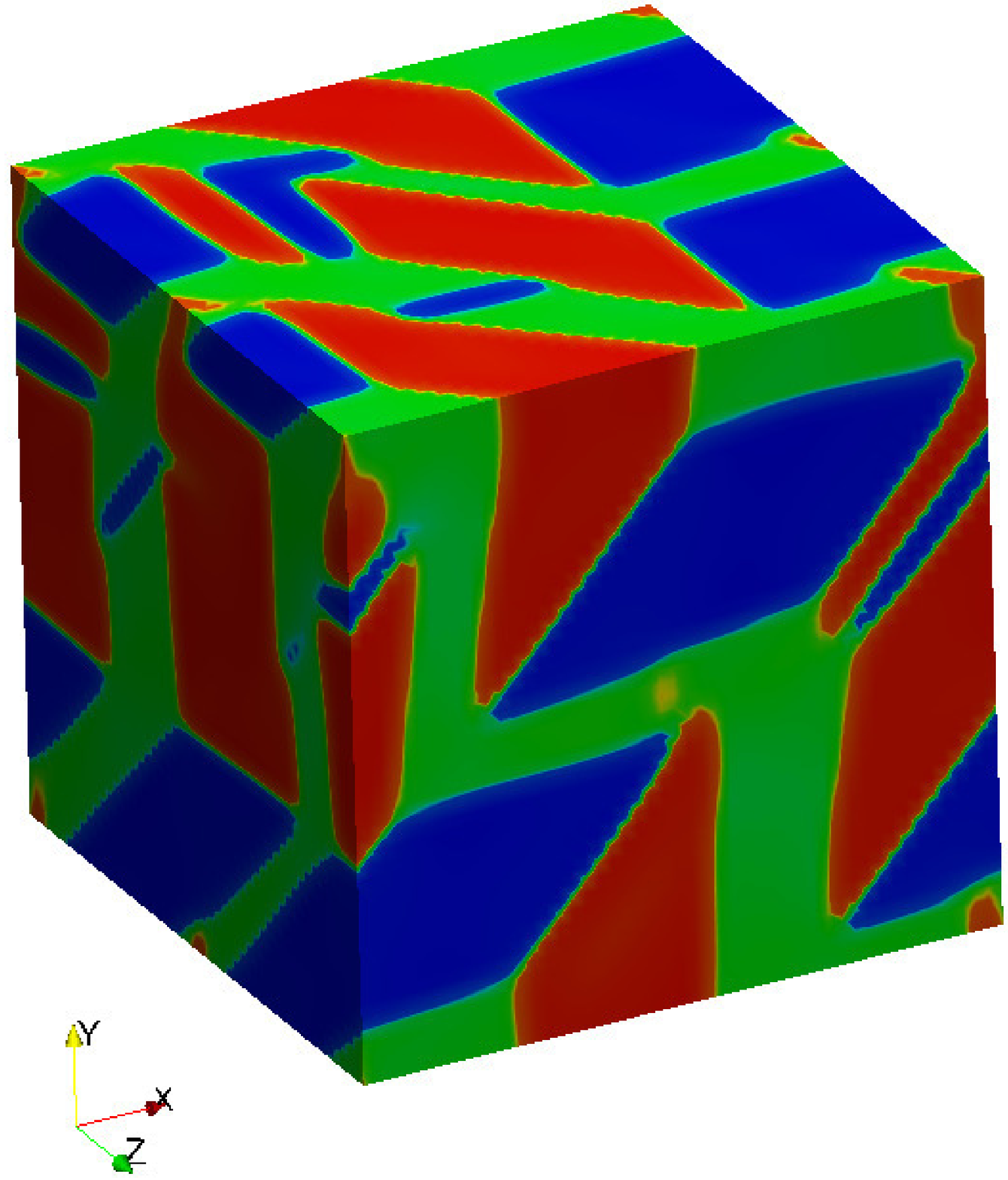}
}
\subfigure[\textit{t} = 1 ns]
{
\includegraphics[trim=20mm 20mm 20mm 40mm,clip, width=0.16\textwidth]{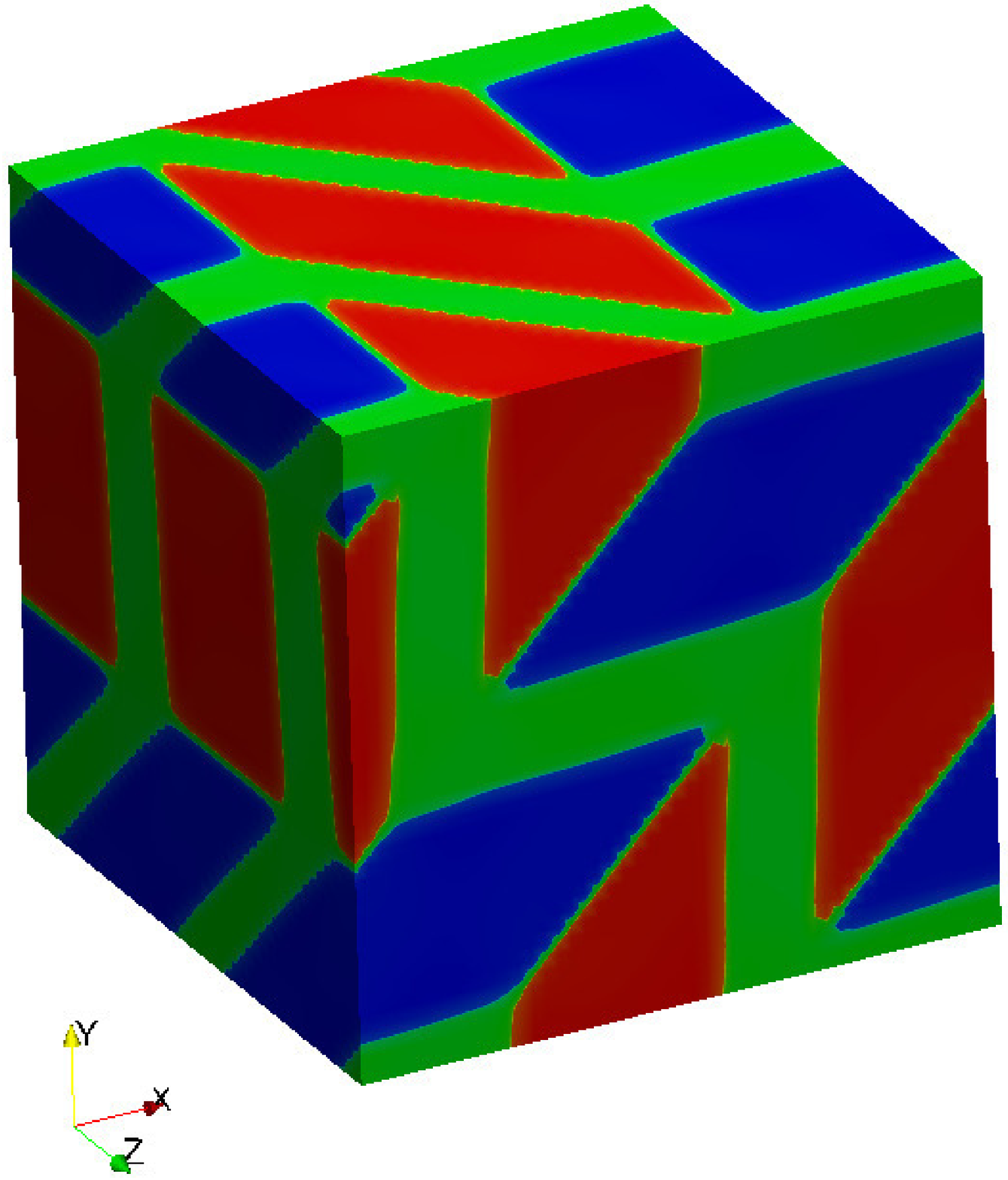}
}
\caption{(Color online) Microstructure morphology evolution in a cube specimen with side 60 nm (red, blue, and green colors represent M$_1$, M$_2$, and M$_3$ variants, respectively).}
\label{fig:DiffGeomCube}
\end{figure}

Fig. \ref{fig:DiffGeomCube} shows snapshots of the transient microstructure morphology evolution in the cube domain (\lx=\ly=\lz= 60 nm). The martensitic variants M$_1$, M$_2$, and M$_3$ are represented in red, blue, and green color, respectively. Martensitic variants nucleate at different places autocatalytically and coalesce to form the domain walls along cubic \pooz planes. Owing to the fully periodic boundary conditions, the three martensitic variants exist in approximately equal proportions. The microstructure morphology results in a competition between dilatation, shear and gradient energies. The stabilized microstructure after sufficiently long time, seen in Fig. \ref{fig:DiffGeomCube}(e)  for $t$ = 1 ns, shows chevron or herringbone or multiply banded structures of martensitic variants. In particular, on the $\Gamma_{x_3}$(+) surface (the front surface in Fig. \ref{fig:DiffGeomCube}(e)), the primary bands of  M$_1$ and M$_2$ are intersected by the M$_3$ variant at 90$^{\circ}$. The ratio of the widths of primary and secondary bands, i.e., M$_1$:M$_3$ and  M$_2$:M$_3$ are approximately 2:1. A similar ratio was found in \cite{jacobs2003simulations} and the experimental references therein. The domain wall orientations  are in accordance with experiments \cite{vlasova2010ferroelastic}, crystallographic theory \cite{Sapriel1975} and other models in the literature \cite{jacobs2003simulations,Idesman2008}. During the microstructure evolution, the temperature increase is observed due to the insulated boundary conditions and thermo-mechanical coupling. The time evolution of the average temperature coefficient $ \tau $ in the cube specimen is shown in Fig. \ref{fig:DiffGeomTauEvolution}(a) in blue color.

\subsubsection{Slab Geometry} \label{sec:DiffGeomSlab}
\begin{figure}[h!]
\centering
\subfigure[\textit{t} = 0.036 ns]
{
\includegraphics[trim=20mm 10mm 20mm 10mm,clip, width=0.16\linewidth]{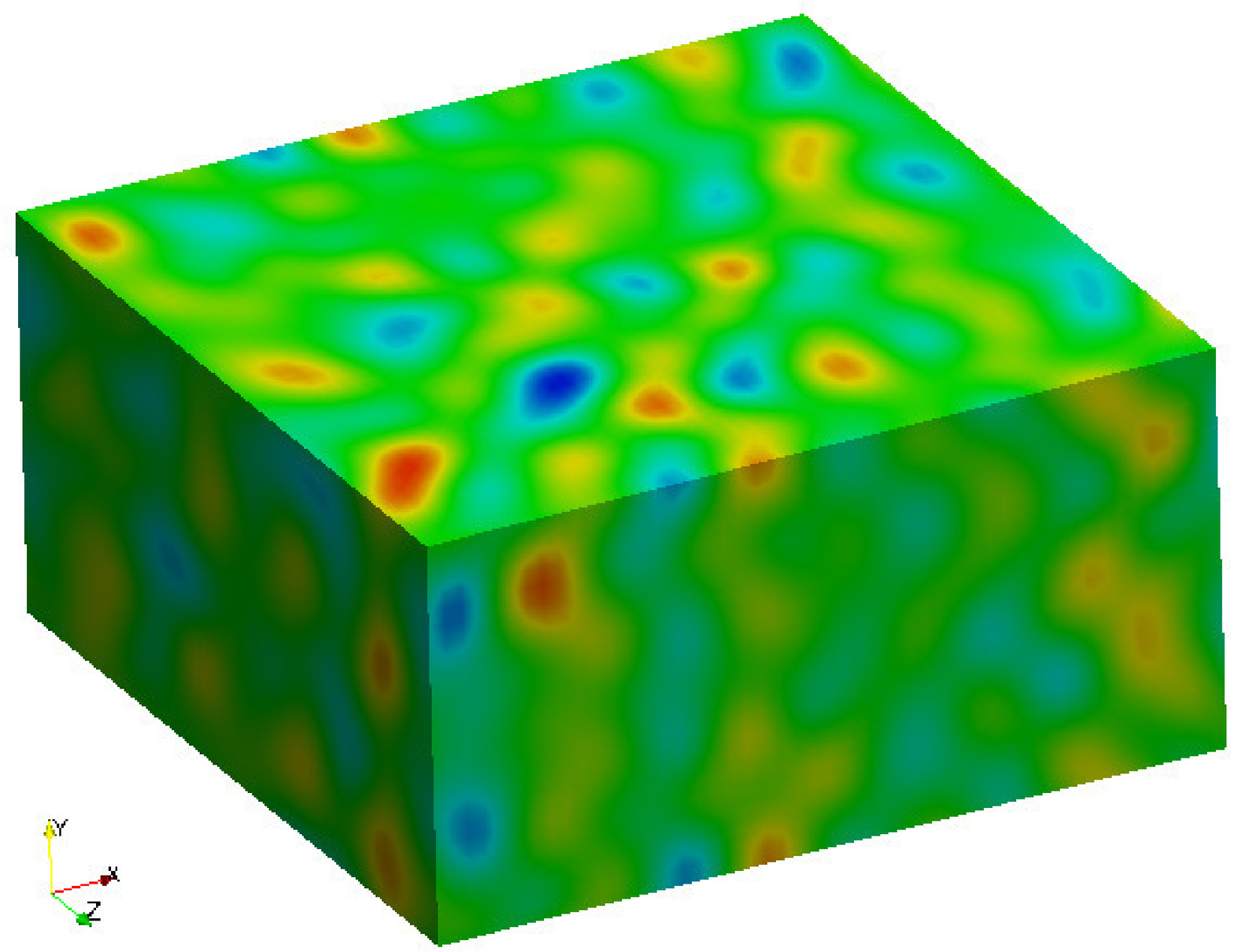}
}
\subfigure[\textit{t} = 0.09 ns]
{
\includegraphics[trim=20mm 10mm 20mm 10mm,clip, width=0.16\textwidth]{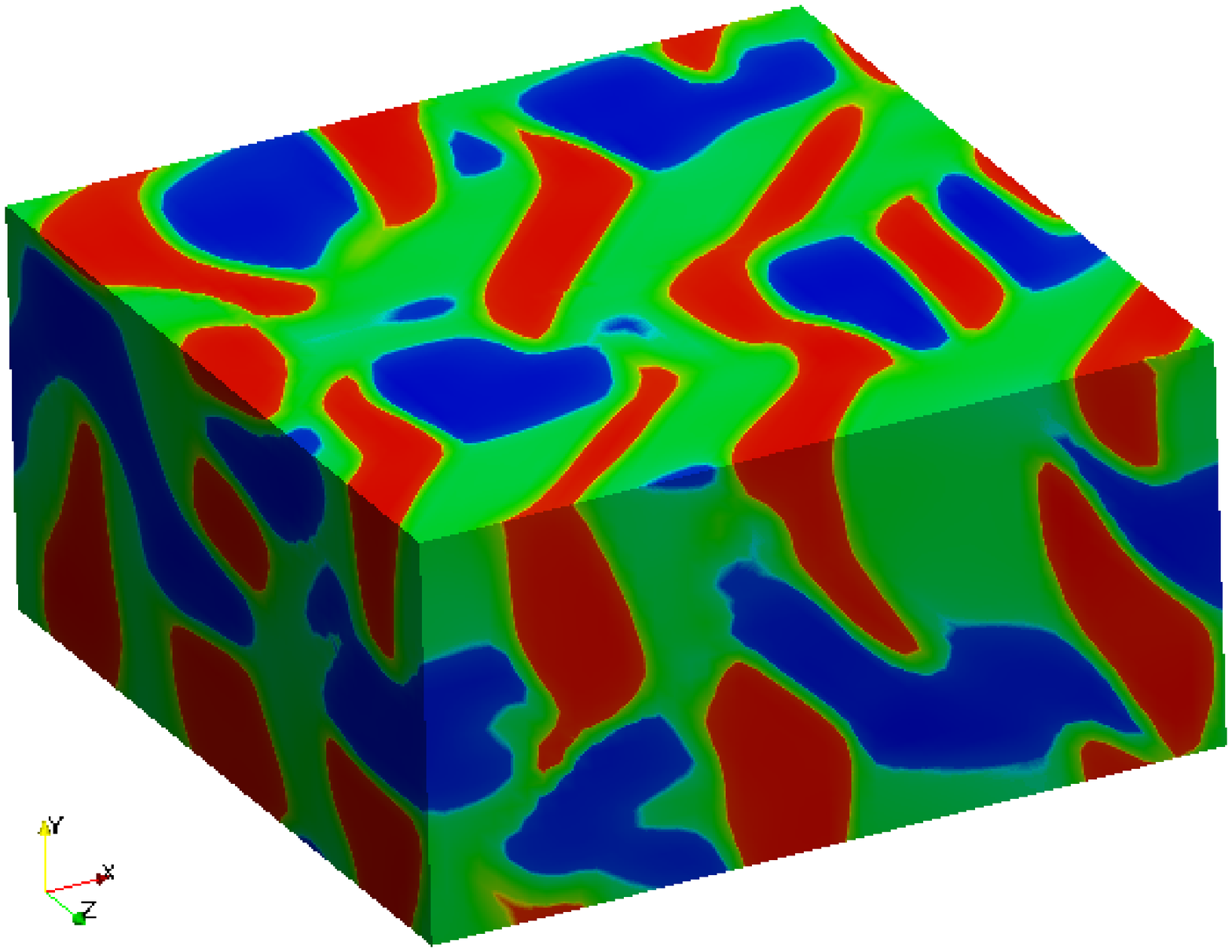}
}
\subfigure[\textit{t} = 0.18 ns]
{
\includegraphics[trim=20mm 10mm 20mm 10mm,clip, width=0.16\textwidth]{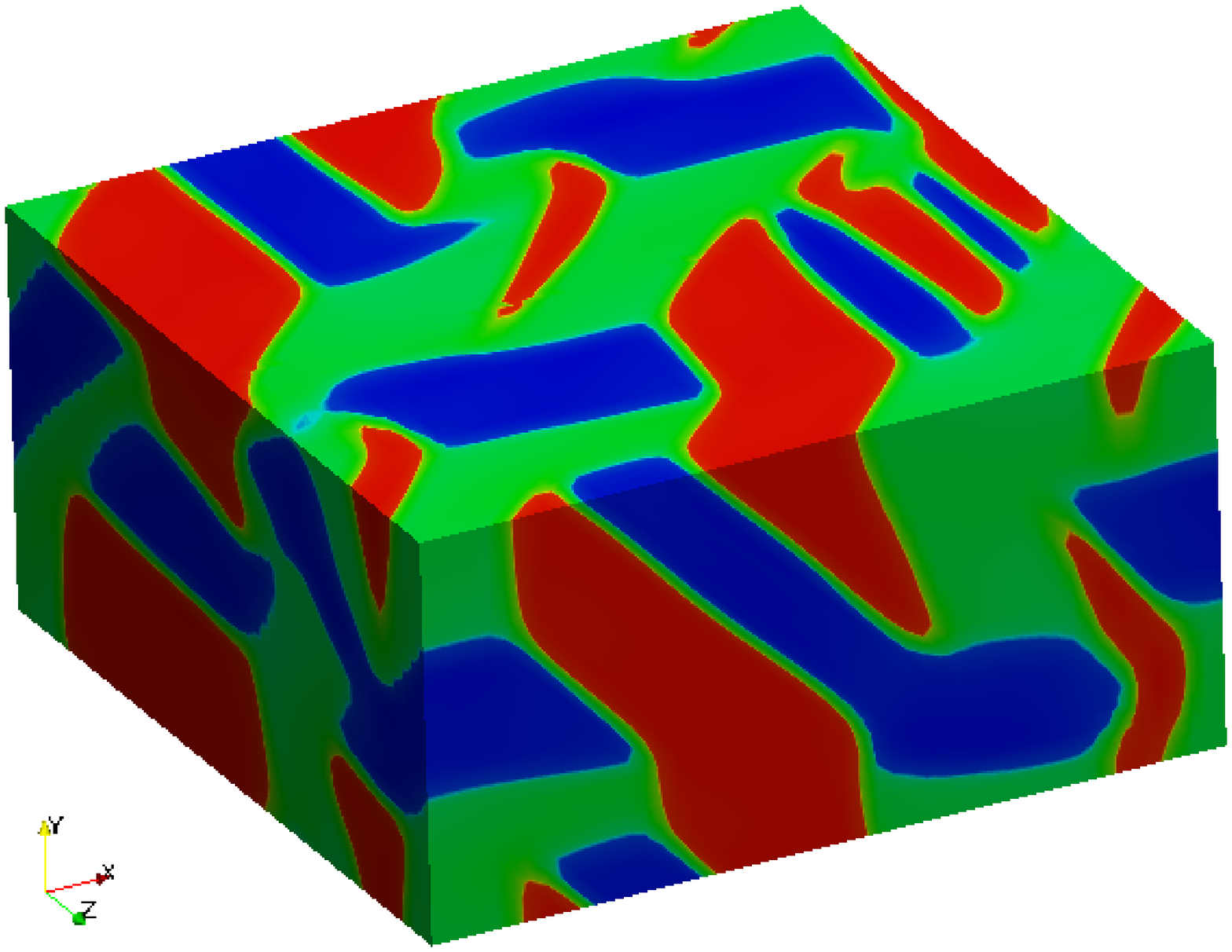}
}
\subfigure[\textit{t} = 0.27 ns]
{
\includegraphics[trim=20mm 10mm 20mm 10mm,clip, width=0.16\textwidth]{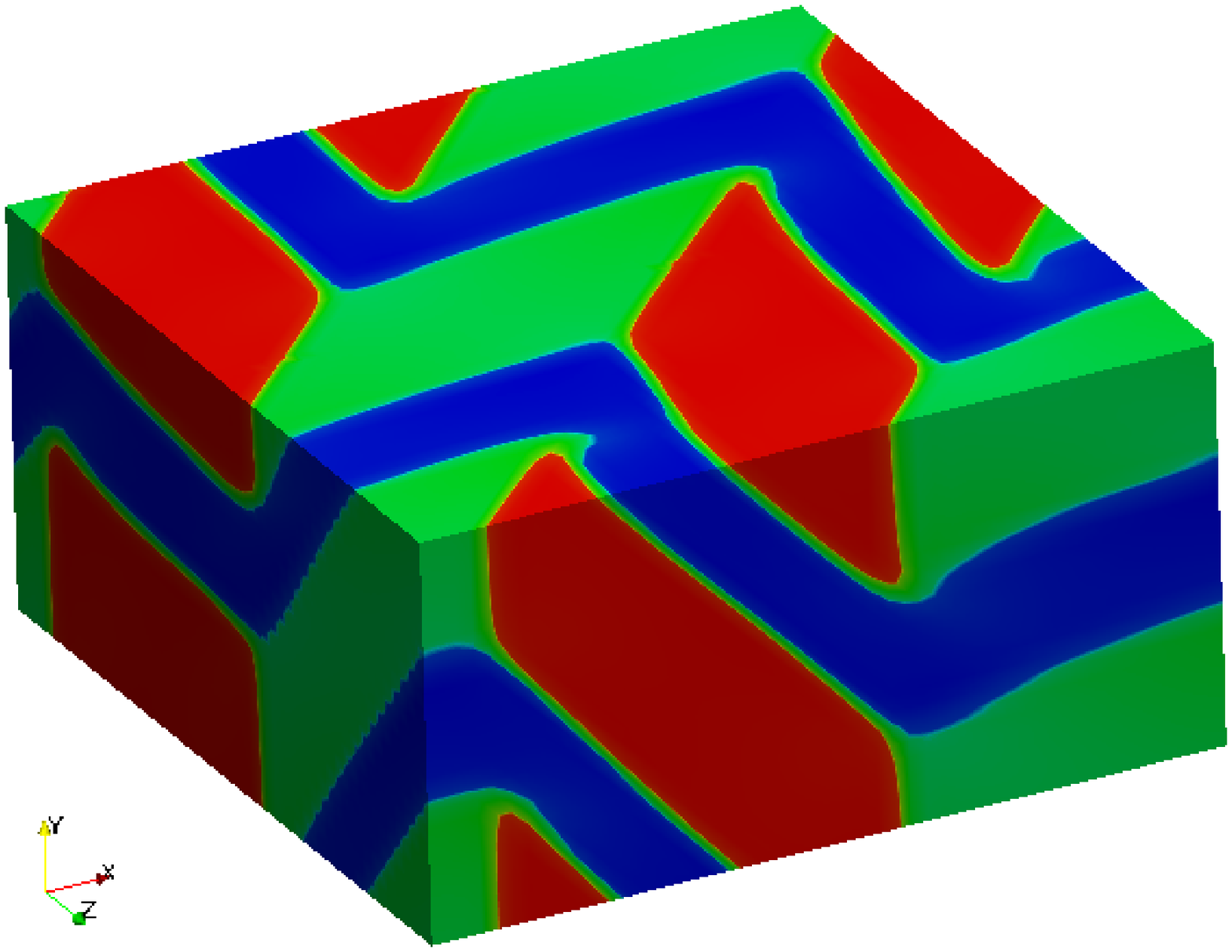}
}
\subfigure[\textit{t} = 1 ns]
{
\includegraphics[trim=20mm 10mm 20mm 10mm,clip, width=0.16\textwidth]{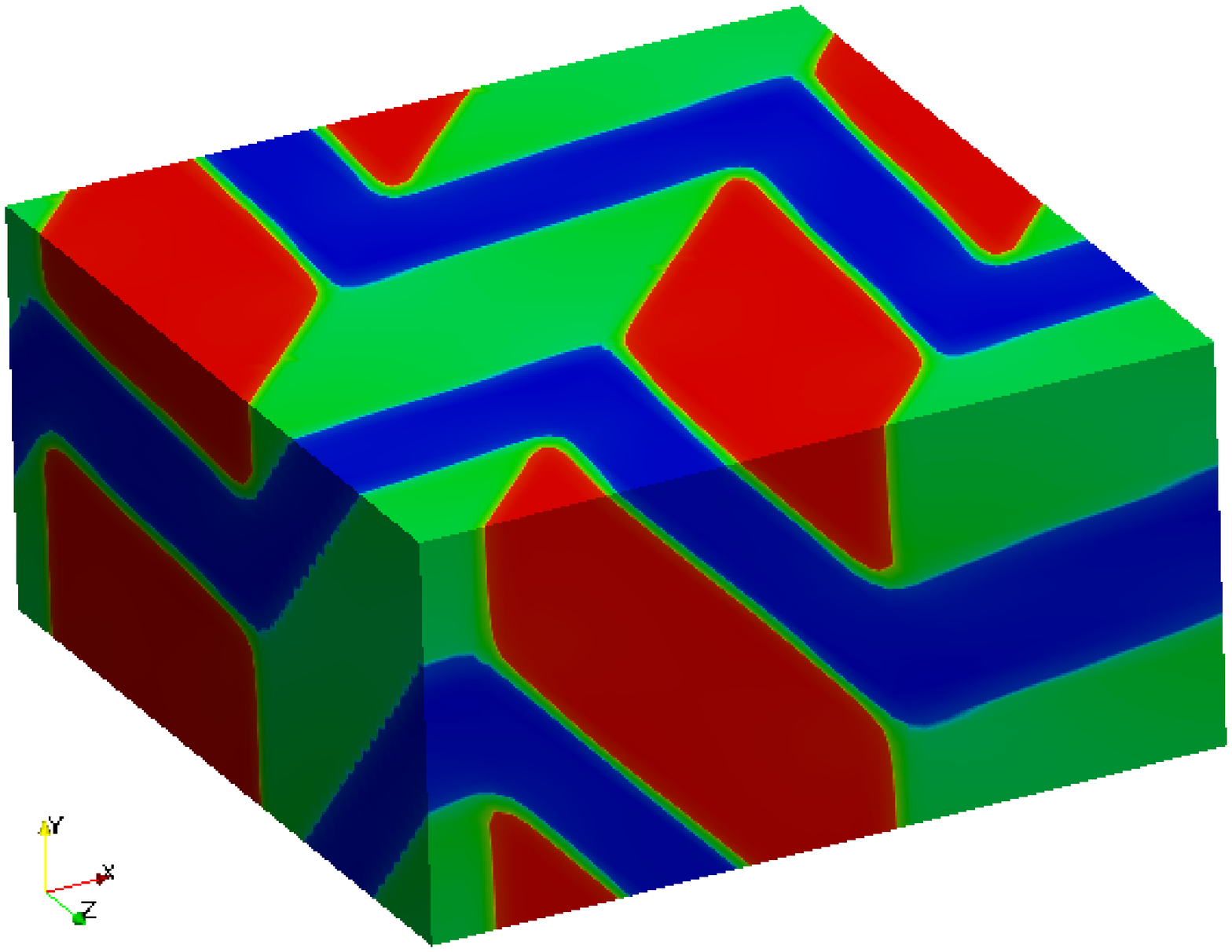}
}
\caption{(Color online) Microstructure morphology evolution in a slab specimen with \lx = \lz = 60 nm and \ly = 30 nm (red, blue, and green colors represent M$_1$, M$_2$, and M$_3$ variants, respectively).}
\label{fig:DiffGeomSlab}
\end{figure}

Now, if the domain size is reduced to half in one of the directions (\lx=\lz= 60 nm and \ly = 30 nm)  and microstructures are allowed to evolve in the slab domain, we obtain morphology evolution as shown in Fig. \ref{fig:DiffGeomSlab}. The three variants exist in approximately equal proportions forming chevron patterns as shown in Fig. \ref{fig:DiffGeomSlab}(e). The microstructure in the slab domain (Fig. \ref{fig:DiffGeomSlab}(e)) is morphologically different than that of the cube domain (Fig. \ref{fig:DiffGeomCube}(e)). The  primary bands of M$_1$ and M$_3$ variants are prominent and intersected by a thin secondary band of the M$_2$ variant at 90$^{\circ}$. The width ratios of primary and secondary bands are 2:1, as in the cube specimen. The time evolution of the average temperature coefficient $ \tau $ in the slab specimen is shown in Fig. \ref{fig:DiffGeomTauEvolution}(a) in red color. The simulations on the cube and slab domains illustrate the influence of specimen geometry on microstructure evolution and morphology. 
 
\subsubsection{Cylindrical Tube Geometry} \label{sec:DiffGeomTube}
\begin{figure}[h!]
\centering
\subfigure[\textit{t} = 0.09 ns]
{
\includegraphics[trim=5mm 5mm 5mm 10mm,clip, width=0.2\linewidth]{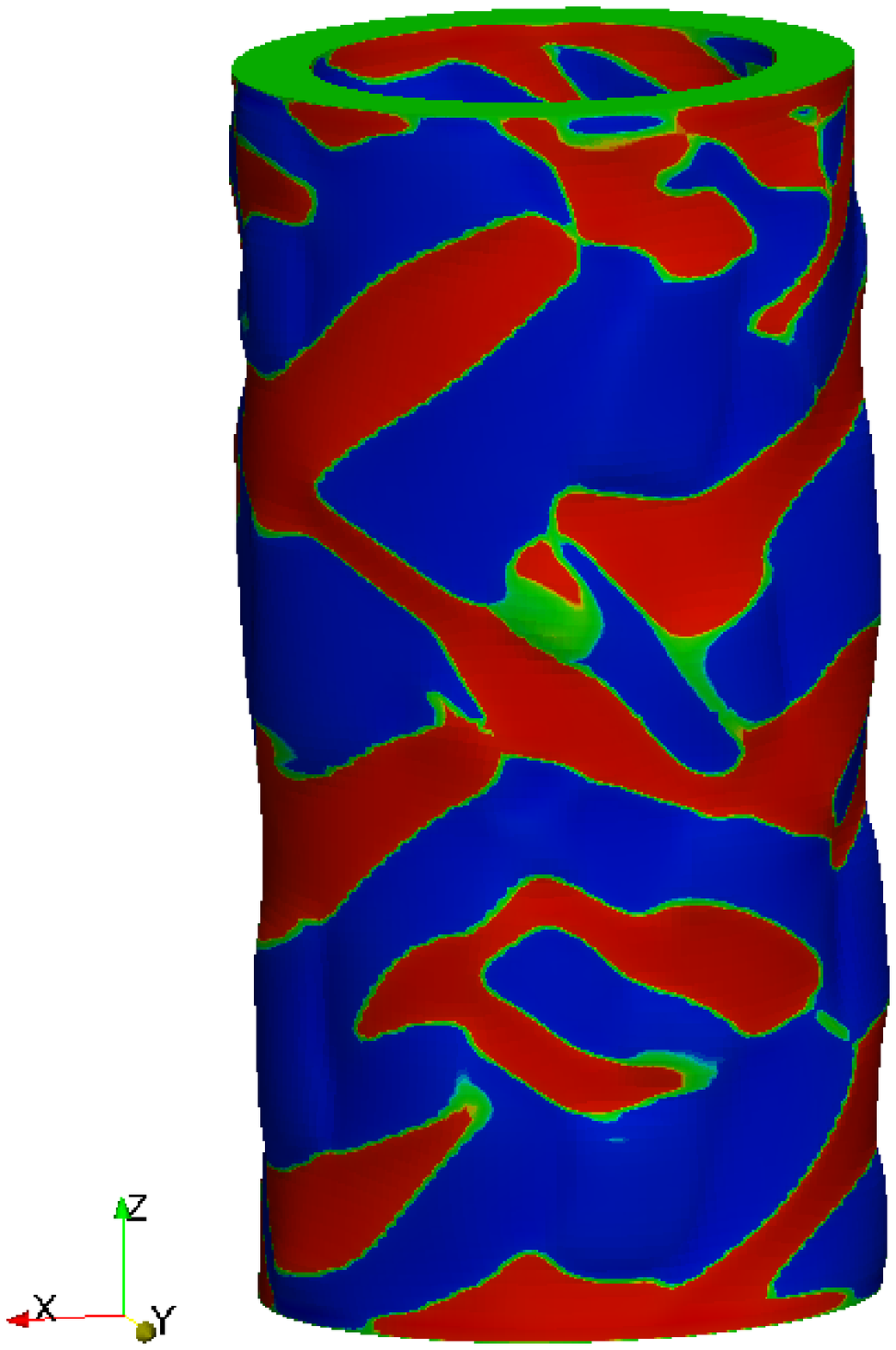}
}
\subfigure[\textit{t} = 0.18 ns]
{
\includegraphics[trim=30mm 5mm 5mm 10mm,clip, width=0.158\textwidth]{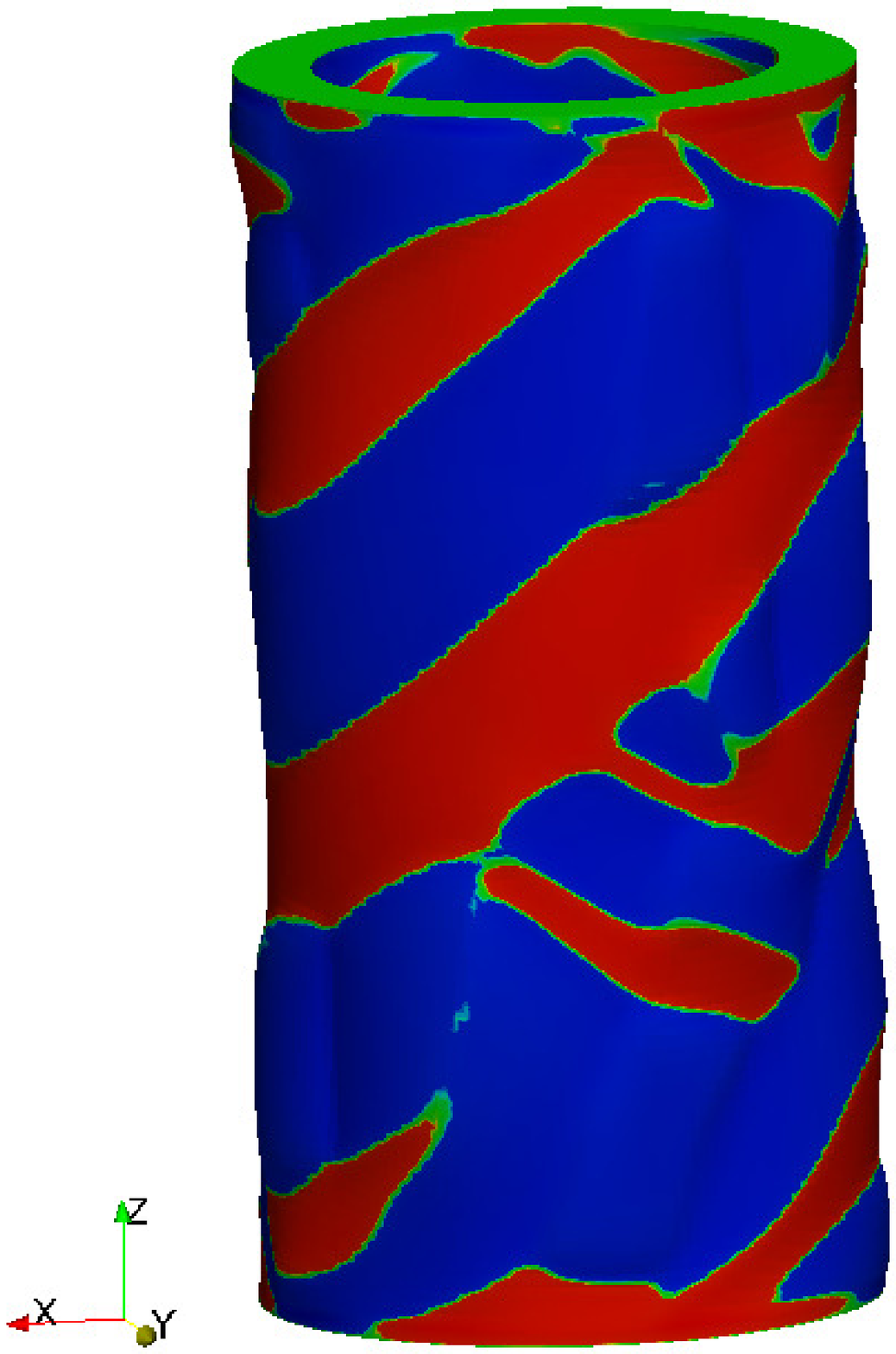}
}
\subfigure[\textit{t} = 0.27 ns]
{
\includegraphics[trim=30mm 5mm 5mm 10mm,clip, width=0.158\textwidth]{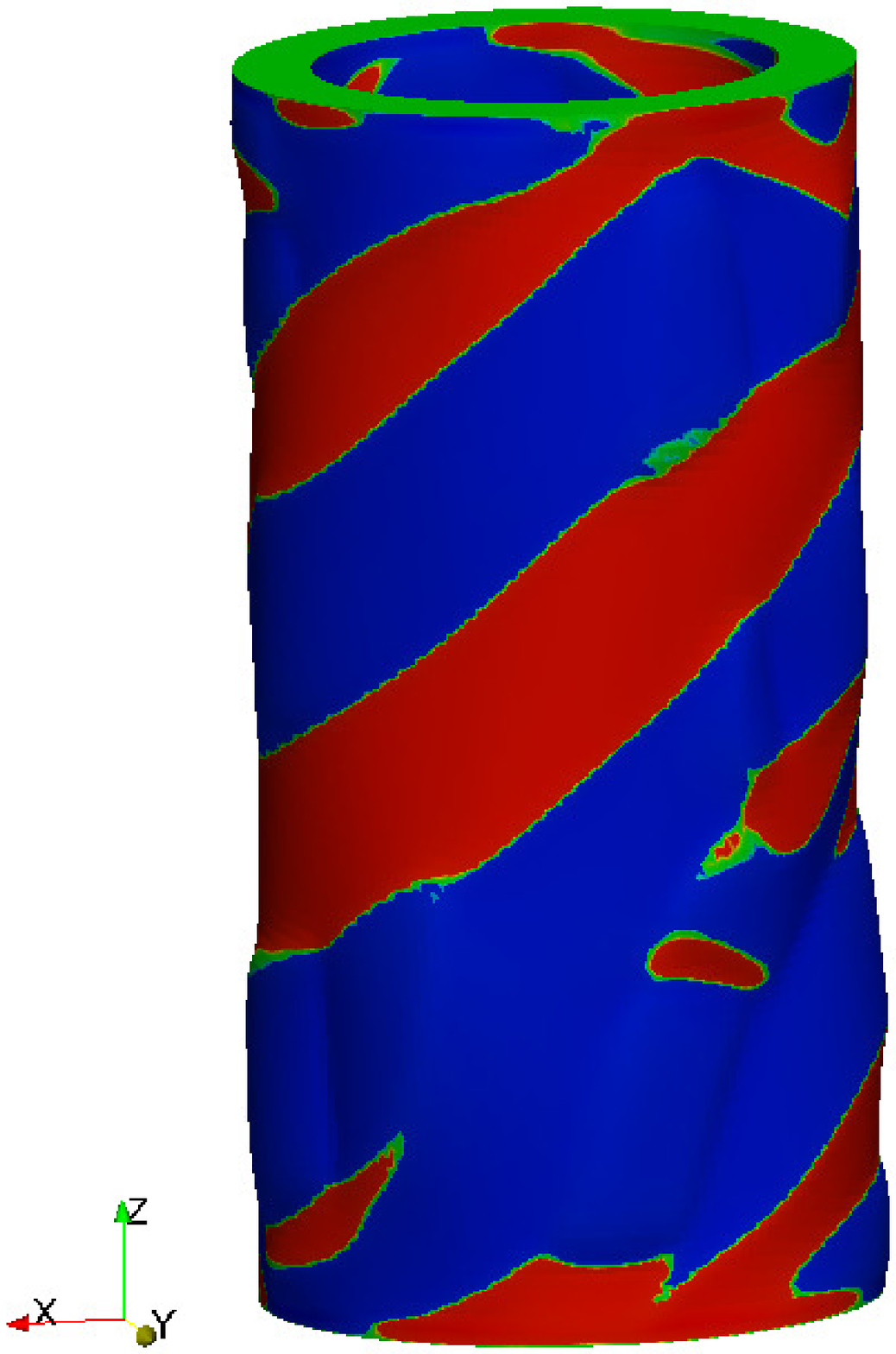}
}
\subfigure[\textit{t} = 0.45 ns]
{
\includegraphics[trim=30mm 5mm 5mm 10mm,clip, width=0.158\textwidth]{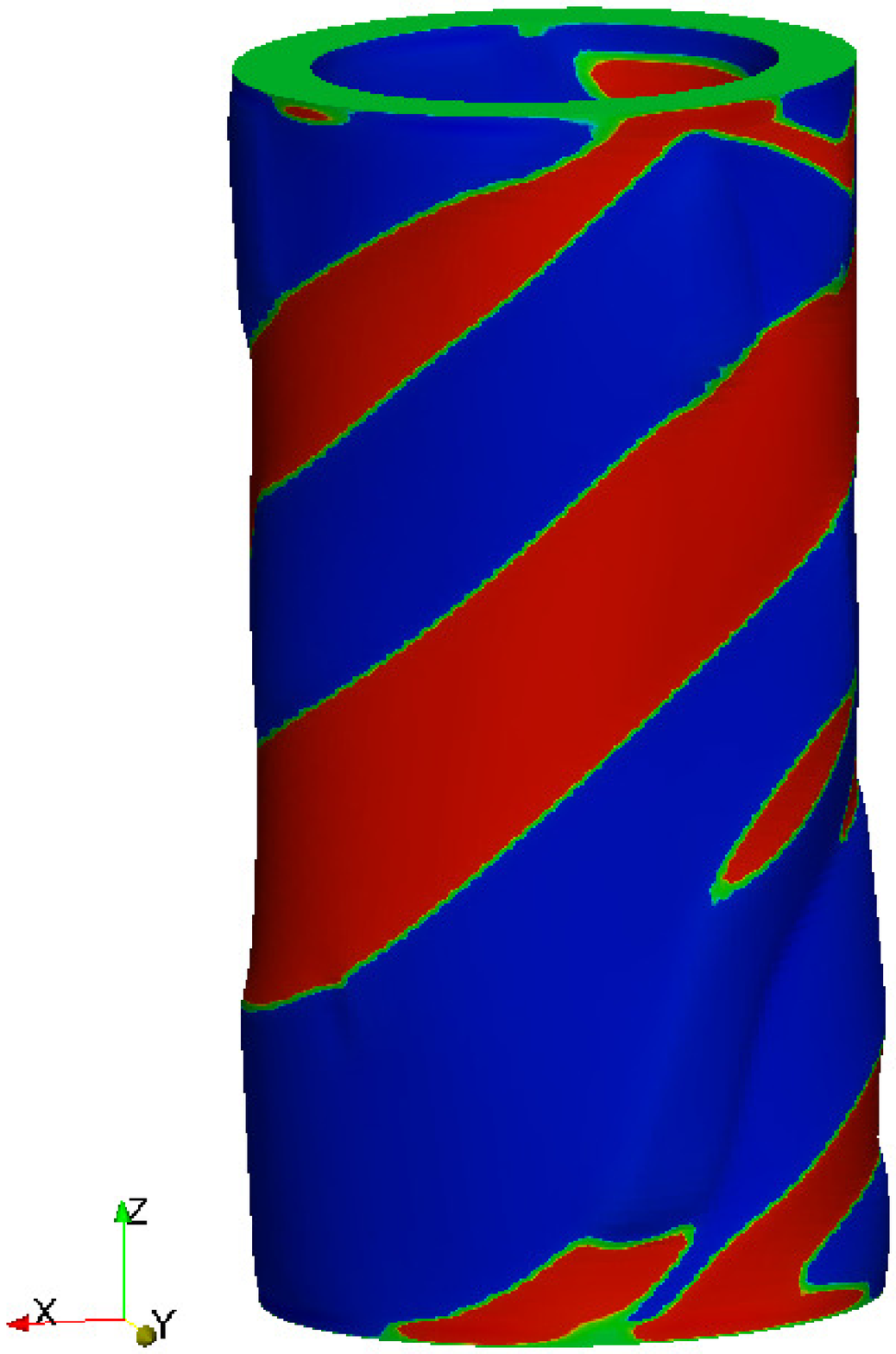}
}
\subfigure[\textit{t} = 0.63 ns]
{
\includegraphics[trim=30mm 5mm 5mm 10mm,clip, width=0.158\textwidth]{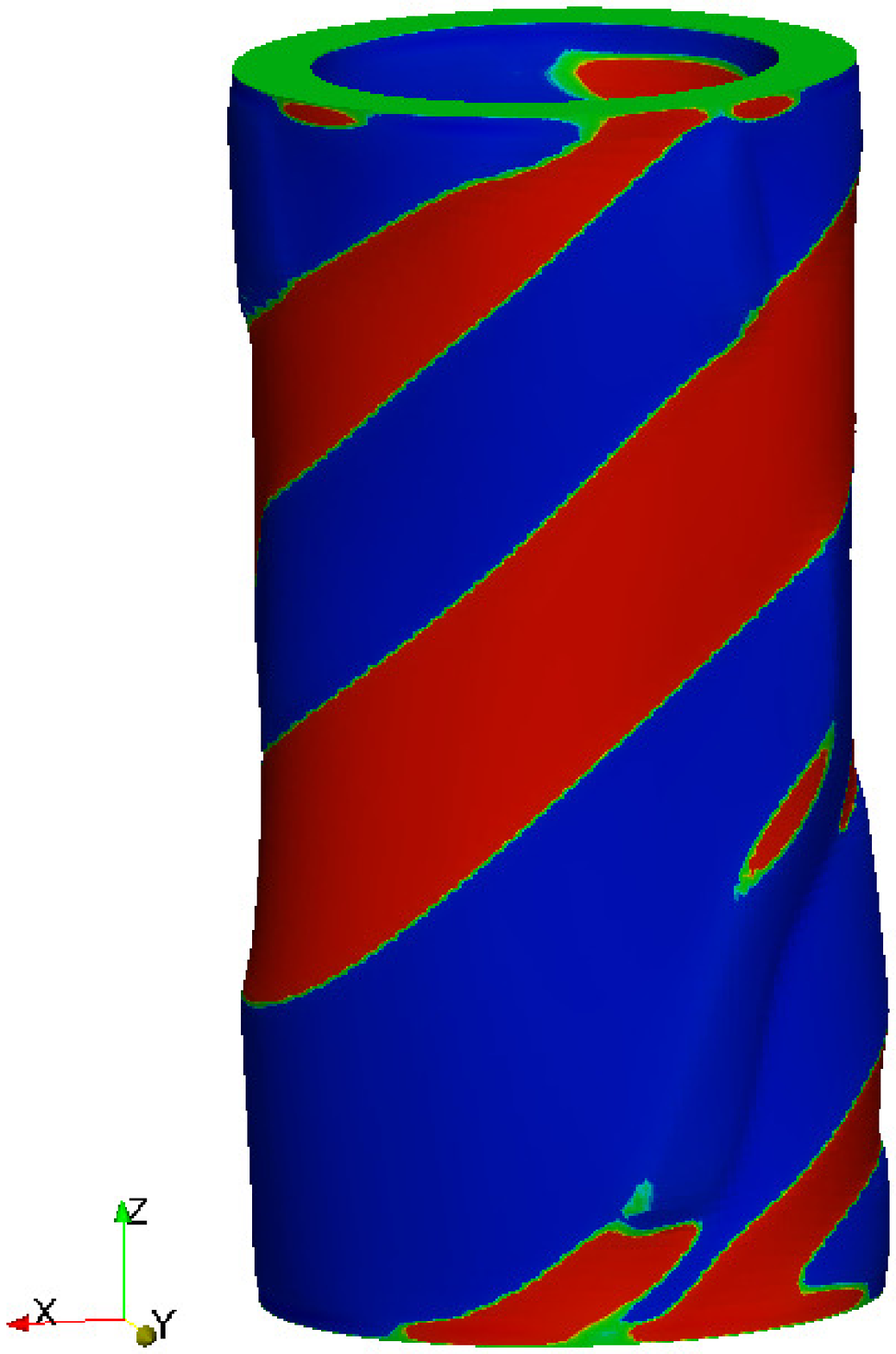}
}
\caption{(Color online) Helix microstructure morphology evolution in a tubular specimen with $R_i$ = 22.5 nm, $R_o$ = 30 nm, and $H$ = 120 nm (red and blue color represent M$_3$, and remaining two variants, respectively). The microstructure is superimposed with the displacement vector $\vect u$ for better clarity.}
\label{fig:DiffGeomTube}
\end{figure}

The numerical simulations on a cylindrical tube domain have a two-fold purpose: (i) to study the microstructure morphology, and (ii) to show the flexibility of IGA to model real-life geometries as those observed in nanotubes for drug delivery \cite{vzuvzek2012electrochemical} and other applications. The simulation has been conducted on a tube specimen with $R_i$ = 22.5 nm, $R_o$ = 30 nm, and $H$ = 120 nm (refer to Fig. \ref{fig:SchematicDomains}(b)) that can be modeled exactly using NURBS. To generate the geometry of the tube we defined first a circular annulus using four NURBS patches as described in \cite{Hughes}. Then, we swept that annulus in the $x_3$ direction using the procedure explained in \cite{Hughes}. To achieve \cone continuity across the different patches we used matching discretizations and built linear constraints into the discrete spaces, so that the solution and weighting functions are automatically \cone across patches. As for boundary conditions, the surfaces $ \Gamma_{x_3}(+) $ and $ \Gamma_{x_3}(-) $ are constrained in displacement with $ \pmb{u}  = \pmb{0}$ and stress-free boundary conditions have been applied on the outer $ \Gamma_{R_o} $ and inner $ \Gamma_{R_i} $ surfaces. The relevant microstructure morphology corresponding to $\partial u_3/\partial x_3$ is  plotted in Fig. \ref{fig:DiffGeomTube}. Note that in Fig. \ref{fig:DiffGeomTube}, the color map that indicates the microstructure has been plotted on the deformed configuration to provide information about the displacements field. In the plots, we observe that the helix microstructure of M$_3$ variant (red color) evolves from the initial random condition. Such helix microstructures have been reported experimentally in tubular SMA specimens \cite{he2009scaling}. Remnant traces of austenite exist near the $ \Gamma_{x_3}(+) $ and $ \Gamma_{x_3}(-) $ surfaces (shown in green) due to the constrained boundary conditions. The time evolution of the average temperature coefficient $ \tau $ is shown in Fig. \ref{fig:DiffGeomTauEvolution}(a) in black color.

\subsubsection{Tubular Torus Geometry} \label{sec:DiffGeomTorus}

\begin{figure}[h!]
\centering
\subfigure[\textit{t} = 0.045 ns]
{
\includegraphics[width=0.22\linewidth]{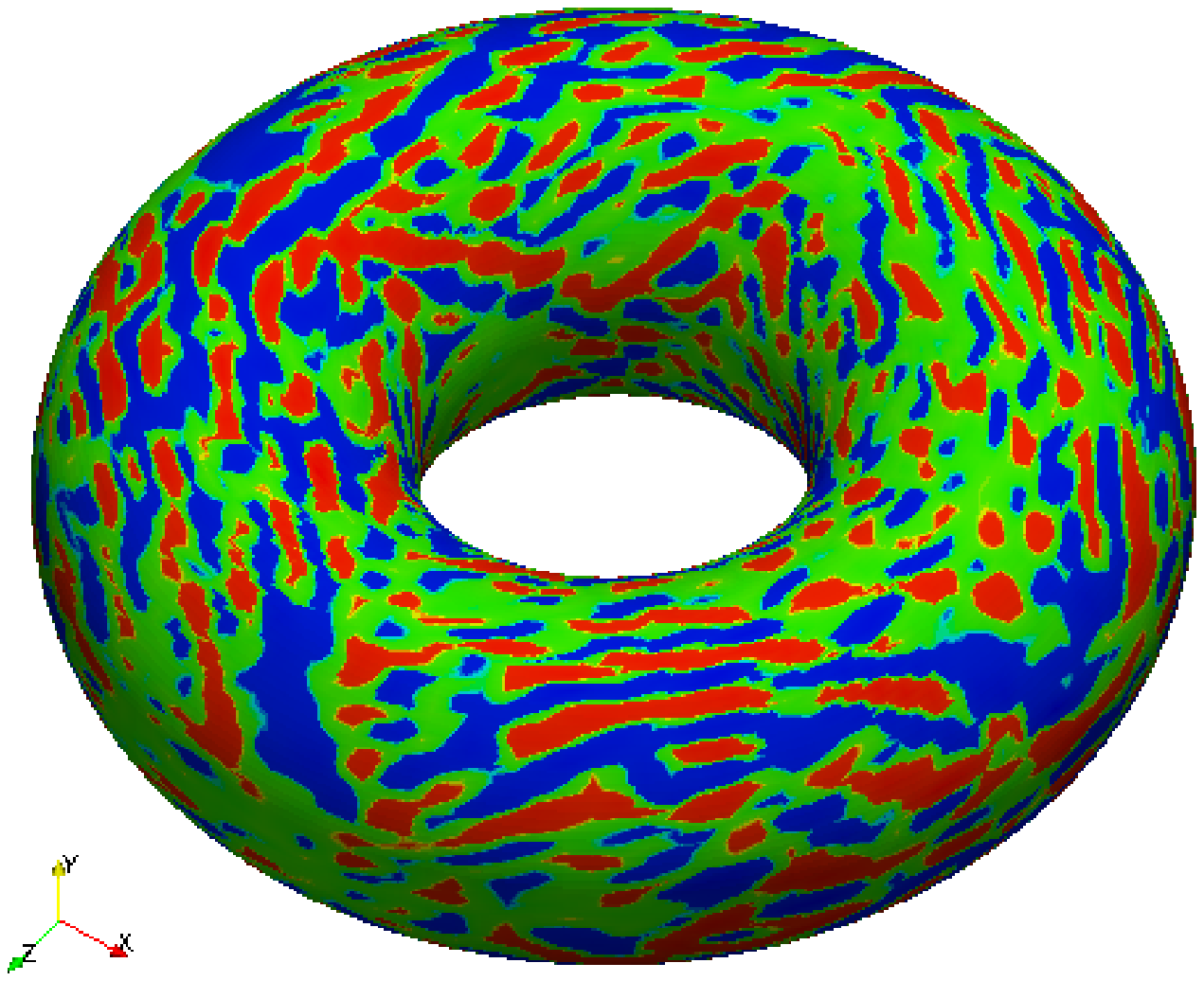}
}
\subfigure[\textit{t} = 0.09 ns]
{
\includegraphics[width=0.22\textwidth]{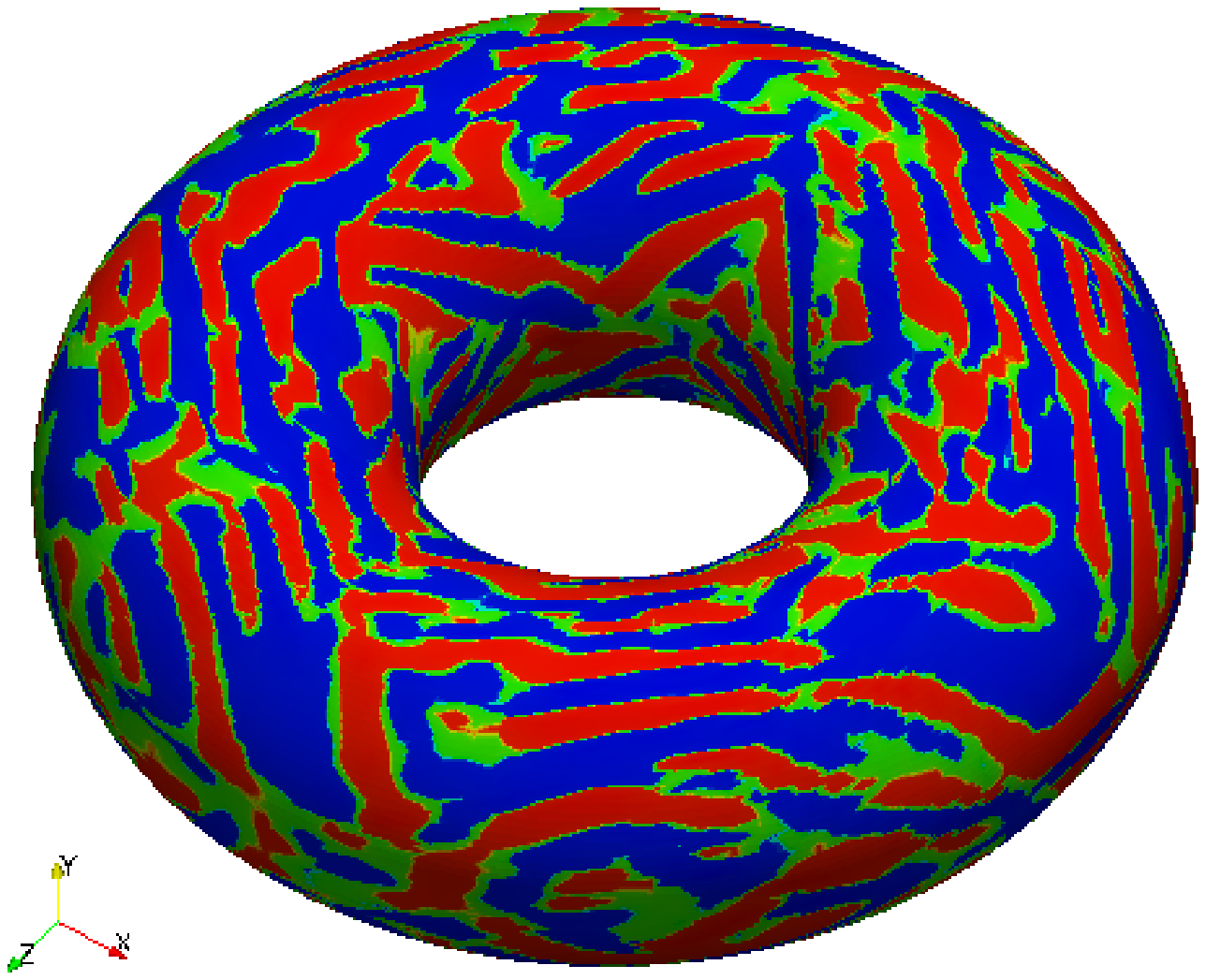}
}
\subfigure[\textit{t} = 0.27 ns]
{
\includegraphics[width=0.22\textwidth]{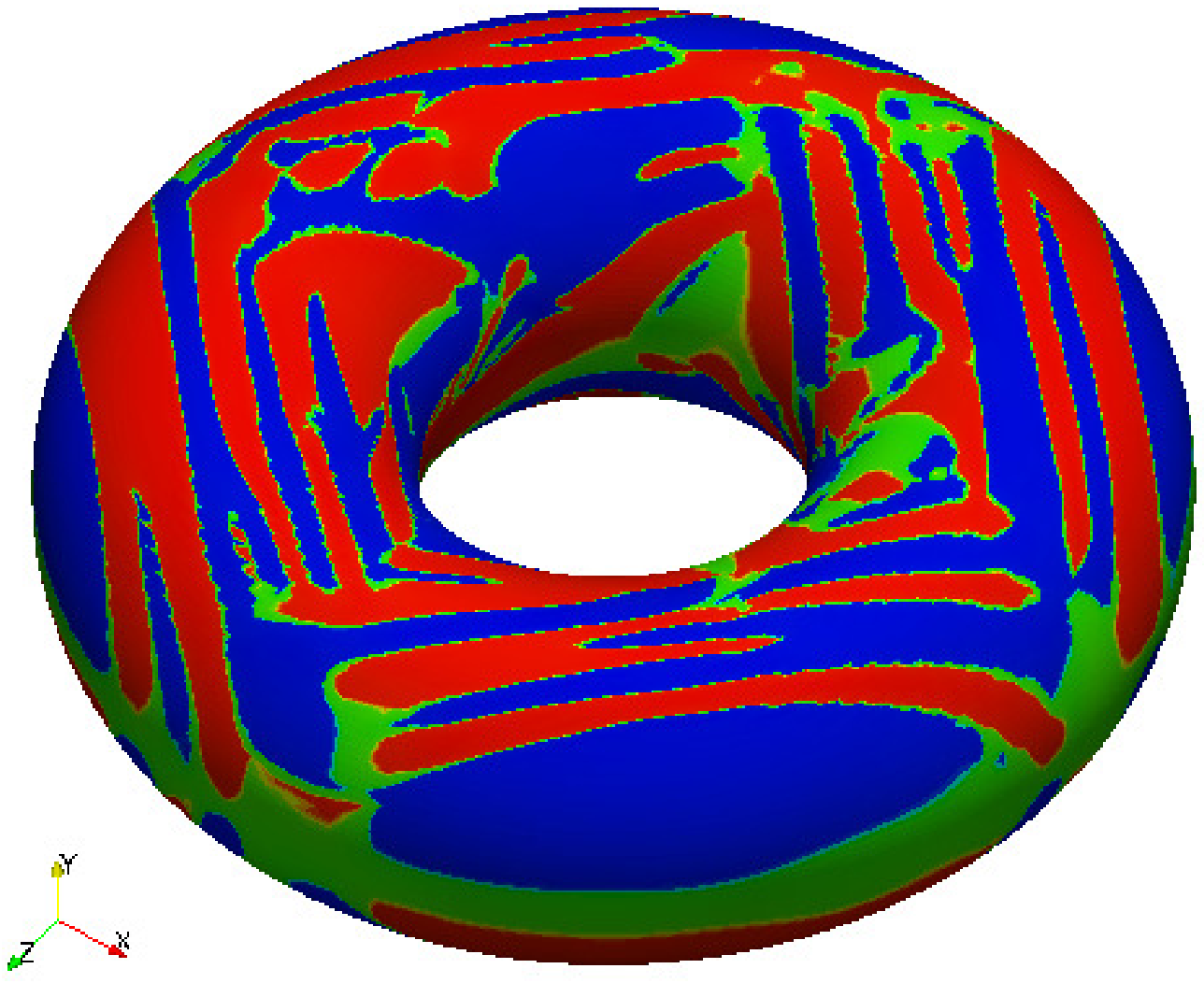}
}
\subfigure[\textit{t} = 1 ns]
{
\includegraphics[width=0.22\textwidth]{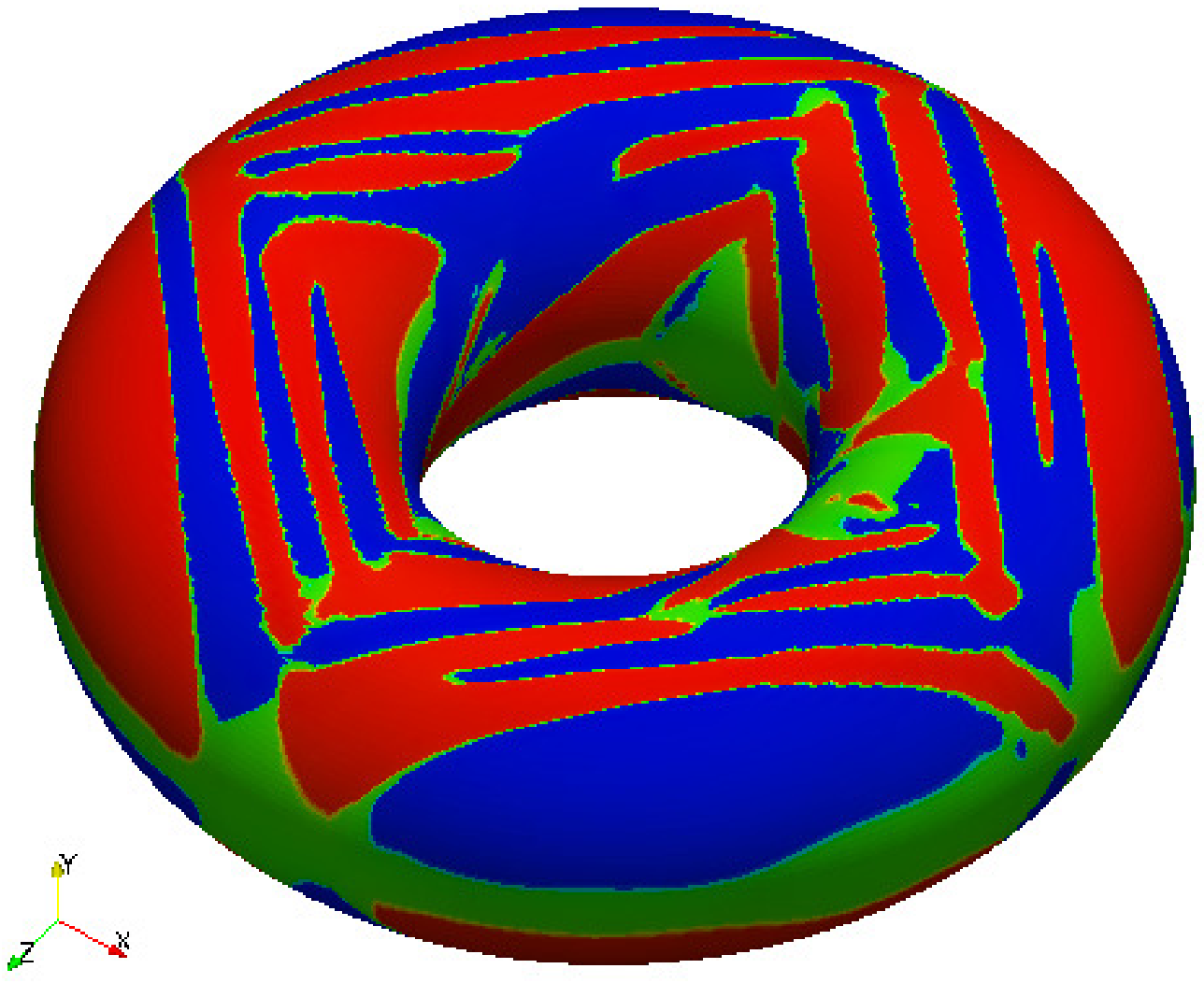}
}
\caption{(Color online) Self-accommodated microstructures in a tubular torus specimen with $R_d$ = 100 nm, $R_{to}$ = 50 nm, and $R_{ti}$ = 25 nm (red, blue, and green colors represent M$_1$, M$_2$, and M$_3$ variants, respectively).  }
\label{fig:DiffGeomTorus}
\end{figure}

\begin{figure}[h!]
\centering
\subfigure[]
{
\includegraphics[width=0.4\textwidth]{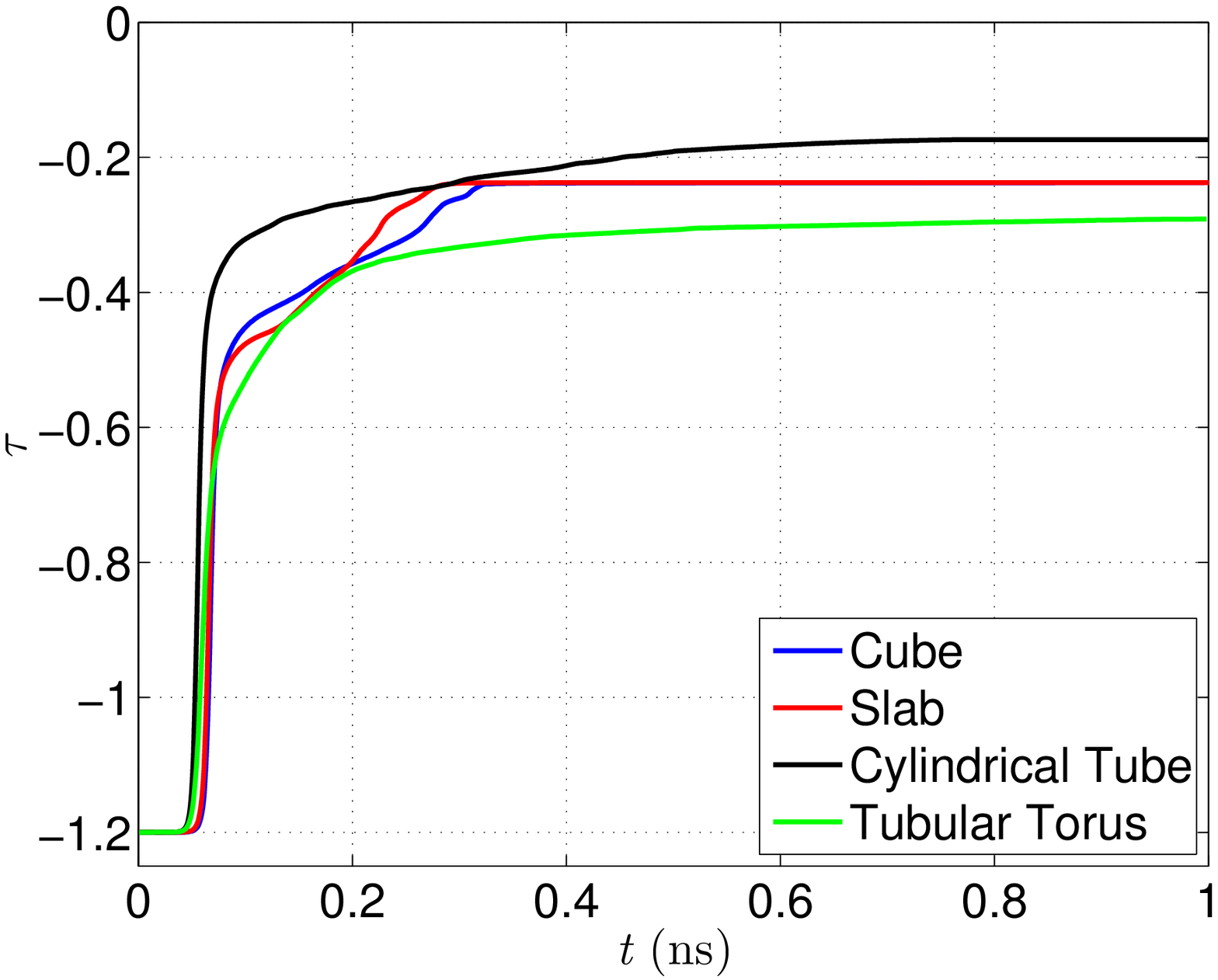}
}
\subfigure[]
{
\includegraphics[width=0.4\textwidth]{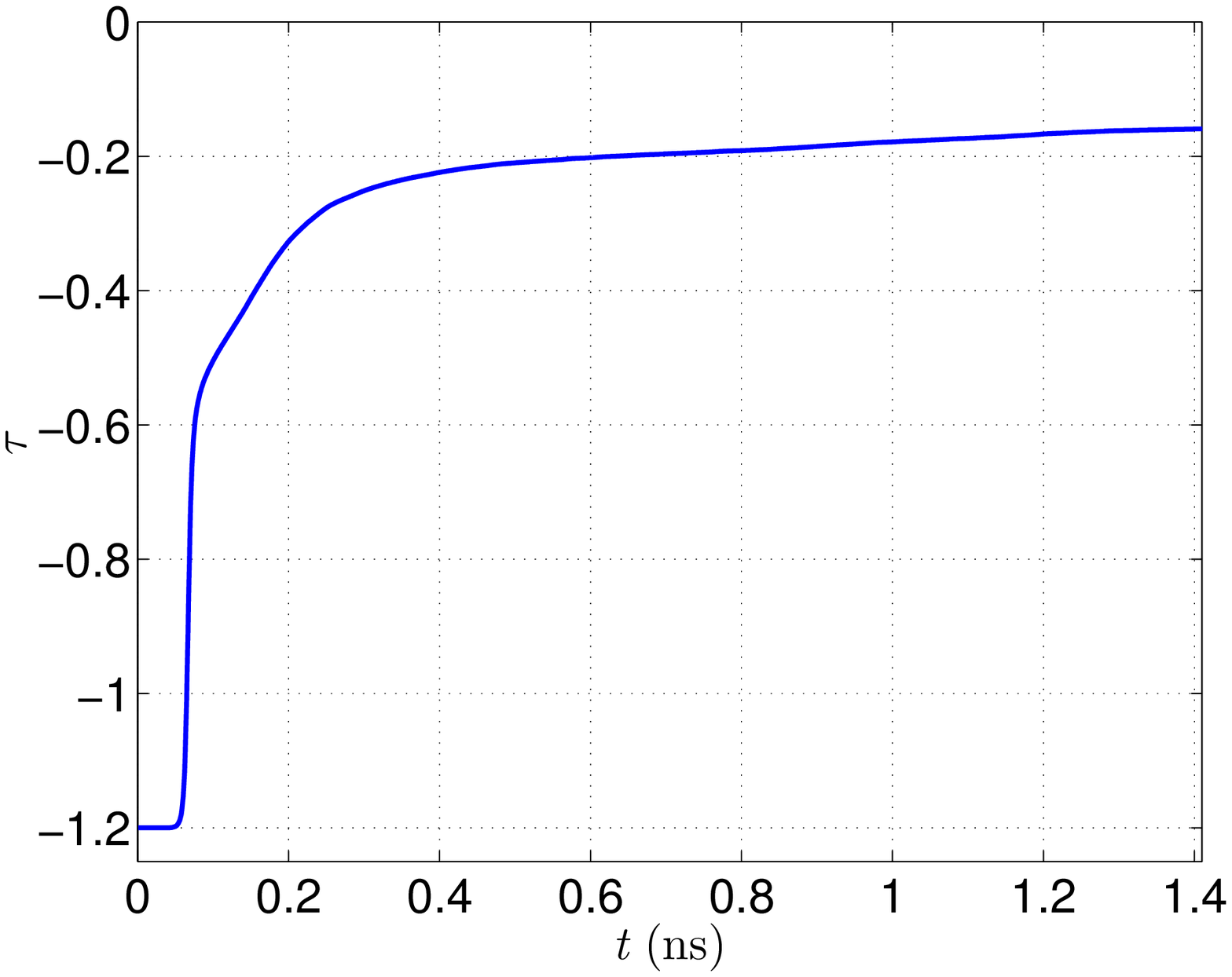}
}
\caption{(Color online) Time evolution of the average temperature coefficient $ \tau $ in a (a) cube (blue), slab (red), cylindrical tube (black) and tubular torus (green) geometries and (b) bigger cube domain.}
\label{fig:DiffGeomTauEvolution}
\end{figure}

The simulation has been conducted on a tubular torus SMA specimen with $ R_d = 100 $ nm, $ R_{to} = 50 $ nm  and $ R_{ti} = 25 $ nm (refer to Fig. \ref{fig:SchematicDomains}(c)), which can be modeled exactly using NURBS. The geometry was generated by sweeping a circular annulus in a circular line as illustrated in \cite{Hughes}. We force the trial solution and weighting functions to be \cone across patches by imposing linear constraints on the spaces. The specimen is constrained in displacements with $ \pmb{u}  = \pmb{0}$ on the inner surface $ \Gamma_{R_{ti}} $. Stress-free boundary conditions have been applied on outer surface $ \Gamma_{R_{to}} $. The self-accommodated microstructure morphology evolution has been presented in Fig. \ref{fig:DiffGeomTorus}. The microstructure evolves into a complex morphology due to the interplay of the phase-transformation physics with the geometry of the computational domain.  The time evolution of the average temperature coefficient $ \tau $ in the tubular torus geometry is shown in Fig. \ref{fig:DiffGeomTauEvolution}(a) in green color.


\subsection{Microstructure Morphology on a Bigger Cube Domain} \label{sec:bigcube}

In Section \ref{sec:DiffGeom}, we reported periodic patterns of self-accommodated martensite variants in a cube specimen. The periodic boundary conditions add artificial constraints by forcefully imposing the same value for all the degrees of freedom on the two opposite surfaces of the cube. Thus, periodic boundary conditions can alter the microstructure evolution on small domains. As the system size becomes larger, we expect the periodic-boundary effect will not be felt far from the boundaries. To reduce the effect of the periodic boundary conditions on the microstructure, we conduct numerical simulations on an SMA cube of side 210 nm, which is probably one of the biggest domains used for microstructure evolution of SMAs. Similar to Section \ref{sec:DiffGeom}, the SMA specimen is quenched to temperature corresponding to $ \tau  = -1.2$ and microstructures are allowed to evolve. The domain is discretized using 170 uniform \cone-quadratic B-spline basis functions (168 elements) in each direction. The total number of degrees of freedom is approximately 19 millions, which would be approximately 12 times larger for a mixed formulation using the same number of quadratic Lagrange elements in the classical finite element method.

\begin{figure}[h!]
\centering
\subfigure[\textit{t} = 0.09 ns]
{
\includegraphics[trim = 25mm 20mm 30mm 40mm, clip, width=0.22\linewidth]{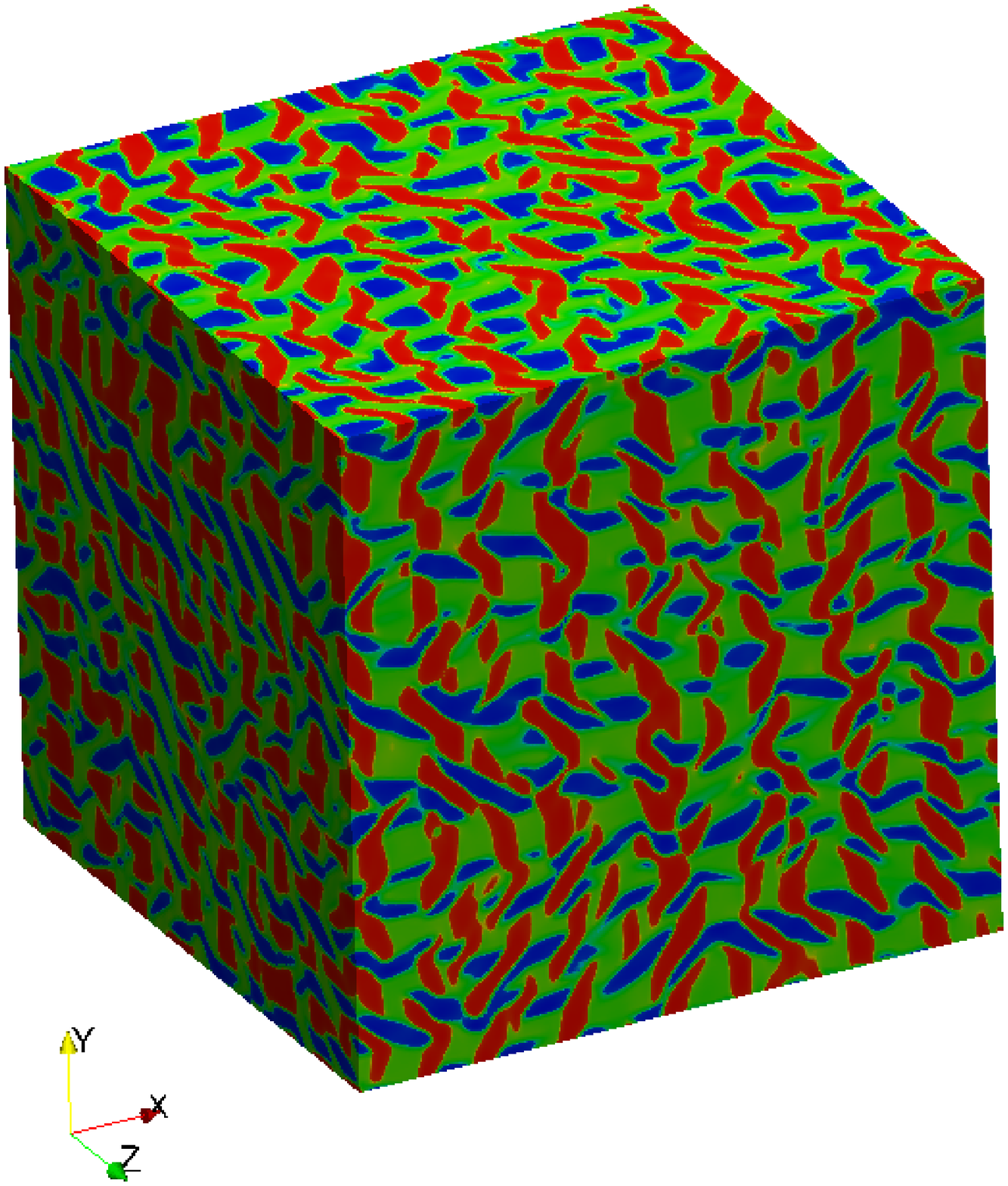}
}
\subfigure[\textit{t} = 0.27 ns]
{
\includegraphics[trim = 25mm 20mm 30mm 40mm, clip, width=0.22\textwidth]{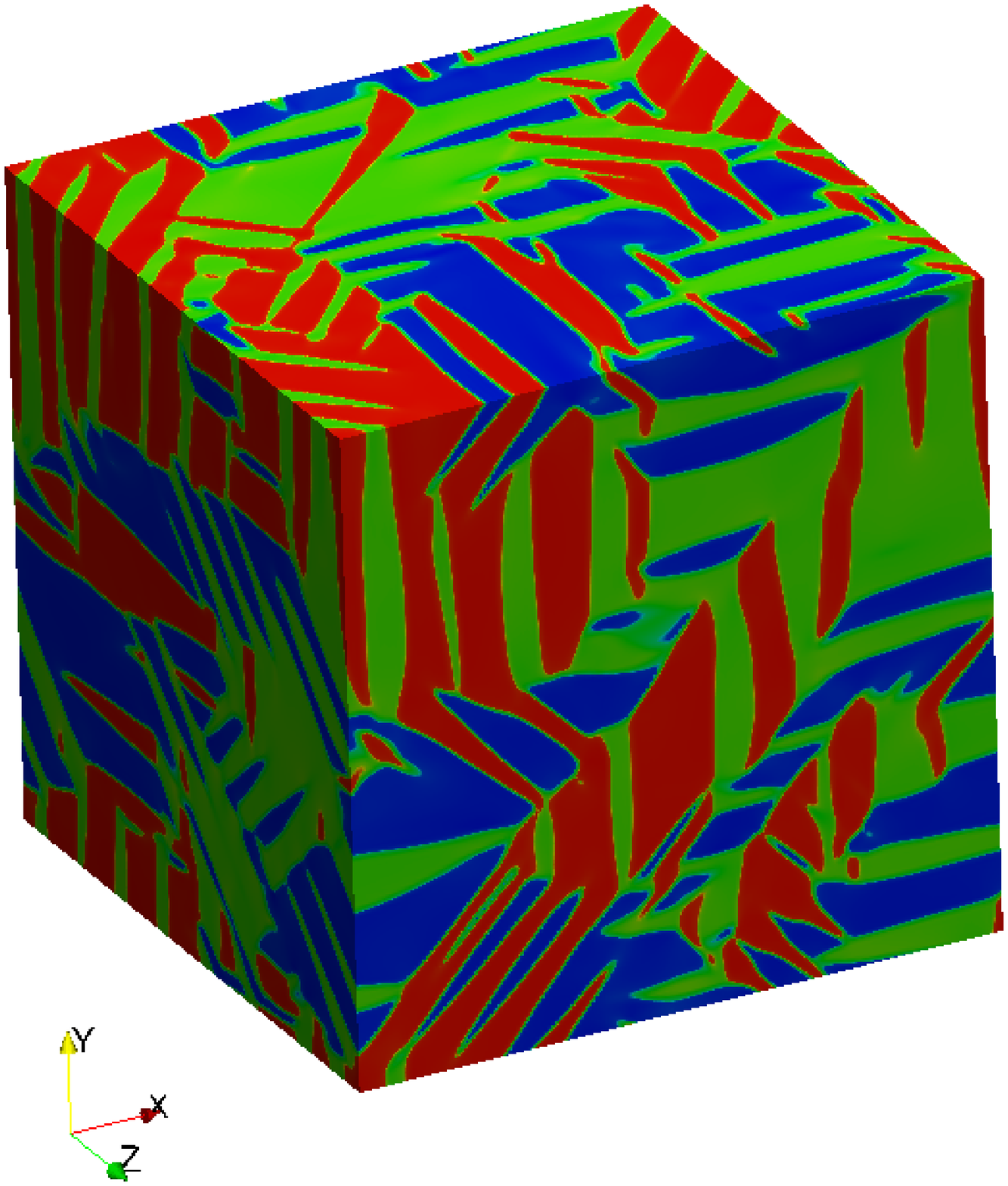}
}
\subfigure[\textit{t} = 0.45 ns]
{
\includegraphics[trim = 25mm 20mm 30mm 40mm, clip, width=0.22\textwidth]{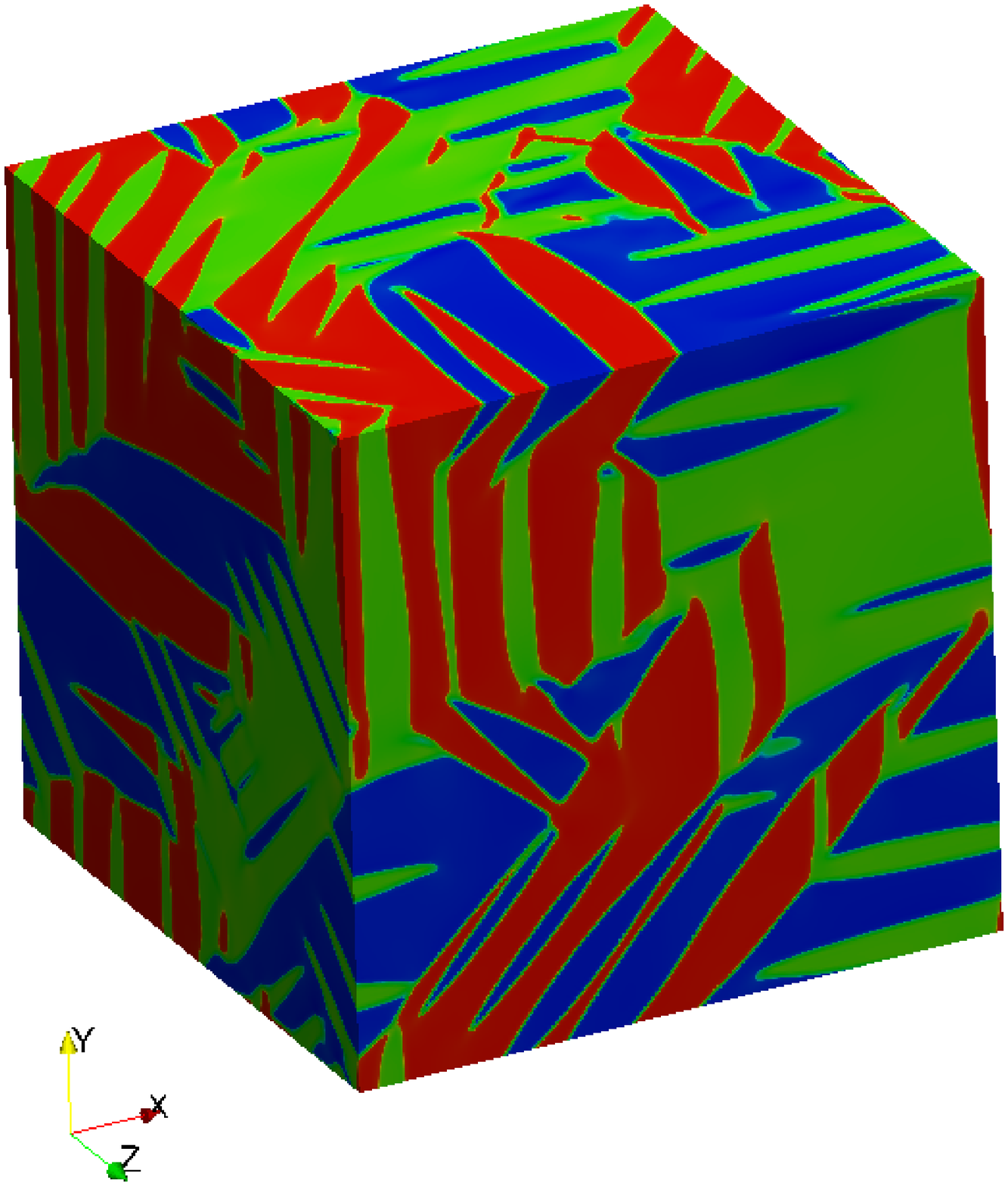}
}
\subfigure[\textit{t} = 1.135 ns]
{
	\includegraphics[trim = 25mm 20mm 30mm 40mm, clip, width=0.22\textwidth]{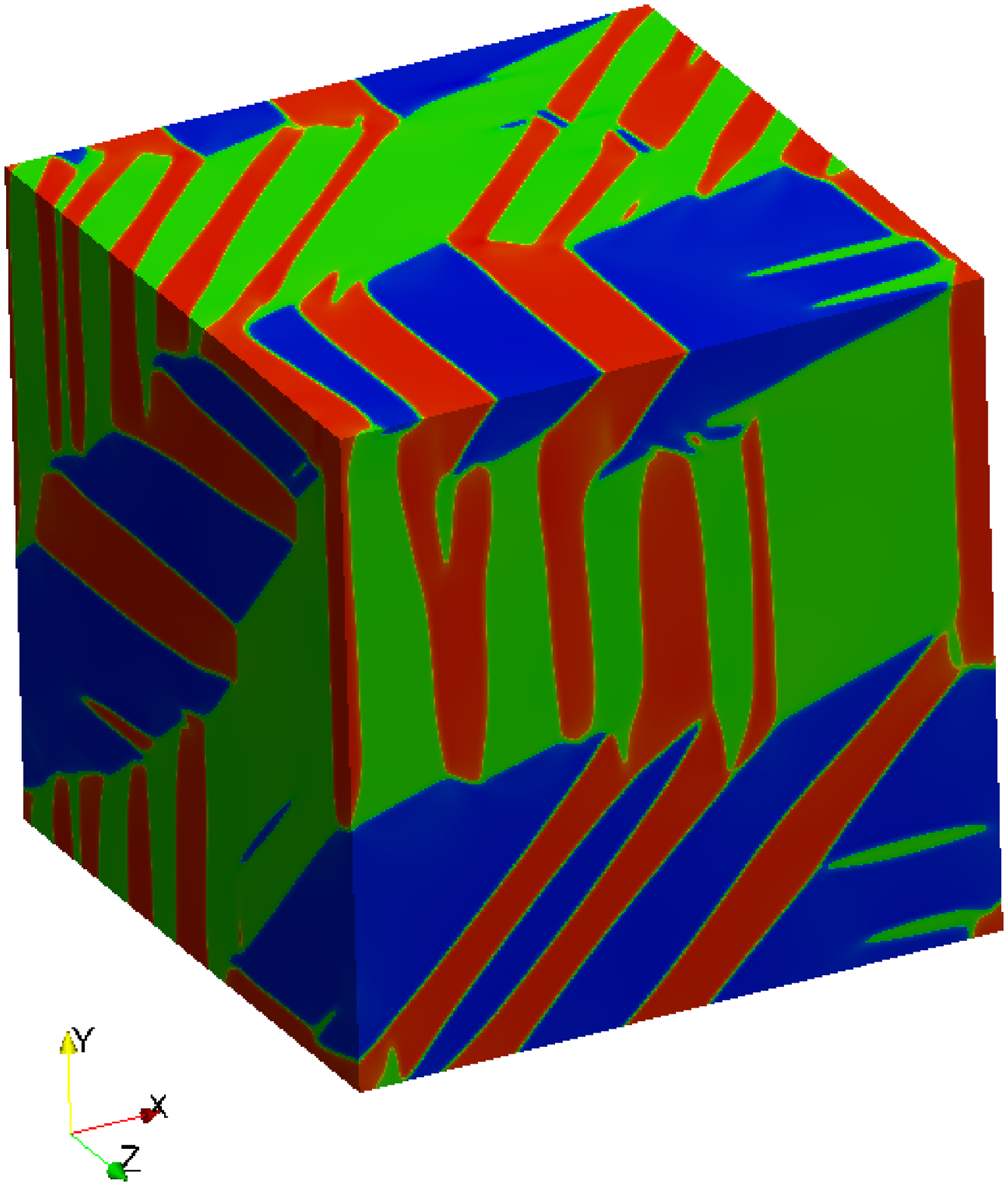}
}	
\caption{(Color online) Self-accommodated microstructure on a 210 nm side cube  (red, blue, and green colors represent M$_1$, M$_2$, and M$_3$ variants, respectively).}
\label{fig:BigCube}
\end{figure}

Fig. \ref{fig:BigCube} shows time snapshots of the transient microstructure evolution in the bigger cube domain. The martensitic variants nucleate at several places and they coalesce to form bigger domains. The smaller domain shows the regular herringbone patterns of variants at the stabilized state (Fig. \ref{fig:DiffGeomCube}(e)), however the big domain does not show regular patterns of the variants. Several features like needle twins, M$_i$-M$_i$ and M$_i$-M$_j$ martensitic variants collision, and a split tip morphology have been revealed. Such features have also been reported in \cite{jacobs2003simulations} and the experimental references therein. The time evolution of the average temperature coefficient $ \tau $ in the bigger cube specimen is shown in Fig. \ref{fig:DiffGeomTauEvolution}(b).  The simulation has also demonstrated the influence of periodic boundary conditions on the morphology evolution.

\section{Conclusions} \label{sec:Conclusions}
We have developed a 3D phase-field theory for modeling cubic-to-tetragonal phase transformations and thermo-mechanical behavior of SMAs. The model is numerically implemented in the IGA framework, which allows the straightforward solution to the fourth-order differential equations. The microstucture morphology and thermo-mechanical behavior is in qualitative agreement with the experiments and previously developed models from the literature. Simulation results indicate that IGA allows the use of relatively coarse meshes, thus permitting modeling bigger domain sizes. These features, along with modeling complex geometries, are useful in understanding microstructure morphology, which influences thermo-mechanical behavior during dynamic loading of SMA nanostructures in real-life application development. 

\section*{Acknowledgments}
RD, RM, and JZ have been supported by the NSERC and CRC program (RM), Canada. HG was partially supported by the European Research Council through the FP7 Ideas Starting Grant program (project \# 307201) and by \emph{Conseller\'ia de Educaci\'on e Ordenaci\'on Universitaria} (\emph{Xunta de Galicia}). Their support is gratefully acknowledged. This work was made possible with the facilities of the Shared Hierarchical Academic Research Computing Network (SHARCNET: www.sharcnet.ca) and Compute/Calcul Canada.
\appendix

\section{Dimensionless form of the governing equations} \label{app:Rescaling}
We rescale the equations Eqs. (\ref{eq:StrongForm3D}) to a dimensionless form by using the following change of variables:
\begin{eqnarray}
e_i = e_c \bar{e_i}, \qquad
u_i = e_c \delta \bar{u_i}, \qquad
x = \delta \bar{x}, \qquad
\mathscr{F} = \mathscr{F}_c \bar{\mathscr{F}}, \qquad
t = t_c \bar{t}, \qquad
\theta = \theta_c \bar{\theta}.
\label{eq:rescaleconst}
\end{eqnarray}

The governing thermo-mechanical Eqs. (\ref{eq:structuralconservation3D})--(\ref{eq:thermalconservation3D}) can now be converted to the dimensionless form as:
\begin{subequations}
\begin{eqnarray}
\rho \bar{u}_{\bar{i},\bar{t}\bar{t}}  && =\bar{\sigma}_{\bar{i}\bar{j},\bar{j}}+  \bar{\eta} \bar{\sigma}^{\prime}_{\bar{i}\bar{j},\bar{j}}  +\bar{\mu}_{\bar{i}\bar{j},\bar{k}\bar{k}\bar{j}} +\bar{f}_{\bar{i}}, \\
\bar{C}_v \bar{\theta}_{,\bar{t}} && =\bar{\kappa} \bar{\theta}_{,\bar{i}\bar{i}}+ \bar{\Xi} \bar{\theta} \left(\bar{u}_{\bar{i},\bar{i}} \bar{u}_{\bar{j},\bar{j} \bar{t}}-3 \bar{u}_{\bar{i},\bar{i}} \bar{u}_{\bar{i},\bar{i} \bar{t}}\right) + \bar{g},
\end{eqnarray}
\end{subequations}
with
\begin{eqnarray}
\delta = \sqrt{\frac{k_g}{a_{0}}}, \qquad \bar{a}_1 = \frac{a_1}{a_{0}}, 
\qquad \bar{a}_2 = \frac{a_2}{a_{0}}, \qquad \bar{a}_4 = 2, \qquad 
\bar{a}_5 = 1, \qquad \bar{k}_g = 1,  \nonumber \\
\mathscr{F}_c = \delta^2 e_c^2 a_{0}, \quad
\bar{\eta} = \frac{\eta}{a_0} \sqrt{\frac{a_0}{\rho \delta^2}}, \quad 
\bar{C_v} = \frac{\rho C_v \tau}{t_c}, \quad
\bar{\kappa} = \frac{\kappa \tau}{\delta^2 \bar{C_v}}, \quad 
\bar{\Xi} = -\frac{2}{3} \frac{a_0 e_c^2}{t_c \tau \bar{C_v}}.
\label{eq:rescaleconstval1}
\end{eqnarray}

The rescaled free energy takes on the form:
\begin{eqnarray}
\bar{\mathscr{F}} &=& \int_{\bar{\Omega}}\Big(\displaystyle 
\frac{\bar{a}_1}{2} \left( \bar{e}_1 \right)^2 
+ \frac{\bar{a}_2}{2} (\bar{e}_4^2 + \bar{e}_5^2 
+\bar{e}_6^2) + \bar{a}_3 \tau  (\bar{e}_2^2 + \bar{e}_3^2) + \bar{a}_4 
\bar{e}_3 (\bar{e}_3^2 - 3 \bar{e}_2^2) \nonumber \\
&& + \bar{a}_5 (\bar{e}_2^2 + \bar{e}_3^2)^2 + \frac{\bar{k_g}}{2} 
\left[ |\nabla \bar{e}_2|^2 + |\nabla \bar{e}_3|^2 \right]\Big)\der\bar{\Omega}.
\label{eq:rescaleFE}
\end{eqnarray}





\begingroup
\setstretch{0.8}
\bibliographystyle{unsrt}
\bibliography{IGAComputationalarXiv}

\begin{thebibliography}{10}

\bibitem{vzuvzek2012electrochemical}
K.~Ro{\v{z}}man, D.~Pe{\v{c}}ko, S.~{\v{S}}turm, U.~Maver, P.~Nadrah, M.~Bele,
  and S.~Kobe.
\newblock {Electrochemical synthesis and characterization of Fe$_{70}$Pd$_{30}$
  nanotubes for drug-delivery applications}.
\newblock {\em Materials Chemistry and Physics}, 133(1):218--224, 2012.

\bibitem{bhattacharya2005material}
K.~Bhattacharya and R.D. James.
\newblock The material is the machine.
\newblock {\em Science}, 307(5706):53--54, 2005.

\bibitem{koig2010micro}
D.~Ko¨nig, M.~Ehmann, S.~Thienhaus, and A.~Ludwig.
\newblock Micro-to nanostructured devices for the characterization of scaling
  effects in shape-memory thin films.
\newblock {\em Journal of Microelectromechanical Systems}, 19(5):1264--1269,
  2010.

\bibitem{bayer2011carbon}
B.~Bayer, S.~Sanjabi, C.~Baehtz, C.~Wirth, S.~Esconjauregui, R.~Weatherup,
  Z.~Barber, S.~Hofmann, and J.~Robertson.
\newblock {Carbon nanotube forest growth on NiTi shape memory alloy thin films
  for thermal actuation}.
\newblock {\em Thin Solid Films}, 519(18):6126--6129, 2011.

\bibitem{clements2003wireless}
K.~Clements.
\newblock Wireless technique for microactivation, 2003.
\newblock {US Patent 6,588,208}.

\bibitem{san2008superelasticity}
J.~Juan, M.~N{\'o}, and C.~Schuh.
\newblock Superelasticity and shape memory in micro-and nanometer-scale
  pillars.
\newblock {\em Advanced Materials}, 20(2):272--278, 2008.

\bibitem{san2009nanoscale}
J.~Juan, M.~N{\'o}, and C.~Schuh.
\newblock Nanoscale shape-memory alloys for ultrahigh mechanical damping.
\newblock {\em Nature Nanotechnology}, 4(7):415--419, 2009.

\bibitem{Khandelwal2009}
A.~Khandelwal and V.~Buravalla.
\newblock {Models for Shape Memory Alloy Behavior: An overview of modelling
  approaches}.
\newblock {\em International Journal of Structural Changes in Solids -
  Mechanics and Applications}, 1(1):111--148, 2009.

\bibitem{mamivand2013review}
M.~Mamivand, M.~Zaeem, and H.~El~Kadiri.
\newblock A review on phase field modeling of martensitic phase transformation.
\newblock {\em Computational Materials Science}, 77:304--311, 2013.

\bibitem{auricchio2009macroscopic}
F~Auricchio, A~Reali, and U~Stefanelli.
\newblock A macroscopic 1d model for shape memory alloys including asymmetric
  behaviors and transformation-dependent elastic properties.
\newblock {\em {Computer Methods in Applied Mechanics and Engineering}},
  198(17):1631--1637, 2009.

\bibitem{auricchio2010shape}
F.~Auricchio, A.~Reali, and A.~Tardugno.
\newblock {Shape-memory alloys: effective 3D modelling, computational aspects
  and design of devices}.
\newblock {\em {International Journal of Computational Materials Science and
  Surface Engineering}}, 3(2):199--223, 2010.

\bibitem{auricchio2011three}
F.~Auricchio, A.~Bessoud, A.~Reali, and U.~Stefanelli.
\newblock A three-dimensional phenomenological model for magnetic shape memory
  alloys.
\newblock {\em {GAMM-Mitteilungen}}, 34(1):90--96, 2011.

\bibitem{sengupta2012multiscale}
A.~Sengupta, P.~Papadopoulos, and R.~L Taylor.
\newblock A multiscale finite element method for modeling fully coupled
  thermomechanical problems in solids.
\newblock {\em International Journal for Numerical Methods in Engineering},
  91(13):1386--1405, 2012.

\bibitem{Ahluwalia2006}
R.~Ahluwalia, T.~Lookman, and A.~Saxena.
\newblock {Dynamic Strain Loading of Cubic to Tetragonal Martensites}.
\newblock {\em {Acta Materialia}}, 54:2109--2120, 2006.

\bibitem{Bouville2008}
M.~Bouville and R.~Ahluwalia.
\newblock {Microstructure and Mechanical Properties of Constrained Shape Memory
  Alloy Nanograins and Nanowires}.
\newblock {\em Acta Mater.}, 56(14):3558--3567, 2008.

\bibitem{Idesman2008}
A.~Idesman, J.~Cho, and V.~Levitas.
\newblock {Finite Element Modeling of Dynamics of Martensitic Phase
  Transitions}.
\newblock {\em Appl. Phys. Lett.}, 93(4):043102, 2008.

\bibitem{Bouville2009}
M.~Bouville and R.~Ahluwalia.
\newblock {Phase field simulations of coupled phase transformations in
  ferroelastic-ferroelastic nanocomposites}.
\newblock {\em Phys. Rev. B}, 79(9):094110, 2009.

\bibitem{wang1997three}
Y.~Wang and AG~Khachaturyan.
\newblock Three-dimensional field model and computer modeling of martensitic
  transformations.
\newblock {\em {Acta Materialia}}, 45(2):759--773, 1997.

\bibitem{Levitas2002a}
V.I. Levitas and D.L. Preston.
\newblock Three-dimensional landau theory for multivariant stress-induced
  martensitic phase transformations. i. austenite $\leftrightarrow$ martensite.
\newblock {\em Physical Review B}, 66(13):134206, 2002.

\bibitem{Barsch1984}
G.~Barsch and J.~Krumhansl.
\newblock {Twin Boundaries in Ferroelastic Media without Interface
  Dislocations}.
\newblock {\em Phys. Rev. Lett.}, 53(11):1069--1072, Sep 1984.

\bibitem{Levitas2002b}
V.~Levitas and D.~Preston.
\newblock {Three-dimensional Landau theory for multivariant stress-induced
  martensitic phase transformations. II. Multivariant phase transformations and
  stress space analysis}.
\newblock {\em Physics Review B}, 66(134206):1--15, 2002.

\bibitem{Levitas2003}
V.~Levitas, D.~Preston, and D.~Lee.
\newblock {Three-dimensional Landau theory for multivariant stress-induced
  martensitic phase transformations. III. Alternative potentials, critical
  nuclei, kink solutions, and dislocation theory}.
\newblock {\em Physics Review B}, 68(134201):1--24, 2003.

\bibitem{ni2003three}
Y~Ni, LH~He, and L~Yin.
\newblock Three-dimensional phase field modeling of phase separation in
  strained alloys.
\newblock {\em {Materials Chemistry and Physics}}, 78(2):442--447, 2003.

\bibitem{seol2003computer}
DJ~Seol, SY~Hu, YL~Li, J~Shen, KH~Oh, and LQ~Chen.
\newblock Computer simulation of spinodal decomposition in constrained films.
\newblock {\em {Acta Materialia}}, 51(17):5173--5185, 2003.

\bibitem{man2011study}
J.~Man, J.~Zhang, Y.~Rong, and N.~Zhou.
\newblock Study of thermoelastic martensitic transformations using a
  phase-field model.
\newblock {\em Metallurgical and Materials Transactions A}, 42(5):1154--1164,
  2011.

\bibitem{yeddu2012three}
H.K. Yeddu, A.~Malik, J.~Ågren, G.~Amberg, and A.~Borgenstam.
\newblock Three-dimensional phase-field modeling of martensitic microstructure
  evolution in steels.
\newblock {\em Acta Materialia}, 60(4):1538--1547, 2012.

\bibitem{jacobs2003simulations}
AE~Jacobs, SH~Curnoe, and RC~Desai.
\newblock Simulations of cubic-tetragonal ferroelastics.
\newblock {\em Physical Review B}, 68(22):224104, 2003.

\bibitem{melnik2000computing}
R.~Melnik, A.~Roberts, and K.~Thomas.
\newblock Computing dynamics of copper-based sma via centre manifold reduction
  of 3d models.
\newblock {\em Computational materials science}, 18(3):255--268, 2000.

\bibitem{melnik1999modeling}
R.~Melnik, A.~Roberts, and K.~Thomas.
\newblock Modeling dynamics of shape memory alloys via computer algebra.
\newblock In {\em 1999 Symposium on Smart Structures and Materials}, pages
  290--301. International Society for Optics and Photonics, 1999.

\bibitem{melnik2001coupled}
R.~Melnik, A.~Roberts, and K.~Thomas.
\newblock Coupled thermomechanical dynamics of phase transitions in shape
  memory alloys and related hysteresis phenomena.
\newblock {\em Mechanics Research Communications}, 28(6):637--651, 2001.

\bibitem{hughes2005isogeometric}
T.~Hughes, J.~Cottrell, and Y.~Bazilevs.
\newblock {Isogeometric analysis: CAD, finite elements, NURBS, exact geometry
  and mesh refinement}.
\newblock {\em {Computer Methods in Applied Mechanics and Engineering}},
  194(39):4135--4195, 2005.

\bibitem{Hughes}
J.~Cottrell, T.~Hughes, and Y.~Bazilevs.
\newblock {\em Isogeometric Analysis: Toward Integration of CAD and FEA}.
\newblock John Wiley \& Sons., 2009.

\bibitem{TS1}
R.~Dimitri, L.~De Lorenzis, M.A. Scott, P.~Wriggers, R.L. Taylor, and
  G.~Zavarise.
\newblock {Isogeometric large deformation frictionless contact using
  T-splines}.
\newblock {\em {Computer Methods in Applied Mechanics and Engineering}},
  269(0):394 -- 414, 2014.

\bibitem{TS2}
M.A. Scott, R.N. Simpson, J.A. Evans, S.~Lipton, S.P.A. Bordas, T.J.R. Hughes,
  and T.W. Sederberg.
\newblock {Isogeometric boundary element analysis using unstructured
  T-splines}.
\newblock {\em {Computer Methods in Applied Mechanics and Engineering}},
  254(0):197 -- 221, 2013.

\bibitem{TS3}
M.A. Scott, X.~Li, T.W. Sederberg, and T.J.R. Hughes.
\newblock {Local refinement of analysis-suitable T-splines}.
\newblock {\em {Computer Methods in Applied Mechanics and Engineering}},
  213–216(0):206 -- 222, 2012.

\bibitem{TS4}
Y.~Bazilevs, V.~Calo, J.~Cottrell, J.Evans, T.~Hughes, S.~Lipton, M.~Scott, and
  T.~Sederberg.
\newblock {Isogeometric analysis using T-splines}.
\newblock {\em {Computer Methods in Applied Mechanics and Engineering}},
  199(5-8):229 -- 263, 2010.

\bibitem{Evans2013141}
J.~Evans and T.~Hughes.
\newblock {Isogeometric divergence-conforming B-splines for the unsteady Navier
  Stokes equations}.
\newblock {\em Journal of Computational Physics}, 241(0):141 -- 167, 2013.

\bibitem{Liu2013321}
J.~Liu, L.~Dede', J.~Evans, M.~Borden, and T.~Hughes.
\newblock Isogeometric analysis of the advective cahn–hilliard equation:
  Spinodal decomposition under shear flow.
\newblock {\em Journal of Computational Physics}, 242(0):321 -- 350, 2013.

\bibitem{Liu201347}
J.~Liu, H.Gomez, J.~Evans, T.~Hughes, and C.~Landis.
\newblock {Functional entropy variables: A new methodology for deriving
  thermodynamically consistent algorithms for complex fluids, with particular
  reference to the isothermal Navier-Stokes-Korteweg equations}.
\newblock {\em Journal of Computational Physics}, 248(0):47 -- 86, 2013.

\bibitem{Bazilevs2007173}
Y.~Bazilevs, V.M. Calo, J.A. Cottrell, T.J.R. Hughes, A.~Reali, and
  G.~Scovazzi.
\newblock Variational multiscale residual-based turbulence modeling for large
  eddy simulation of incompressible flows.
\newblock {\em Computer Methods in Applied Mechanics and Engineering},
  197(1–4):173 -- 201, 2007.

\bibitem{Cottrell20065257}
J.~Cottrell, A.~Reali, Y.~Bazilevs, and T.~Hughes.
\newblock Isogeometric analysis of structural vibrations.
\newblock {\em Computer Methods in Applied Mechanics and Engineering},
  195(41–43):5257 -- 5296, 2006.

\bibitem{Cottrell20074160}
J.~Cottrell, T.~Hughes, and A.~Reali.
\newblock Studies of refinement and continuity in isogeometric structural
  analysis.
\newblock {\em Computer Methods in Applied Mechanics and Engineering},
  196(41–44):4160 -- 4183, 2007.

\bibitem{Hughes20084104}
T.J.R. Hughes, A.~Reali, and G.~Sangalli.
\newblock Duality and unified analysis of discrete approximations in structural
  dynamics and wave propagation: Comparison of p-method finite elements with
  k-method nurbs.
\newblock {\em Computer Methods in Applied Mechanics and Engineering},
  197(49–50):4104 -- 4124, 2008.

\bibitem{Auricchio2007160}
F.~Auricchio, L.~Beirao da~Veiga, A.~Buffa, C.~Lovadina, A.~Reali, and
  G.~Sangalli.
\newblock A fully “locking-free” isogeometric approach for plane linear
  elasticity problems: A stream function formulation.
\newblock {\em Computer Methods in Applied Mechanics and Engineering},
  197(1–4):160 -- 172, 2007.

\bibitem{NME:NME3159}
L.~De~Lorenzis, I.~Temizer, P.~Wriggers, and G.~Zavarise.
\newblock A large deformation frictional contact formulation using nurbs-based
  isogeometric analysis.
\newblock {\em International Journal for Numerical Methods in Engineering},
  87(13):1278--1300, 2011.

\bibitem{laura}
L~De~Lorenzis, P~Wriggers, and G~Zavarise.
\newblock {A mortar formulation for 3D large deformation contact using
  NURBS-based isogeometric analysis and the augmented Lagrangian method}.
\newblock {\em Computational Mechanics}, 49(1):1--20, 2012.

\bibitem{fsi1}
Y.~Bazilevs, V.M. Calo, T.J.R. Hughes, and Y.~Zhang.
\newblock Isogeometric fluid-structure interaction: theory, algorithms, and
  computations.
\newblock {\em Computational Mechanics}, 43(1):3--37, 2008.

\bibitem{fsi2}
Y.~Bazilevs, V.~Calo, Y.~Zhang, and T.~Hughes.
\newblock Isogeometric fluid--structure interaction analysis with applications
  to arterial blood flow.
\newblock {\em Computational Mechanics}, 38(4-5):310--322, 2006.

\bibitem{cmp1}
H.~Gomez and X.~Nogueira.
\newblock An unconditionally energy-stable method for the phase field crystal
  equation.
\newblock {\em Computer Methods in Applied Mechanics and Engineering},
  249–252(0):52 -- 61, 2012.

\bibitem{cmp2}
H.~Gomez and J.~Paris.
\newblock {Numerical simulation of asymptotic states of the damped
  Kuramoto-Sivashinsky equation}.
\newblock {\em Phys. Rev. E}, 83:046702, Apr 2011.

\bibitem{cmp3}
U.~Thiele, A.~Archer, M.~Robbins, H.~Gomez, and E.~Knobloch.
\newblock {Localized states in the conserved Swift-Hohenberg equation with
  cubic nonlinearity}.
\newblock {\em Phys. Rev. E}, 87:042915, Apr 2013.

\bibitem{cmp4}
H.~Gomez, L.~Cueto-Felgueroso, and R.~Juanes.
\newblock Three-dimensional simulation of unstable gravity-driven infiltration
  of water into a porous medium.
\newblock {\em {Journal of Computational Physics}}, 238(0):217 -- 239, 2013.

\bibitem{Evans20091726}
J.~Evans, Y.~Bazilevs, I.~Babuška, and T.~Hughes.
\newblock n-widths, sup–infs, and optimality ratios for the k-version of the
  isogeometric finite element method.
\newblock {\em {Computer Methods in Applied Mechanics and Engineering}},
  198(21–26):1726 -- 1741, 2009.

\bibitem{Schillinger2013170}
D.~Schillinger, J.~Evans, A.~Reali, M.~Scott, and T.~Hughes.
\newblock {Isogeometric collocation: Cost comparison with Galerkin methods and
  extension to adaptive hierarchical NURBS discretizations}.
\newblock {\em {Computer Methods in Applied Mechanics and Engineering}},
  267(0):170 -- 232, 2013.

\bibitem{borden2012phase}
M.~Borden, C.~Verhoosel, M.~Scott, T.~Hughes, and C.~Landis.
\newblock A phase-field description of dynamic brittle fracture.
\newblock {\em Computer Methods in Applied Mechanics and Engineering},
  217:77--95, 2012.

\bibitem{dede2012isogeometric}
L.~Ded{\`e}, M.~Borden, and T.~Hughes.
\newblock Isogeometric analysis for topology optimization with a phase field
  model.
\newblock {\em Archives of Computational Methods in Engineering},
  19(3):427--465, 2012.

\bibitem{vilanovaisogeometric}
G.~Vilanova, I.~Colominas, and H.~Gomez.
\newblock Capillary networks in tumor angiogenesis: From discrete endothelial
  cells to phase-field averaged descriptions via isogeometric analysis.
\newblock {\em International Journal for Numerical Methods in Biomedical
  Engineering}, 2013.

\bibitem{ho1}
J.~Kiendl, K.-U. Bletzinger, J.~Linhard, and R.~Wuchner.
\newblock {Isogeometric shell analysis with Kirchhoff-Love elements}.
\newblock {\em {Computer Methods in Applied Mechanics and Engineering}},
  198(49–52):3902 -- 3914, 2009.

\bibitem{ho2}
J.~Kiendl, Y.~Bazilevs, M.-C. Hsu, R.~Wuchner, and K.-U. Bletzinger.
\newblock {The bending strip method for isogeometric analysis of
  Kirchhoff–Love shell structures comprised of multiple patches}.
\newblock {\em {Computer Methods in Applied Mechanics and Engineering}},
  199(37-40):2403 -- 2416, 2010.

\bibitem{ho3}
F.~Auricchio, L.~Beirao da~Veiga, A.~Buffa, C.~Lovadina, A.~Reali, and
  G.~Sangalli.
\newblock {A fully locking-free isogeometric approach for plane linear
  elasticity problems: A stream function formulation}.
\newblock {\em {Computer Methods in Applied Mechanics and Engineering}},
  197(1-4):160 -- 172, 2007.

\bibitem{ho4}
P.~Fischer, M.~Klassen, J.~Mergheim, P.~Steinmann, and R.~M{\"u}ller.
\newblock Isogeometric analysis of 2d gradient elasticity.
\newblock {\em Computational Mechanics}, 47(3):325--334, 2011.

\bibitem{anders2012computational}
D.~Anders, C.~Hesch, and K.~Weinberg.
\newblock Computational modeling of phase separation and coarsening in solder
  alloys.
\newblock {\em International Journal of Solids and Structures},
  49(13):1557--1572, 2012.

\bibitem{anders2012isogeometric}
D.~Anders, K.~Weinberg, and R.~Reichardt.
\newblock Isogeometric analysis of thermal diffusion in binary blends.
\newblock {\em Computational Materials Science}, 52(1):182--188, 2012.

\bibitem{Gomez2014}
H.~Gomez, A.~Reali, and G.~Sangalli.
\newblock Accurate, efficient, and (iso) geometrically flexible collocation
  methods for phase-field models.
\newblock {\em {Journal of Computational Physics}}, 2014.

\bibitem{dhote2013PCS}
R.~Dhote, H.~Gomez, R.~Melnik, and J.~Zu.
\newblock Isogeometric analysis of coupled thermo-mechanical phase-field models
  for shape memory alloys using distributed computing.
\newblock {\em Procedia Computer Science}, 18:1068--1076, 2013.

\bibitem{Dhote2013CM}
R.~Dhote, H.~Gomez, R.~Melnik, and J.~Zu.
\newblock {Isogeometric Analysis of a Dynamic Thermo-Mechanical Phase-Field
  Model Applied to Shape Memory Alloys}.
\newblock {\em Computational Mechanics}, pages 1--16, 2014.

\bibitem{RPD_physics_paper}
R.~Dhote, H.~Gomez, R.~Melnik, and J.~Zu.
\newblock {Modeling Shape Memory Alloy Nanostructures with 3D Coupled Dynamic
  Phase-Field Models}.
\newblock 2014.
\newblock (to be submitted).

\bibitem{chung1993time}
J.~Chung and G.~Hulbert.
\newblock A time integration algorithm for structural dynamics with improved
  numerical dissipation: the generalized-$\alpha$ method.
\newblock {\em {Journal of Applied Mechanics}}, 60(2):371--375, 1993.

\bibitem{Jansen2000305}
K.~Jansen, C.~Whiting, and G.~Hulbert.
\newblock {A generalized-$\alpha$ method for integrating the filtered
  Navier-Stokes equations with a stabilized finite element method}.
\newblock {\em Computer Methods in Applied Mechanics and Engineering},
  190(3–4):305 -- 319, 2000.

\bibitem{akkerman2008role}
I.~Akkerman, Y.~Bazilevs, VM~Calo, T.J.R. Hughes, and S.~Hulshoff.
\newblock The role of continuity in residual-based variational multiscale
  modeling of turbulence.
\newblock {\em Computational Mechanics}, 41(3):371--378, 2008.

\bibitem{bazilevs2007variational}
Y.~Bazilevs, V.~Calo, J.~Cottrell, T.~Hughes, A.~Reali, and G.~Scovazzi.
\newblock Variational multiscale residual-based turbulence modeling for large
  eddy simulation of incompressible flows.
\newblock {\em Computer Methods in Applied Mechanics and Engineering},
  197(1-4):173--201, 2007.

\bibitem{Gomez2008}
H.~Gomez, V.~Calo, Y.~Bazilevs, and T.~Hughes.
\newblock {Isogeometric analysis of the Cahn-Hillard phase-field model}.
\newblock {\em Computer Methods in Applied Mechanics and Engineering},
  197(49-50):4333--4352, 2008.

\bibitem{gomez2010isogeometric}
H.~Gomez, T.J.R. Hughes, X.~Nogueira, and V.M. Calo.
\newblock {Isogeometric analysis of the isothermal Navier--Stokes--Korteweg
  equations}.
\newblock {\em Computer Methods in Applied Mechanics and Engineering},
  199(25):1828--1840, 2010.

\bibitem{provably}
H.~Gomez and T.~Hughes.
\newblock Provably unconditionally stable, second-order time-accurate, mixed
  variational methods for phase-field models.
\newblock {\em {Journal of Computational Physics}}, 230(13):5310 -- 5327, 2011.

\bibitem{vlasova2010ferroelastic}
N.~Vlasova, N.~Shchegoleva, A.~Popov, and G.~Kandaurova.
\newblock {Ferroelastic domains and phases in ferromagnetic nanostructured FePd
  alloy}.
\newblock {\em The Physics of Metals and Metallography}, 110(5):449--463, 2010.

\bibitem{Sapriel1975}
J.~Sapriel.
\newblock Domain-wall orientations in ferroelastics.
\newblock {\em Physical Review B}, 12(11):5128, 1975.

\bibitem{he2009scaling}
Y.~He and Q.~Sun.
\newblock {Scaling relationship on macroscopic helical domains in NiTi tubes}.
\newblock {\em International Journal of Solids and Structures},
  46(24):4242--4251, 2009.

\end{thebibliography}
\endgroup

\end{document}